# HUMAN FACTORS IN AGILE SOFTWARE DEVELOPMENT

**[Book Draft]**

**LIN JUN**

**Junlin@ntu.edu.sg / linjun@buaa.edu.cn**

**www.linjun.net.cn**

**2015**

# PREFACE

Software is made by humans and works for humans. During the dynamic, continuous, complex and chaotic software development process, human is at the core. Through observing historical evolutions and empirical studies of software engineering methodologies, we can see that as more human factors are considered in the development process, the methodology has become more successful in the real world.

*Agile Software Development* (ASD) is one of the methodologies with successful use of human factors. ASD is a relatively new software development paradigm that has gained popularity over the last decade. It is a group of methodologies based on iterative and incremental software development, where requirements and functions evolve through collaboration between self-organizing and cross-functional teams. Existing ASD methodologies, which have focused on some human factors, such as openness, collaboration, proactivity, self-organizing, and communication etc., are more people oriented than other plan-driven methodologies. However, they lack good quantitative methods to analyze the impacts of human factors on the process, which rely heavily on project managers and team members' intuition and feeling. Moreover, according to our empirical studies, this intuitive and subjective decision making process without data analytics affects the progress, quality and productivity of ASD.

Through our four years experiments on students' Scrum based ASD process, we have gained deep understanding into the human factors of agile methodology. We designed an agile project management tool (APM) – the *Human-centered Agile Software Engineering* (HASE) collaboration development platform to support more than 400 students who self-organized into 80 teams to practice ASD. In these experiments, we focus on two research directions. The first research direction is to study the agile software development process, through data collections in the lab and interactions with the software developers during software engineering course studies. The second research direction is to design new collaboration tools for agile team members to deal with difficult situations such as detecting users' real goals from specific stories, allocating tasks to a novice or distributed team members based on the current contexts, and dealing with possible task delay or team member's emotional crisis. These two research directions are mutually reinforcing: the analysis of agile software development process inspires better development of the HASE platform, and the problems discovered through the usage of HASE platform shed light on the issues of the current agile software development process. These two aspects can be integrated as a human computation system (HCS), in which human and computer work together to solve problems in the agile software development process.

In this book, we are the first to treat agile process as a HCS and explore the usage of goal-oriented method, agent-oriented method and emotion modeling method to allow agile teams to quantitatively handle human factors during development process with fewer burdens. Our research mainly focuses the following questions in ASD practice:

- How to help product owners or project managers to distinguish stakeholders’ real goals hidden in user stories described in nature language and intuitively present them to the team?
- How to properly suggest developers to accept or refuse incoming tasks based on the current situations before a new iteration/sprint start?

- How does the developers' mood or team morale swing affect the task performing process and the quality of artifacts?

To address these three questions, we have conducted a series of experiments, simulations and analysis, and contributed a series of solutions and insights in this researches, including 1) a Goal Net based method to enhance goal and requirement management for ASD process, 2) a novel Simple Multi-Agent Real-Time (SMART) approach to enhance intelligent task allocation for ASD process, 3) a Fuzzy Cognitive Maps (FCMs) based method to enhance emotion and morale management for ASD process, 4) a large-scale in-depth empirical analysis of human factors in the agile development process through the continuous observation of student ASD teams, and 5) the first to identify ASD process as a human-computation system that exploit human efforts to perform tasks that computers are not good at solving. On the other hand, computers can assist human decision making in the ASD process.

# TABLE OF CONTENTS

# LIST OF FIGURES

# LIST OF TABLES

# NOMENCLATURE

| | |
|---|---|
| **AM** | Agile Modeling |
| **ASD** | Agile Software Development |
| **ASE** | Agile Software engineering |
| **AUP** | Agile Unified Process |
| **CMM** | Capability Maturity Model |
| **CMMI** | Capability Maturity Model Integration |
| **CSCW** | Computer Supported Collaborative Work |
| **DSDM** | Dynamic Systems Development Method |
| **EssUP** | Essential Unified Process |
| **FDD** | Feature Driven Development |
| **GET** | Goal-Environment-Task |
| **GO** | Goal-oriented |
| **HCS** | Human Computation System |
| **HCSE** | Human-Centered Software Engineering |
| **HF** | Human Factor |
| **LSD** | Lean Software Development |
| **OpenUP** | Open Unified Process |
| **RAD** | Rapid Application Development |
| **RUP** | Rational United Process |
| **SE** | Software engineering |
| **SMART** | Simple Multi-Agent Real-Time |
| **SPI** | Software Process Improvement |
| **SSE** | Social Software Engineering |
| **SWEBOK** | Software Engineering Body of Knowledge |
| **UCD** | User Centered Design |
| **USDP** | United Software Development Process |
| **VSD** | Velocity Software Development |
| **XP** | eXtreme Programming |

# CHAPTER 1

# INTRODUCTION

Human is the key determinant of success to software development [1]. Software development consists of a lot of human being's activities, including analyzing, thinking, decision making, communicating, designing, implementing, collaborating, and even personal morale swing etc. For decades, to solve the software crisis, a variety of software engineering methodologies have evolved over years, each with its own recognized strengths and weaknesses. One software development methodology framework is not necessarily suitable for use by all projects. Since then, researchers and practitioners have started to know that there is no silver bullet for software engineering [2]. After the turn of the century, with a boom of cloud-based applications and mobile development technologies, as well as the emergence of e-commerce extending people's daily life into the cyberspace, human factor research has become an important topic in software engineering research as well, especially for the field of agile methodologies research [3-11]. In this chapter, we provide an introduction to the evolutions of modern software engineering, current main researches on agile software development, the scope of research documented in this book and the contributions to this area of research.

## 1.1 The Concept and Evolutions of Modern Software Engineering

*Software engineering* (SE) is an engineering discipline that is concerned with all aspects of software production, which applies engineering theories and methodologies to the process of software development. Use of these SE methodologies enables software engineers and project managers to visualize the dynamic process of software development, and ultimately transform the human activities into a working set of code, data and documents. In 2004 version of *Software Engineering Body of Knowledge* (SWEBOK) guide [12], *IEEE Computer Society* listed the following four types of popular methods.

- *Heuristic Methods* are those experience-based software engineering methods, which have been and are fairly widely practiced in software industry, including structured analysis and design methods, data modeling methods, and object-oriented analysis and design methods.

- *Formal Methods* are those software engineering methods used to specify, develop, and verify the software through application of a rigorous mathematically based notation and language to avoid ambiguity, including specification languages, program refinement and derivation, formal verification, and logical inference.

- *Prototyping Methods* are the software engineering method that generally creates incomplete or minimally functional versions of a software application, usually for evaluating specific new features, soliciting feedback on requirements or user interfaces, further exploring requirements, design, or implementation options, and/or gaining some other useful insight into the software. Typically, the prototype does not become the final software product without extensive development rework or code refactoring.

- *Agile Methods* are considered lightweight methods in that they are characterized by short, iterative development cycles, self-organizing teams, simpler designs, code refactoring, test-driven development, frequent customer involvement, and an emphasis on creating a demonstrative working product with each development cycle, including *Pair*

*Programming*, *Rapid Application Development* (RAD), *eXtreme Programming* (XP), *Scrum*, and *Feature-Driven Development* (FDD) etc.

In practice, the *Software Development Process* (SDP) is a dynamic, continuous, incremental, and chaotic process that is hard to control every aspect by people. In order to assure the quality of software produced from the process, researchers and practitioners proposed a set of SDP models, which are given high hope to solve the software crisis since 1970s.

At the early stage, the plan-driven methodologies, which require strictly order and plan, such as waterfall model and spiral model etc., can be able to handle some large projects without frequently requirement changes. However, excessive planning and controlling bring high cost and risk, even led to the inhibition of users' requirements and rejection of new changes [13, 14]. After that, the research of use case-driven methodologies, which focuses on user case and incremental iteration, such as *United Software Development Process* (USDP) and *Rational United Process* (RUP) etc. [15], as well as the research of process improvement methodologies, which emphasizes organization and team maturity, such as *Capability Maturity Model* (CMM) and *Capability Maturity Model Integration* (CMMI) etc. [16-18], became very active and popular in both academia and industry. Those methodologies have been successful applied to some big projects. However, they are still too planned and costly to small and medium projects.

After the turn of new century, with the increasing number of small size projects, such as component/plugin based web applications, mobile applications and cloud-based SaaS applications etc., the *Agile Software Development* (ASD) methodologies, e.g. *XP* [19] and *Scrum* [20], have attracted strong interests of developers and researchers. As their lean, agile, and flexible characteristics are very suitable for responding to continuous changes and fast releases, especially in today's turbulent economic environment.

By studying the historical transition of software engineering methodologies, we can see that from the strictest plan-driven waterfall to the flexible human-centered agile, the trend is that with more human's characteristics and human factors are considered in the software development process, the methodology can become more successful in the real world [3]. Our motivation for the research on *Human Factors in Agile Software Development* is to follow this trend to get more empirical insights into current agile methodologies and try to improve them from the aspect of human factor.

## 1.2 The Concept of Agile Software Development

*Agile Software Development* (ASD) is a group of software development methods based on iterative and incremental development, where requirements and solutions evolve through collaboration between self-organizing, cross-functional teams. It promotes adaptive planning, evolutionary development and delivery, a time-boxed iterative approach, face to face communication and encourages rapid and flexible response to changes [21, 22].

In February 2001, 17 software developers met in Utah of USA to discuss lightweight development methods, and then they published the famous *Manifesto for Agile Software Development* (see at http://agilemanifesto.org):

> *"We are uncovering better ways of developing software by doing it and helping others do it. Through this work we have come to value:*
>
> - ***Individuals and interactions** over Processes and tools*
> - ***Working software** over Comprehensive documentation*

- ***Customer collaboration*** *over Contract negotiation*
- ***Responding to change*** *over Following a plan*

*That is, while there is value in the items on the right, we value the items on the left more.*

| | | |
|---|---|---|
| *Kent Beck* | *James Grenning* | *Robert C. Martin* |
| *Mike Beedle* | *Jim Highsmith* | *Steve Mellor* |
| *Arie van Bennekum* | *Andrew Hunt* | *Ken Schwaber* |
| *Alistair Cockburn* | *Ron Jeffries* | *Jeff Sutherland* |
| *Ward Cunningham* | *Jon Kern* | *Dave Thomas* |
| *Martin Fowler* | *Brian Marick"* | |

Since 2001, the year of announcement of the agile manifesto, the research community has devoted a great deal of attention to agile methodologies. A literature search in the *ISI Web of Science2* identified 1551research papers that were published between 2001 and 2010 on *ASD* [23].

During this period, Abrahamsson et al. (2002) [24], Cohen et al. (2004) [25], Erickson et al. (2005) [26], Dybå et al. (2008) [27], and Dingsøyr et al. (2010) [28] (2012) [23] gave the introductions to and overviews of agile methodologies respectively. These six reports describe the state-of-the-art and state-of-the-practice in terms of characteristics of the various agile methods, as well as lessons learned from applying such methods to industry at different stages.

### *1.2.1 Iterative and Incremental Software Development Process*

Current ASD methodologies provide a conceptual framework that promotes foreseen interactions and communications throughout the development cycle. Most of ASD methodologies are adaptive, active, iterative and incremental [29]. Those features help ASD process increase productivity and reduce risks. ASD involves more customer interaction and testing effort, which tries to satisfy customer through quickly and continuous delivery of working software. This is useful when developer don't have a clear idea of customer's goals.

Development activities in ASD process can be carried out using the iterative actions. Since ASD process has more iteration, so developer can assure if a small modification meets customer's goal or not. This is better than one build system in the plan-driven process. So it is more effective where customer frequently changes the requirement.

ASD methodologies attempt to provide many opportunities to assess the direction of a project throughout the development cycle. This is achieved through regular cadences of work, known as iteration (or sprint). At end of iteration the team must present a shippable increment of work. Thus, as the features of focusing on the repetition of abbreviated work cycles as well as the functional software, agile methodologies are described as "iterative" and "incremental". Figure 1 shows a general development cycle in iteration.

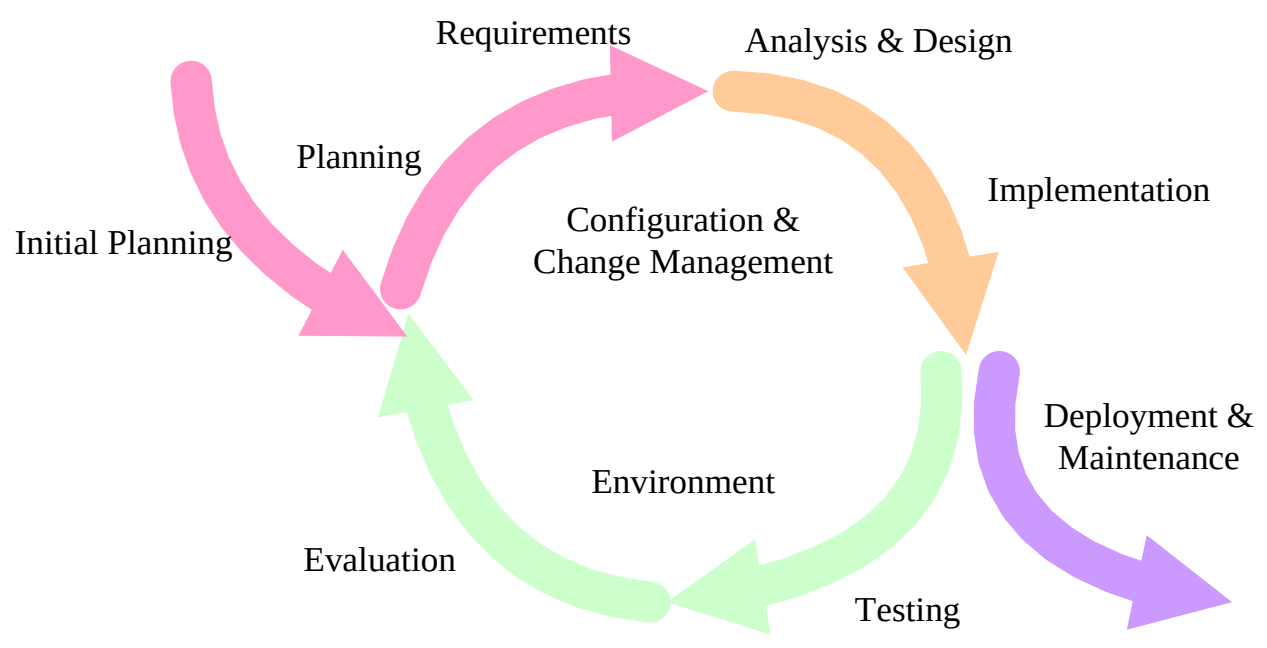

Figure 1. Iterative development model (source from wikipedia website)

In waterfall model or V model, shown in Figure 2, development teams only have one or a few chance to refine each stages of process. But for ASD process, every aspect of development, such as requirement, design, implementation, testing etc., is continually revisited throughout the iteration. When a team stops and re-evaluates the direction of a project every week or two weeks, there's always time to steer it into another direction.

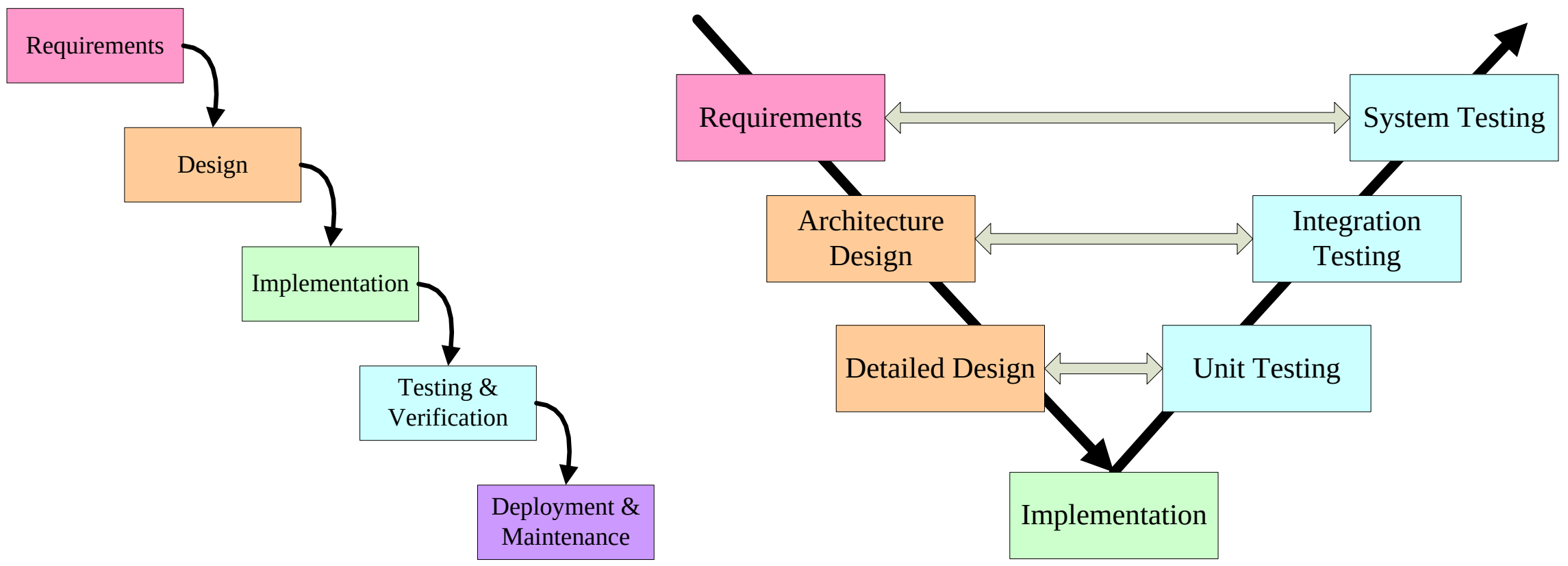

Figure 2. Traditional waterfall model (left) and V model (right)

A typical ASD process looks like figure 3. The "*Inspect-and-Adapt*" approach of ASD greatly reduces both development cost and time to market. Because the work cycle is limited to short time, it gives stakeholders recurring opportunities to calibrate goals for success, which helps team create the product at right direction.

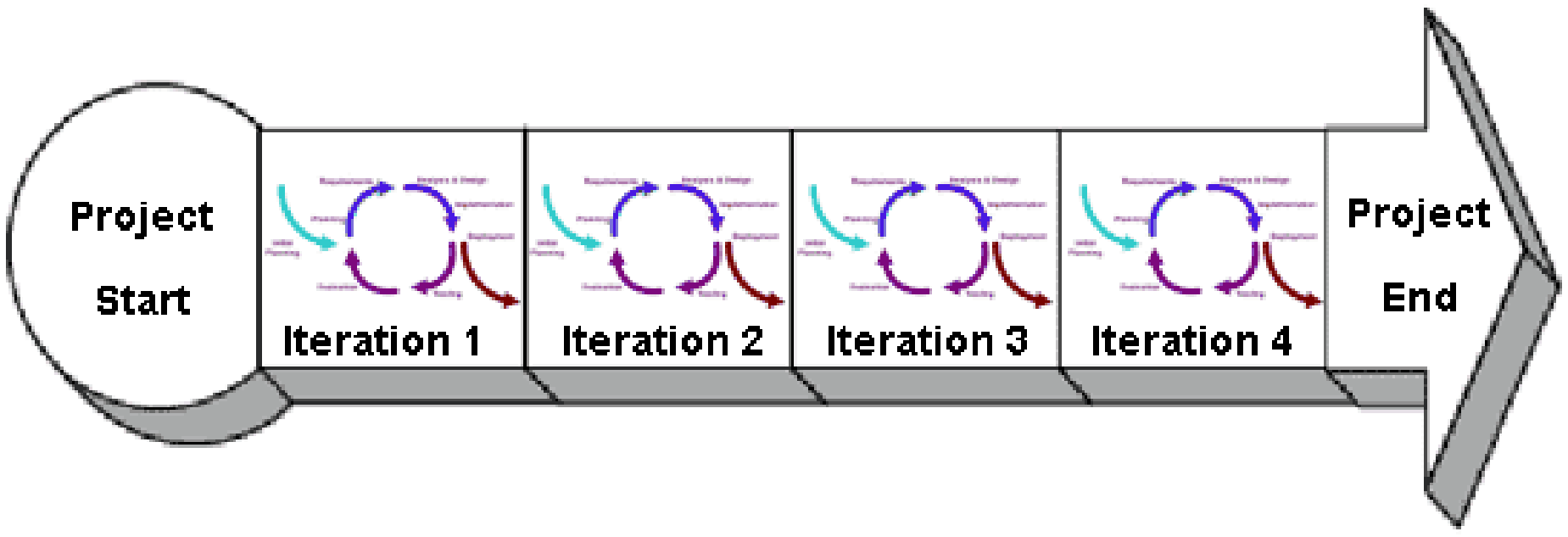

Figure 3. A typical ASD process

We can compare the *Software Development Life Cycle* (SDLC) between traditional waterfall and ASD process in Figure 4.

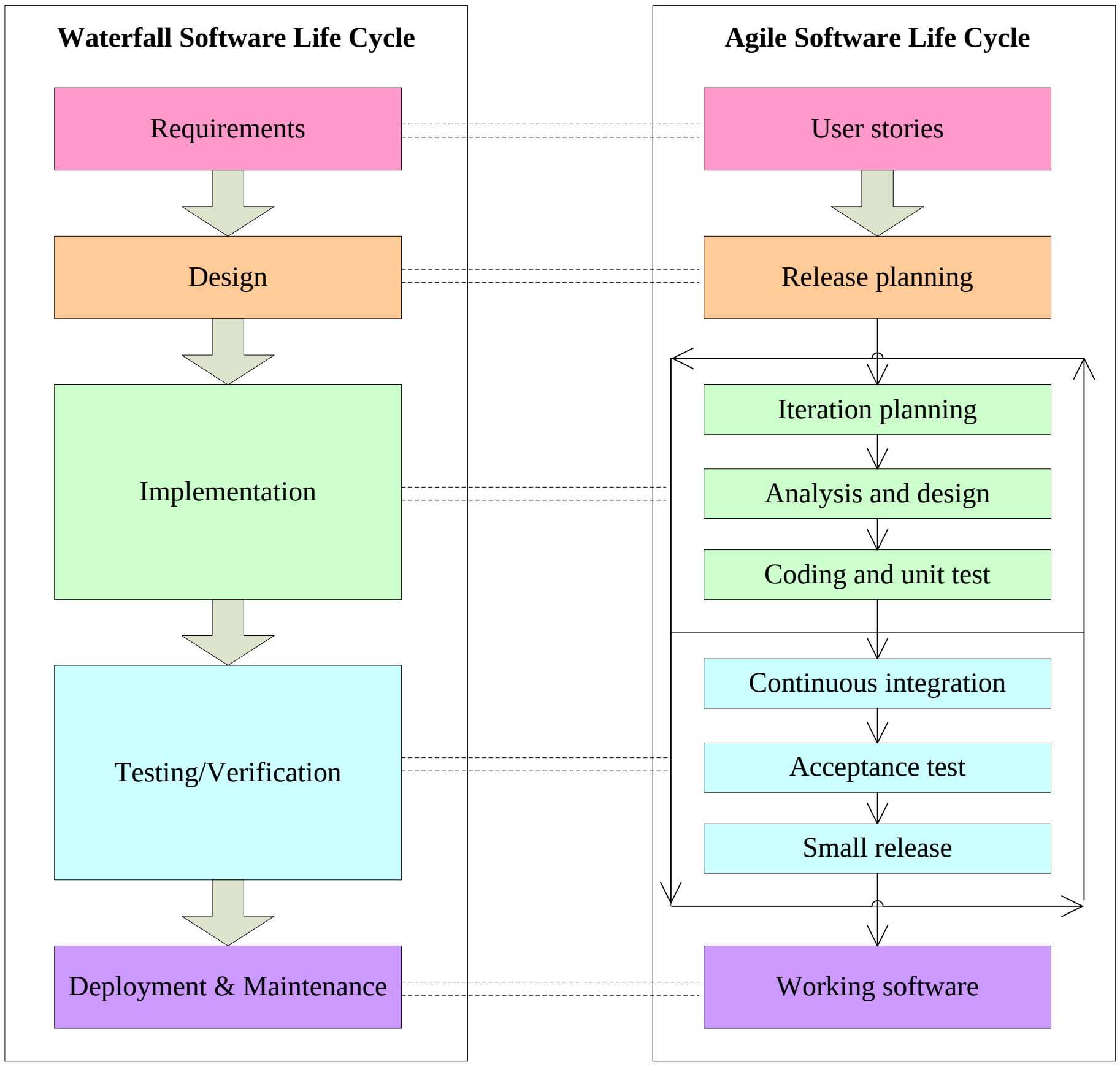

Figure 4. Comparison of waterfall and ASD

The stage of creating user stories in *ASD* is equivalent to the stage of analyzing requirements in *Waterfall*. The stage of planning release in *Agile* (such as the sprint planning in *Scrum*) is equivalent to the stage of designing system in *Waterfall*, the former focuses on planning but the latter focuses on designing. The stage of implementation and the stage of testing in *Waterfall* are split into *Agile*'s small iterative development cycle consisted of six activities, the first three activities, including iteration planning, analysis and design, coding and unit test, are similar to the works in implementation stage of *Waterfall*, and the last three activities, including continuous integration, acceptance test, and small release, are similar to the works in testing stage of *Waterfall*. These six activities are iterative and incremental to cope with continuous requirements and changes. Finally, the stage of releasing working software in *Agile* is equivalent to the stage of maintenance in *Waterfall*.

Current ASD methodologies are suitable for small projects with co-located and self-organized teams. With the increase of team size and distance between team members, more and more issues caused by human factors will be brought into the process, because of its loose and self-driven management mechanism [30].

As a conceptual framework, ASD may also be integrated with traditional methods, for example *Agile Unified Process* (AUP) [31], a simplified version of the *Rational Unified Process* (RUP) [32], the best-known and extensively documented refinement of the *Unified Software Development Process* (USDP) [33], which describes a simple, easy to understand approach to develop business application software using agile methods and concepts yet still remaining true to the *RUP* [34], as shown in Figure 5.

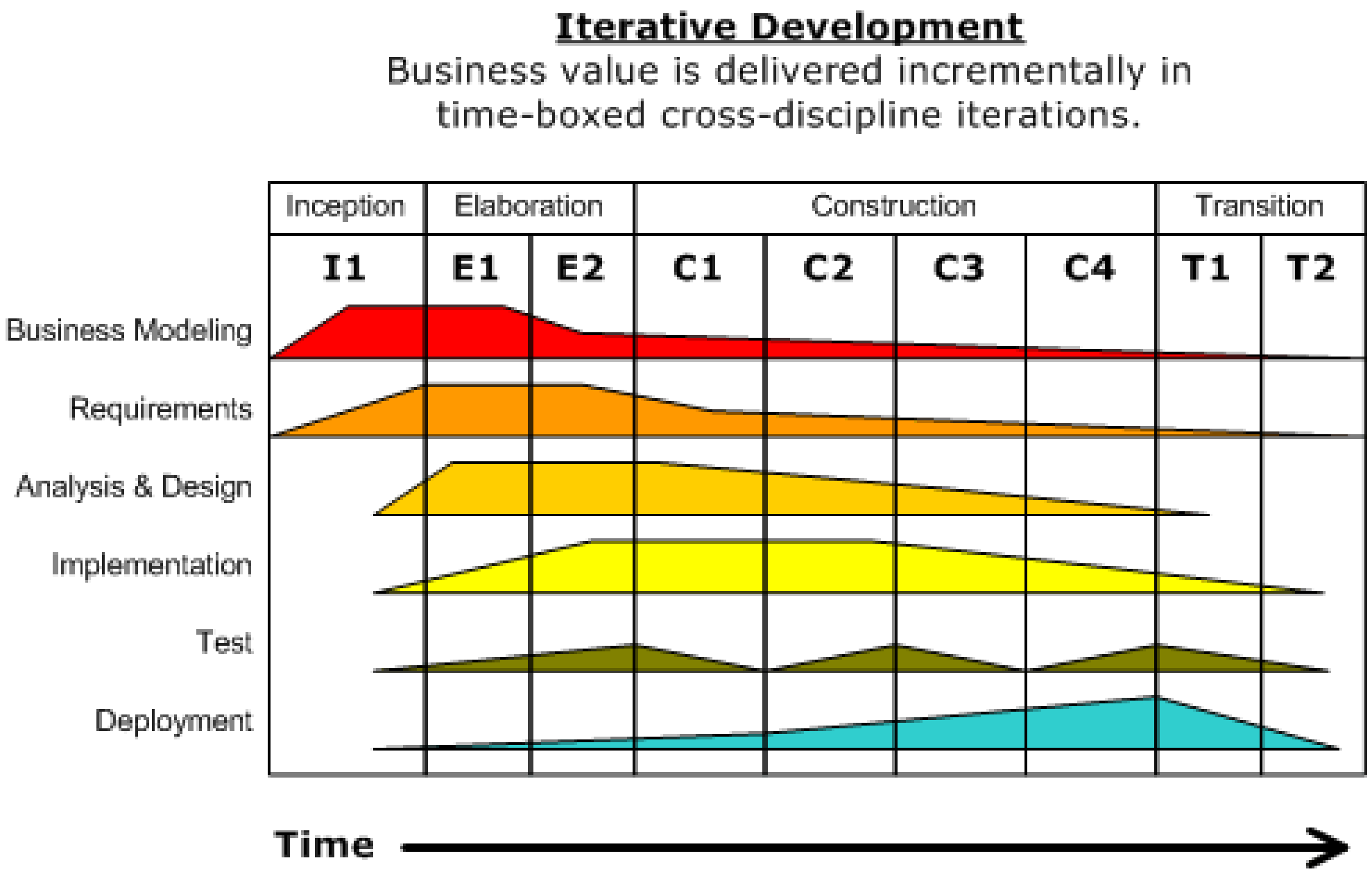

Figure 5. AUP, RUP, USDP also use incremental and iterative development model [29]

### *1.2.2 Principles and Characteristics of Agile Methodologies*

To evaluate the agility of software development process, agile manifesto give us twelve principles below [21]:

1) Customer satisfaction by rapid delivery of useful software – *Agile* asks delivery frequently with more iterations, many small builds are delivered in iteration process, as it emphasizes that working software over comprehensive documentation

2) Welcome changing requirements, even late in development – *Agile* accepts change of requirement at any stage, as it emphasizes that responding to change over following a plan

3) Working software is delivered frequently (weeks rather than months) – *Agile* believes which will lead to less defects

4) Working software is the principal measure of progress – *Agile* believes which will bring good ROI for customer

5) Sustainable development, able to maintain a constant pace – including test frequently, it requires continuous testing

6) Close, daily co-operation between business people and developers – *Agile* advocates collaborative approach, as it emphasizes that customer collaboration over contract negotiation

7) Face-to-face conversation is the best form of communication – *Agile* requires close communication between peoples

8) Projects are built around motivated individuals, who should be trusted – *Agile* emphasizes that individuals and interactions over process and tools

9) Continuous attention to technical excellence and good design – *Agile* advocates standing on the "shoulders of giant", that will reduce risk and time to develop

10) Simplicity: the art of maximizing the amount of work not done – *Agile* advocates avoidance of things that waste time, that's why agile products less documentation works compared to other methodologies

11) Self-organizing teams – *Agile* believes which will bring higher efficiency and lower communication cost

12) Regular adaptation to changing circumstances – *Agile* requires easily moved, lean, agile, active software processes, to fit the process to the project

Table 1 compares the advantages and disadvantages between ASD process and other plan-driven processes [21].

Table 1. Advantage and disadvantage comparison of Agile and Plan-driven process

| **Agile Process** | **Plan Driven Processes** |
|---|---|
| It's suitable for small products and teams, as its limited scalability | It's suitable for large products and teams, as it's hard to scale down |
| It's untested on safety-critical products | It can handle highly critical products |
| It's good for dynamic, but expensive for stable environments. | It's good for stable, but expensive for dynamic environments |
| It requires experienced Agile personnel throughout | It requires experienced personnel only at start if environment is stable |
| Personnel thrive on freedom and chaos | Personnel thrive on structure and order |

Some principles, such as having customer perspective, collaborating effectively, managing by fact, and focusing on execution are recognized main characteristics of *Agile*. *Agile* team that embodied these principles would be well positioned for success. To meet them, members in *Agile* team should have some good behaviors that include 1) asking questions to customers, 2) thinking like customers, 3) being willing to ask for help, 4) being willing to help others, 5) making decisions with concrete facts instead of personal opinions, and 6) striving to ship finished code.

We have surveyed some major human factors related ASD research efforts during the past decade as shown in Table 2.

Table 2. Selected main human factors related agile researches during last decade

| **Year** | **Researcher** | **Emphases and Points** |
|---|---|---|
| 2003 | Abrahamsson, P., J. Warsta, et al. [35] | comparative analysis of methodologies |
| 2003 | Constantine, L. L. and L. A. D. Lockwood [36] | usage-centered design |
| 2004 | Boehm, B. and R. Turner [37] | methodology integration |
| 2004 | Keenan, F. [38] | process tailoring |
| 2004 | Kontio, J., M. Hoglund, et al. [39] | distributed management |
| 2004 | Manhart, P. and K. Schneider [40] | agile for embedded software |
| 2004 | Tichy, W. F [41] | methodology evaluation |
| 2005 | Hirsch, M. [42] | comparative analysis of methodologies |
| 2005 | Maurer, F. and G. Melnik[43] | comparative analysis of methodologies |
| 2005 | McCarey, F. [44] | software reuse |
| 2005 | Melnik, G. and F. Maurer [45] | agile education |

| Year | Researcher | Emphases and Points |
|---|---|---|
| 2007 | Capiluppi, A., J. Fernandez-Ramil, et al. [46] | evolution of software |
| 2007 | Maurer, F. and G. Melnik[47] | methodology analysis |
| 2007 | Poppendieck, M. [48] | agile contacts |
| 2012 | Imtiaz, S. [30] | distributed architectural task allocation |
| 2012 | Lemos, O. A. L., F. C. Ferrari, et al. [49] | comparative analysis of methodologies |

### *1.2.3 Well-known ASD Methodologies*

As *Agile* is just a conceptual framework with some agile principles, so there are a number of specific agile methods espoused by the industry and *Agile Alliance* (see at: http://www.agilealliance.org). According to the characteristics of agile listed above, the following methods are generally considered as agile methods:[24]

- *Scrum* – an iterative and incremental *ASD* method for managing software projects or application development. Scrum has not only reinforced the interest in project management, but also challenged the conventional ideas about such management. Scrum focuses on project management institutions where it is difficult to plan ahead. Mechanisms of empirical process control, where feedback loops that constitute the core management technique are used as opposed to traditional command-and-control oriented management. It represents a radically new approach for planning and managing projects, bringing decision-making authority to the level of operation properties and certainties [20, 50] .

- *Extreme Programming (XP)* – a software development methodology which is intended to improve software quality and responsiveness to changing customer requirements. As a type of *ASD*, it advocates frequent "releases" in short development cycles (time-boxing), which is intended to improve productivity and introduce checkpoints where new customer requirements can be adopted [19, 51, 52].

- *Crystal Clear* – a member of the *Crystal* family of methodologies as described by Alistair Cockburn and is considered an example of an agile or lightweight methodology. It can be applied to teams of up to 6 or 8 co-located developers working on systems that are not life-critical. The Crystal family of methodologies focuses on efficiency and habitability as components of project safety. Crystal Clear focuses on people, not processes or artifacts [4, 53].

- *Agile Modeling (AM)* – a practice-based methodology for modeling and documentation of software-based systems. It is intended to be a collection of values, principles, and practices for modeling software that can be applied on a software development project in a more flexible manner than traditional modeling methods [54, 55].

- *Agile Unified Process (AUP)* – a simplified version of the IBM Rational Unified Process (RUP) developed by Scott Ambler. It describes a simple, easy to understand approach to developing business application software using agile techniques and concepts yet still remaining true to the RUP. The *AUP* applies agile techniques including test driven development (*TDD*), Agile Modeling, agile change management, and database refactoring to improve productivity [31, 34].

- *Dynamic Systems Development Method (DSDM)* – an agile project delivery framework, primarily used as a software development method. First released in 1994, *DSDM* originally sought to provide some discipline to the rapid application development (*RAD*) method. In 2007 *DSDM* became a generic approach to project management and solution delivery. *DSDM* is an iterative and incremental approach that embraces principles of *Agile* development, including continuous user/customer involvement [56-58].

- *Essential Unified Process (EssUP)* – it was invented by Ivar Jacobson as an improvement on the *Rational Unified Process*. It identifies practices, such as use cases, iterative development, architecture driven development, team practices and process practices, which are borrowed from *RUP, CMMI* and *Agile*. The idea is that you can pick those practices that are applicable to your situation and combine them into your own process. This is considered an improvement with respect to *RUP*, because with *RUP* the practices are all intertwined and cannot be taken in isolation [59, 60].

- *Feature Driven Development (FDD)* – an iterative and incremental software development process. It is one of a number of *Agile* methods for developing software and forms part of the *Agile Alliance*. *FDD* blends a number of industry-recognized best practices into a cohesive whole. These practices are all driven from a client-valued functionality (feature) perspective. Its main purpose is to deliver tangible, working software repeatedly in a timely manner [61, 62].

- *Kanban Development* – a method for developing software products and processes with an emphasis on just-in-time delivery while not overloading the software developers. It emphasizes that developers pull work from a queue, and the process, from definition of a task to its delivery to the customer, is displayed for participants to see. It can be divided into two parts: *Kanban* – A visual process management system that tells what to produce, when to produce it, and how much to produce, and the *Kanban* method–an approach to incremental, evolutionary process change for organizations [63, 64].

- *Lean Software Development (LSD)* – a translation of lean manufacturing and lean IT principles and practices to the software development domain. Adapted from the Toyota Production System, a pro-lean subculture is emerging from within the *Agile* community [65].

- *Open Unified Process (OpenUP)* – a part of the *Eclipse Process Framework (EPF)*, an open source process framework developed within the Eclipse Foundation. Its goals are to make it easy to adopt the core of the *RUP/Unified Process*. The *OpenUP* began with a donation to open source of process content known as the *Basic Unified Process (BUP)* by IBM. It was transitioned to the Eclipse Foundation in late 2005 and renamed *OpenUP/Basic* in early 2006. It is now known simply as *OpenUP* [66].

- *Velocity Software Development (VSD)* – a measure of productivity sometimes used in *ASD*. Velocity tracking is the act of measuring said velocity. The velocity is calculated by counting the number of units of work completed in a certain interval, determined at the start of the project.[67]

- *Adaptive Software Development (ASD)* – focuses mainly on the problems in developing complex, large systems. The method strongly encourages incremental, iterative development, with constant prototyping. Fundamentally, *ASD* is about “balancing on the edge of chaos”; its aim is to provide a framework with enough guidance to prevent projects from falling into chaos, but not too much, which could suppress emergence and creativity [61].

In industry, *Agile* principles and those features have led agile methods to achieve a tremendous success in projects. Today, more and more companies have embraced and joined into agile community, including almost all software giants such as *Microsoft* [68-72], *Google* [73, 74], *IBM* [75, 76], *Facebook* [77, 78], *SAP* [79], *Oracle* [80, 81], *Salesforce* [82, 83] and so on.

## 1.3 Scope, Methods, Problems and Motivations of Research

### *1.3.1 Scope of Research*

The term human factors engineering is the scientific discipline concerned with the understanding of interactions among humans and other elements of a system, which study how humans behave physically and psychologically in relation to particular environments, products, or services.

As for Agile software development, *Sallyann* and *Helen* listed the top 10 burning research issues voted by more than 300 practitioners and researchers in the Agile conference 2010 [10].

1) Is *Agile* suitable for large projects?
2) What factors can break self-organization?
3) Do teams really need to always be collocated to collaborate effectively?
4) Architecture and agile – how much design is enough for different classes of problem?
5) What are hard facts on costs of distribution?
6) What is the correlation between release length and success rate?
7) What metrics can we use with minimal side-effects?
8) Distributed agile and trust – what happens around 8–12 weeks?
9) Statistics and data about how much money/time is saved by agile.
10) Sociological studies – what were the personalities in successful/failed agile teams?

The issue #2, #3, #8, and #10 are directly related to human factors in agile development process, including collaboration, confidence, trust, emotion, and social factor etc. Issue #1 and #5 are related to collaboration problem with distributed teams. Issue #6, #7, and #9 are related to the evaluation metrics for an agile development process.

### *1.3.2 Research Methods*

For empirically studying and analyzing the impact of human factors to the ASD process, we have conducted four years *Scrum* based agile educational experiments in China since 2011. According to *VersionOne*'s surveys in 2012 and 2013 [84, 85], the *Scrum* and *Scrum* variants continuously remain the most popular agile methodologies being used in industry (adopted by 72% of practitioners in 2012 and 73% of practitioners in 2013). Figure 6 shows the using percentages of different ASD methods in companies.

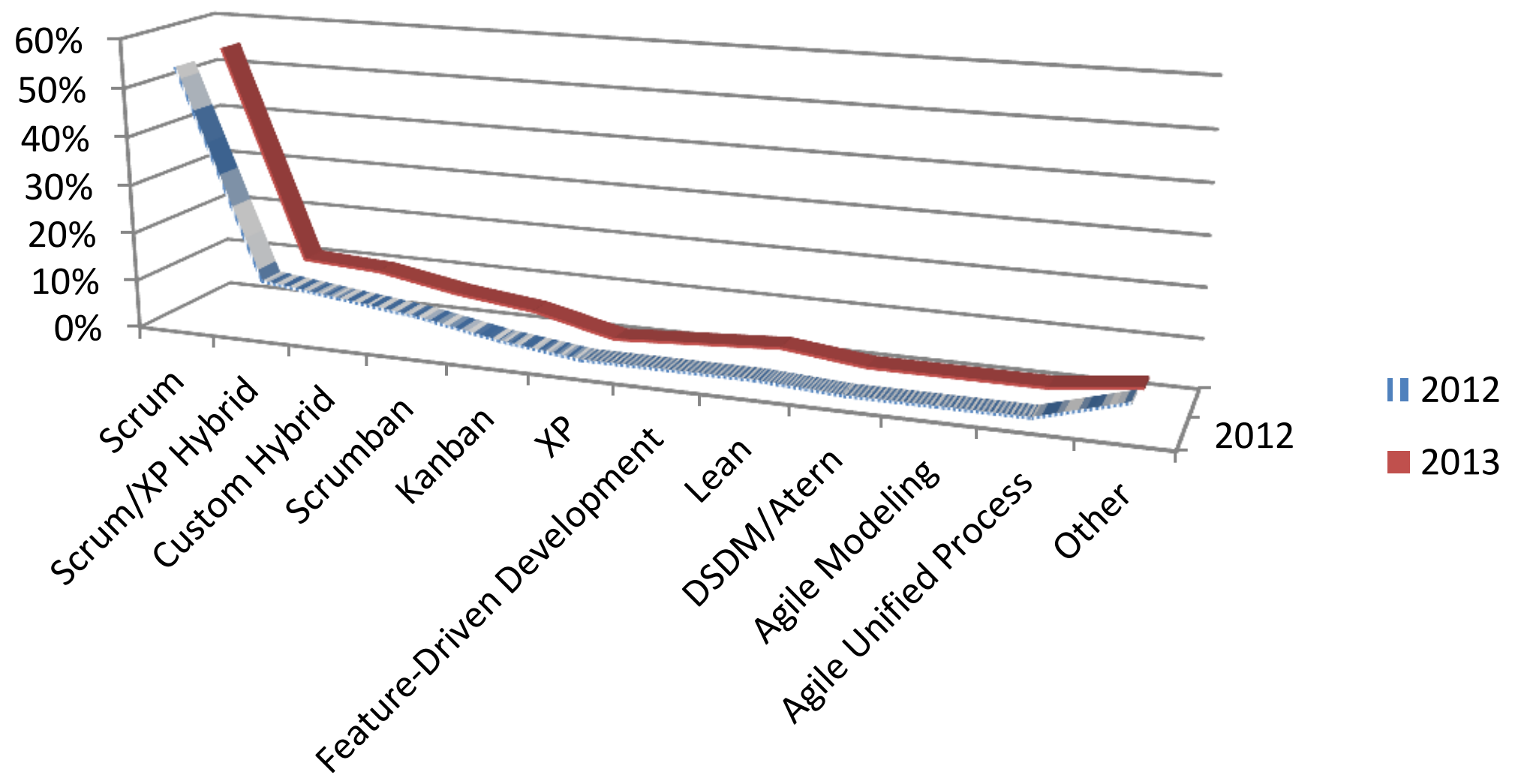

Figure 6. The using percentages of different ASD methods in companies (source: [84, 85])

During those experiments, we introduced *Scrum* into practical software engineering courses to train development and management skills for undergraduate students. *Scrum* consists of six phases: 1) conceptualization defines the high level deliverables and project roadmap; 2) release planning assigns deliverables into different releases; 3) sprint planning breaks down selected deliverables into technical tasks; 4) in each sprint (e.g., a 7 day period), software development tasks are to be completed by ASD team members; 5) in sprint review/retrospective, team members demonstrate the product increments and reflect on experience gained from the last sprint; and 6) during release the working software is delivered to the customers [86]. The students were divided into teams with an average team size of 5-7 persons to carry out an 8-13 week group-based software development project. There are typically around 17-25 teams each year. All teams possess similar skill levels and backgrounds and adopt the *Scrum* process during the project. For collecting the data of process, we develop an online *Agile Project Management* (APM) tool - the *Human-centered Agile Software Engineering* (HASE) platform (http://www.linjun.net.cn/hase/). The activities for each team member supported by HASE mainly occur during the sprint planning and sprint review/retrospective phases. They include proposing tasks, estimating the priority, difficulty and time required for each task, deciding how to allocate tasks, collaboration information, reviewing the timeliness and quality of completed tasks, and providing feedback about each team member’s mood at different points in time during a sprint.

After all students’ project finished, we will firstly adopt the *Exploratory Data Analysis* (EDA) [87, 88] approach to analyze the data collected from APM. EDA is an approach for analyzing data sets to summarize their main characteristics, often with visual methods. It is primarily for understanding what can be learnt from the data beyond the formal modeling or hypothesis testing task. Based on the findings and our hypothesis, we then design some new observed factors into the next round data collection.

### *1.3.3 Research Problems and Motivations*

1) Through the data collected from 18 teams in 2011 and 26 teams in 2012, we found the problem that the granularities of user stories proposed by student teams' product owner mostly are not good. By further analyzing and interviewing with them, we have known this is caused by unclear or misunderstanding users' goals when team defining those requirements to user stories. And then we introduced a light-weight goal modeling tool into the class of 2013 and gained a significant improvement. Chapter 3 will give us more details.
2) Through the data collected from 20 teams in 2013, we observed three types of task allocation strategy among these teams: 1) equality based strategy; 2) competence based strategy; and 3) mixed strategy. Different strategy means different collaboration model between team members and decision making mechanism that will lead to different task execution result. How to find a good task allocation mechanism based on the situations of team member and task execution become a problem for those Agile teams. More detailed analysis and simulation will be discussed in Chapter 4.
3) Developer's mood swing and team's morale are very important to the success of software project. Current ASD project managers or scrum masters attempt to user *Happiness Chart* to monitor developer's mood swing and track the team morale [11]. Figure 7 shows an example in which a scrum master tries to use *Happiness Chart* to encourage developer emotional openness by asking them to write their mood directly on a white board. An obvious issue with this approach is that it is difficult to know whether one developer's mood will be influenced by the emotions wrote down by another developer. Moreover, although *Happiness Chart* is visible, visibility is only one aspect of transparency for the process. The other aspects including accuracy, authenticity, and congruence etc. also should be considered into the method, for example, what kind of happiness is a banana or a square in Figure 7? Does each developer expand on their drawing in any way? For investigating the impact of team member's mood swing to task execution quality, we added more emotional factors to be collected in the experiment of 19 teams in 2014 class. Chapter 5 will discuss our proposed emotional theoretical model, simulation result and the analysis. Chapter 6 will also analyze more empirical insights including mood factors from the data of those teams.

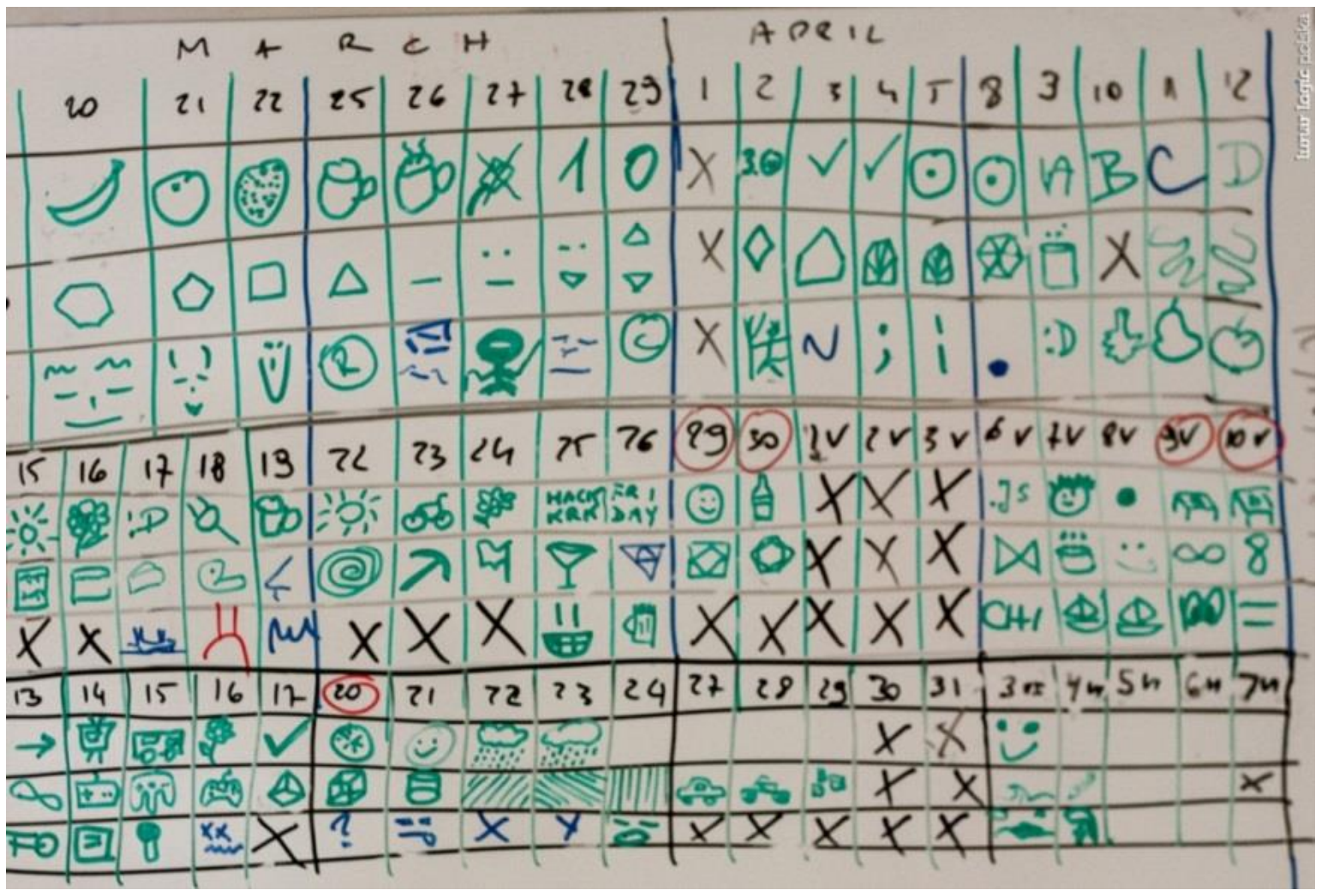

Figure 7. A Happiness Chart for monitoring mood in an Agile team (source: [11])

## 1.4 Summary of Contributions

By achieving the research objectives set out in the previous section, this book makes the following important contributions to the state of the art in the area of ASD research.

- **Integrate Goal Net model into current agile process.** Although there are a large number of research works over the years focused on goal oriented requirement engineering for software development. Many of these approaches generally suffer from three main types of shortcomings: these approaches 1) require the product owner and team member to spend too much extra efforts on studying new modeling theories and tools; 2) lack a hierarchical or layered structure (diagram) to link goals between different stages or iterations; and 3) are difficult to be integrated into the current ASD smoothly and leanly to form a new type of chart. After surveying current GORE methods (in section 3.1), we propose a novel goal-oriented method to model user stories in ASD processes. By modeling user stories into a Goal Net diagram, we provide a clear overview to allow ASD team members to see requirements from a new perspective. Through studies with student developers during a software engineering course, we observe that more user stories are formed and the quality of these user stories is significantly improved through the use of the proposed goal-net method.

- **Propose the *SMART* Approach for task allocation in ASD process.** During task allocations for ASD teams, it is challenging to efficiently exploit team members' skills based on their personal conditions under changing environment. In this book, we propose a context-aware task allocation decision support algorithm that improves the overall utility of an ASD project through the smart tradeoff between product quality and time limitation. We formulate the ASD process as a distributed constraint optimization problem, and propose a technology framework that assesses individual developers' situations based on data collected from a *Scrum*-based agile process, and helps individual developers make situation-aware decisions on which tasks from the backlog to select in real-time. Our analysis and simulation results show that it can achieve close to optimally efficient utilization of the developers' collective capacity.
- **Propose a *FCM*-based affective model for predicting developer's mood swing in ASD process.** A vital drawback of existing *Happiness Chart* method is that the chart is built based on purely intuitions without the analysis of historical process data. So that it requires us to find a better quantification method to do this. In order to find such a quantification method, we propose a Fuzzy Cognitive Map-based emotion model which is based on the OCC emotion theory and Fuzzy cognitive maps (FCMs) model. This model computes developer's mood according to the causal relationship between mood status and consequence of task execution. Two simulated case studies are provided to illustrate the modeling and simulating process in Chapter 5. Comparing to *Happiness Chart* method, it’s the first trial of modeling and predicting developer’s mood curve in a computable method.

- **Design and implement the *Human-centered Agile Software Engineering* (HASE) collaboration development platform and a new data collection method using the APM tool.** Different from traditional survey/interview-based empirical studies, our study is based on participants’ ASD activity trajectory data collected unobtrusively during normal ASD processes through our HASE APM platform. This type of data objectively reflects users’ ASD activities and performance at fine granularities. With the help of this form of data, we report key findings in Chapter 6 that reveal important insights into the *Scrum* ASD processes when practiced by novice student teams. These results offer new insights into the aspects of agile team collaboration and team morale which have not yet been well studied by existing research. Furthermore, to the best of our knowledge, we are the first to claim that ASD process is a human computation system that using human effort to perform tasks that computers are not good at solving, as well as using computers to assist agile teams to better serve human’s decision making.

## 1.5 Outline of the Book

The rest of this book is organized as follows:

- In Chapter 2, literature review of most related theories, methods and applications in recent two decades are presented.
- Chapter 3 firstly analyzes the requirement management in existing main ASD process, and then the proposed *Goal Net* based method and its case studies are presented and discussed.
- Chapter 4 firstly analyzes the task management and allocation in current main ASD process, and then the proposed *Simple Multi-Agent Real-Time (SMART)* approach and its simulation results are presented and discussed.
- Chapter 5 firstly analyzes the human management and emotion factors in current main ASD process, and then the proposed *FCMs* based affective model and its simulated case studies are presented and discussed.
- Chapter 6 presents more empirical findings regarding the aspects of task allocation decision-making, collaboration, and team morale related to the Scrum ASD process, which have not yet been well studied by existing research.
- Chapter 7 discusses and outlines potential future research directions.

# CHAPTER 2

# LITERATURE REVIEW

In this chapter, we will firstly go through current main PhD. researches on ASD methodologies during past decade, and then we will review current works on *Goal-Oriented Requirement Engineering* (GORE), *Human-Centered Software Engineering* (HCSE), *Social Software Engineering* (SSE), *User Centered Design* (UCD) and *Human Computation System* (HCS), which are related to requirement and human management aspects of software development process. The state of art for some related modeling theories and methods we try to use in our research works, such as *Goal Net*, will also be reviewed.

## 2.1 Current main PhD. theses and research directions on Agile

Over the past decade, research on the *ASD* methodologies has become more and more popular in *SE* discipline. Before investigating a specific research direction, we have surveyed the general PhD. theses on ASD in recent decade shown in Table 4, as this survey can tell us those young researchers and their advisors' main new ideas and research trends on ASD research.

Table 3. Survey of selected main PhD. theses on ASD in decade

| **Year** | **Author/ Univ.** | **Title** | **Objective** | **Methods** |
|---|---|---|---|---|
| 2005 [89] | Cao, Lan. ( Georgia | Modeling dynamics in | To improve the understanding of and | A well-established methodology in the field |
| | State University) | agile software development | gain insights into the effectiveness and applicability of agile methods. | of systems dynamics, exemplified by the work of Abdel-Hamid, 1991. |
| 2005 [90] | Fernandez, Marco Gero. (Georgia Institute of Technology) | A Framework for Agile Collaboration in Engineering | To establish a consistent Framework for Agile Collaboration in Engineering that more accurately represents the mechanics underlying product development on one hand and supports interacting stakeholders in achieving their respective objectives in light of system level priorities on the other. | (1) establishing and assessing collaborative design spaces, (2) identifying and exploring regions of acceptable performance, and (3) preserving stakeholder dominion over design sub-system resolution throughout the duration of a given design process. |
| 2006 [91] | Cao, Dac-Buu. (Capella University) | An empirical investigation of critical success factors in agile software development projects | To explore and determine the critical success factors of Agile software development projects using quantitative approach. | Reliability analysis and factor analysis; A survey was conducted among Agile professionals, gathering empirical data from 109 Agile projects from 25 countries across the world. Multiple regression techniques were used. |
| 2006 [92] | Johnson, Melanie. (Pace University) | Multi-Project Staffing: An Agile based Framework | To offer guidelines for project staffing that increases the likelihood of efficient and effective development by focusing on a project's | Multi-Project Staffing Model based on Mathematical Staffing Model and Linear programming method. |
| | | | most valuable resource, its people, and to examine the allocation of developers to tasks using a common and shared human resource pool across all the projects. | |
| 2006 [93] | Maruping, Likoebe Mohau. (University of Maryland) | Essays on agility in software development teams: Process and governance perspectives | To examine the processes and governance mechanisms that can potentially enable software project teams to achieve greater flexibility. | Case study: A longitudinal field study of 56 software development teams; |
| 2006 [94] | Zandoli, Robert. (Pace University) | An Agile Framework for an Information Technology Outsourcing Decision process (ITODP) | To identify an efficient and effective method to improve the Information Technology Outsourcing ITO decision process. | Information Technology Outsourcing Decision Process (ITODP); interviews with industry experts; a survey instrument and case study analysis. |
| 2007 [95] | Kile, James F. (Pace University) | An Investigation into the Effectiveness of Agile Software Development with a Highly Distributed Workforce | To investigate the practice of agile software development in a highly distributed (non-collocated) environment to understand its efficacy. | A qualitative case study approach; some quantitative and mixed method approaches to cross-validate the information gathered. |
| 2007 [96] | Rico, David F. (University of Maryland University College) | Effects of agile methods on website quality for electronic commerce | To study the relationships between the use of agile methods to manage the development of Internet websites and website quality. | Literature review, conceptual model, survey instruments, measurement data, and data analysis for agile methods |
| 2007 [97] | Sidky, Ahmed. (Virginia Polytechnic Institute and State University) | A structured approach to adopting agile practices | To propose a structured process to guides organizations in adopting agile practices. | The Sidky Agile Measurement Index (SAMI) and a 4 Stage process framework |
| 2008 [98] | Roden, Patricia L. (The University of Alabama in Huntsville) | An examination of stability and reusability in highly iterative software | To examine the stability and reusability of agilely developed software | Total Quality Index (TQI) of the QMOOD Quality Model; Chidamber and Kemerer metrics; the expert reusability evaluations |
| 2009 [99] | Feng, Kunwu. (The University of Texas at Dallas) | Towards an Agile Product Line Requirements Engineering Framework: Knowledge acquisition and process definition | To define or select a requirements engineering (RE) process and a set of RE techniques that is well suited for a project in terms of the degree of agility for product line projects. | APLE-RE framework |
| 2009 [100] | Ganis, Matthew Robert. (Pace | A study of the agile whole team and its | To understand how the composition of the agile team affects the | Project retrospectives and product backlog analysis. |
| | University) | effectiveness in the software development process | resultant output from the perspective of the customer as well the team itself. | |
| 2010 [101] | Bird, Michael Stephen. (Capella University) | Utilizing agile software development as an effective and efficient process to reduce development time and maintain quality software delivery | To utilize agile software development as an effective and efficient process to reduce development time and maintain quality software delivery | Instrumentation/Measures |
| 2010 [102] | Lee, Jason Chong. (Virginia Polytechnic Institute and State University) | Integrating scenario-based usability engineering and agile software development | To integrate scenario-based usability engineering and agile software development | eXtreme Scenario-Based Design (XSBD) process |
| 2011 [103] | Boehm, Raymond E. (Pace University) | An Approach to Early Lifecycle Estimating for Agile Projects | To estimate the size for agile projects | Early Lifecycle Functionality Estimating (ELFE) process; Elf Poker |
| 2011 [104] | Farid, Weam Mohamed. (Nova Southeastern University) | The NORMAP Methodology: Non-functional Requirements Modeling for Agile Processes | To develop a conceptual framework for Non-Functional Requirements (NFR) modeling in agile processes. | Non-functional Requirements Modeling for Agile Processes (NORMAP) Methodology |
| 2011 [105] | Green, RonzelleL. (The George Washington University) | Understanding the role of synchronous and asynchronous communication in agile software development and its effect on quality | To explore the relationship between synchronous and asynchronous communication within agile teams and its effects on quality of software produced. | General survey and analysis |
| 2012 [106] | Ashmore, Sondra. (Iowa State University) | The impact of process on virtual teams: A comparative analysis of waterfall and agile software development teams | To explore the impact of process on waterfall and agile virtual software development teams | Case study approach; interview; data analysis; comparative analysis |
| 2012 [107] | Nis Ovesen (Aalborg University) | The Challenges of Becoming agile - Implementing and Conducting Scrum In Integrated product development | To apply Scrum and Agile principles to integrated product development process. | Case study approach; data analysis |
| 2013 [108] | Anuradha Sutharshan (Edith Cowan University) | Human Factors and Cultural Influences in Implementing Agile Philosophy and | To provide an analytical comparative framework for implementing agile methods in different cultures, and insight into how cultural differences | Action research approach, case study approach, ethnography, grounded theory, comparison and selection of suitable |
| | | Agility in Global Software Development | may affect a software project and how these challenges can be addressed through agile principles. | research method |

Through the surveying, we can see most of ASD researches for PhD. works try to find quantitative approaches or modeling methods to improve the process from different aspects, such as effectiveness, applicability, stability, quality and reusability etc. In recent decade a few of young researchers have seen the importance of human factors in agile process [90, 100, 105, 108], although human factors and social factors have become hotspot of SE research in decades. As human factors are hard to collect data and to do quantitatively analysis during the process, but actually they are vital to the success of agile teams, especially for the management of stakeholders' goals, decision making of task allocation and developers' mood swing etc.

## 2.2 Goal Oriented Requirement Engineering (GORE) and Goal Net Theory

*Requirement Engineering (RE)* is a sub-discipline of *SE*, which is a very important and vital phase in the overall development process, because that software as a final product will be deemed to fail if it does not fulfill the needs of customers and users. A number of methods have been proposed to improve the requirements engineering process, such as *KAOS* [109] and *GBRAM* [110] etc. Most of them attempt to link requirements to users' goals. This research direction is called as *Goal-Oriented Requirement Engineering* (GORE) [111].

Goal is objective that software system under stakeholders' consideration should achieve, which is a very important human factor in software development. Goals may be formulated at different levels of abstraction, ranging from high-level, the strategic concerns (such as “enhance end-users’ satisfactions” for an elderly entertainment system or “provide ubiquitous payment service” for an online B2C marketplace system) to low-level, the implemental concerns (such as “show messages in big font on the screen” for an elderly entertainment system or “all *VISA* credit card should be supported to pay online” for an online B2C marketplace system). Goals also cover different types of concerns: functional concerns associated with provided services, and non-functional concerns associated with quality of service, e.g. safety, security, accuracy, performance, and so on. The multidimensional hierarchy characteristics of goals bring complexity and difficulty for modeling goals in software development process.

### *2.2.1 Literature Review of GORE*

A number of *GORE* methods have been subsequently derived by various researches. We reviewed some of the significant *GORE* research works that have been done in past decade. The comparative study may serve as a guide for us to select an appropriate goal-oriented method.

In 2002, Kaiya et al. proposed *AGORA* method for requirements elicitation, analysis and estimation, which is an extended version of a goal-oriented method by attaching preference and contribution attributes to an *AND-OR* graph. Furthermore, they also proposed a method to estimate the quality of a requirements specification by means of the structural characteristics and attribute values of an *AND-OR* graph. Their *AGORA* method can support activities of selecting the goals to be decomposed, prioritizing and solving the conflict, choosing and adopting a goal as a requirements specification, analyzing the impacts when requirements change, and improving the quality of the requirements based on a quality of *AGORA* graph [112] .

In 2004, Shen et al. presented *Goal Net*, a goal-oriented modeling method to model the goals of an agent and to model agent coordination in a multi-agent environment. *Goal Net* also serves as a practical methodology for engineering agent oriented software systems [113].The next year, they refined the methodology for multi-agent system development based on *Goal Net* model. The new methodologies cover the whole life cycle of the agent system development, from requirement analysis, architecture design, and detailed design to implementation [114]. Based on this, in 2007, Yu et al. proposed a *Goal Net Designer* which is an integrated tool and development environment for modeling agent behavior based on *Goal Net* model. The *Goal Net Designer* provides a way for users to simplify the various stages of agent design. It also can be used by the *Multi-Agent Development Environment (MADE)*automatically to create intelligent agents [115].

In 2005, Khallouf et al. proposed a refined *Prometheus* method to have a stronger focus on goals in the design process. Their method infuses goals into the interaction diagrams, interaction protocols, and process diagrams [116].

In 2007, Lei et al. presented a goal-oriented process for database requirements analysis, based on existing *GORE* frameworks, which consists of eight steps, including identifying stakeholder goals and their quality requirements, generating a goal model, selecting a design alternative, identifying initial set of domain notions from goals, identifying and select plans, expanding the set of domain notions using plans, constructing the domain model, and constructing the conceptual schema. They also proposed a goal-oriented design strategy to structure the transformation from the domain model to the conceptual schema, according to a set of user defined design issues [117].

In 2007, Sen et al. proposed a methodology involving maximum participation of stakeholders for eliciting the soft goals, where the soft goals are elicited by the stakeholders from the high-level goals through the process of goal refinement. The elicited goals are compiled through a software program called *Activity Card Compiler*, with the result presented back to the stakeholders for adjustment. During successive iteration of this process, the stakeholders will elicit many goals, creating a set of goals which are complete, accurate, and ready to be converted into a high quality software requirements specification [118].

In 2008, Tanabe et al. presented two topics related to supporting techniques and methods for requirements changes in goal oriented analysis. One is version control for goal graphs and another is impact analysis of adding and deleting goals in a goal graph. In the version control system, they extract the differences between successive versions of a goal graph by means of monitoring modification operations performed through a goal graph editor; the impact analysis can detect conflicts that arise when a new goal is added, and can investigate the achievability of the other goals when the existing goal is deleted. They also have implemented the supporting tool based on their proposed techniques [119].

In 2009, Zhang et al. proposed an agent planning system based on the *Goal Net* model. In their system, the agent's goals are identified and organized in a composite goal hierarchy. Three kinds of relations between goals are defined: choice, concurrency and synchronization. Actions between goals are designed to accomplish subsequent goals. The agent's desire is satisfied by accomplishing a serial of intermediary goals and finally achieving the ultimate goal that is satisfying the desire. The agent's action plan is a list of actions to accomplish the intermediary goals in the solution. Because *Goal Net* is designed by considering agent's possible desires directly, their works bridged the distance between *BDI* agent design and the planning system. They also proposed a searching algorithm to select goals in *Goal Net* [120].

In 2009 and 2010, Saeki et al. presented an integrated supporting tool for *Attributed Goal-Oriented Requirements Analysis (AGORA)*, which is an extended version of goal-oriented analysis. The tool assists seamlessly requirements analysts and stakeholders in their activities throughout *AGORA* steps including constructing goal graphs with group work, utilizing domain ontologies for goal graph construction, detecting various types of conflicts among goals, prioritizing goals, analyzing impacts when modifying a goal graph, and version control of goal graphs. Their work focuses on integration of gathering, evaluating, and versioning requirements [121, 122].

In 2010, Zhang et al. applied reinforcement learning algorithms for goal selection in a *Goal Net* to convert an original *Goal Net* to its counterpart that learning algorithm can operate on. They developed a reorganization algorithm to convert a refined Goal Net to a partially ordered network. The algorithm can convert concurrency and synchronization relationships to the choice relationship without losing any information in the original goal net. And then a reinforcement learning algorithm is applied to train the goal selection of the converted goal net. Their work showed that the goal net model can simulate motivated learning of goal selections [123].

In 2011, Alimazighi et al. did some research for proposing the most appropriate method by adapting goal oriented analysis for collaborative information system, especially for *Inter-Organizational Information System (IOIS)* that can work beyond the typical borders of organizations. When partners want to work together, they already know the goal of the project through collaborative information system. Therefore, a collaborative network allows a goal to be met that cannot be completed by one partner. Due to the complexity of this type of systems, the requirements analysis stage is crucial; check the adequacy between expected objectives with the needs expressed by the all users of the several organizations in collaboration and the satisfaction of various goals generated by such collaboration. Their work seems can be referred by a distributed agile team [124].

In 2011, Birkhölzer et al. presented a framework to analyze the complex network of relationships between process fragments and objectives using weighted dependency graphs based on interval arithmetic. Their work can be used for a goal-driven analysis of possible or feasible improvement strategies. Their approach has been demonstrated using a knowledgebase of evidential data to evaluate agile process fragments [125].

During the past two decades, the goal-oriented requirements engineering proposed a set of approaches using the concept of goal to specify and explore the different objectives of systems, organizations, and users: *KAOS* [109], *1** [126], *GBRAM* [110], *Framework NFR* [127], *GQM* [128], *AGORA* [112], *Goal! Strategy MAP* [129], *Goal-scenario coupling: CREWS-L'Ecritoire* [130] etc. These approaches try to model the goals of the actors and the actions required to achieve them.

All in all, using goal-oriented requirement analysis for *ASD* have many advantages, such as:

- Object models and requirements can be derived systematically from goals
- Goals provide the rationale and humanity for requirements
- A goal graph provides vertical traceability from high-level strategic concerns to low-level technical details; it allows evolving versions of the system under consideration to be integrated as alternatives into one single framework
- Goal Nets and other graphs can provide right abstraction level at which decision makers can be involved for important decisions
- The goal hierarchy model provides a comprehensible structure for the requirements document;

- Alternative goal refinements and task assignments allow alternative system proposals to be explored;
- Goal formalization allows refinements to be proved correct and complete.

Table 5 compares those researches during the past decade, including their methods, key points and applications.

Table 4. Comparative study of GORE research during the past decade

| Year | Author | Methodology | Key Points | Applications |
|---|---|---|---|---|
| 2002 | Kaiya et al. [112] | *AGORA* | Attach preference and contribution attributes to an AND-OR graph | Requirements elicitation, analysis and estimation stage for a user account system development |
| 2004/ 2005 | Shen et al. [113, 114] | *Goal Net* | model the goals and coordination of agent; practical methodology covering the whole life cycle of the agent system development | Requirements analysis stage of agent-based e-learning system development |
| 2005 | Khallouf et al. [116] | refined Prometheus | Focus on goal-oriented design process | Requirements analysis stage of proactive software agents design |
| 2007 | Lei et al. [117] | Goal Model (AND/OR) UML | 8 steps spanning the spectrum from high-level stakeholder goal analysis to detailed conceptual schema design | Database requirements analysis |
| 2007 | Sen et al. [118] | *ABGR* | Activity Cards | Requirement analysis stage of a Web Based News |
| | | | | system |
| 2007 | Yu et al. [115] | *Goal Net* | A tool and IDE for modeling Goal Net | Agent system design |
| 2008 | Tanabe1et al. [119] | *AGORA* | change management of goal graphs | Requirement Change Management |
| 2009 | Zhang et al. [120] | *Goal Net* | composite goal hierarchy with action plan; goal selection algorithm | Requirement analysis stage of agent planning system and e-learning agent system |
| 2009/ 2010 | Saeki et al. [122] | *AGORA* | constructing goal graphs; utilizing domain ontologies; detecting various types of conflicts; prioritizing goals; analyzing impacts, and version control | Requirements gathering, evaluating, and versioning |
| 2010 | Zhang et al. [123] | *Goal Net* | reinforcement learning algorithm | goal selection algorithm for agent research |
| 2011 | Zaia et al. [124] | Early exploring | There is not specific method proposed by this research | Requirement analysis stage of Inter-organizational Information Systems |
| 2011 | Birkhölzer et al. [125] | weighted dependency graph | search for sets of process fragments to meet actual objectives | Requirement analysis of the complex network of relationships |

### 2.2.2 Goal Net Theory

After surveying, we found most of current GORE methods are ad hoc that need too much effort to be applied on practice, especially for ASD. However, as a practical and lightweight tool, *Goal Net* theory is very suitable for autonomous system design. *Goal Net* theory was proposed by Shen et al. in 2004 [113, 131], which is designed to model and design goal-oriented agents at first. *Goal Net* model consists of four basic objects or concepts: states, transitions, arcs and branches. There are two types of states in *Goal Net*, composite state and atomic state. An atomic state accommodates a single state which cannot be split. A composite state, represented by a shadowed circle, represents a goal and may be split into sub states. States are interconnected by transitions. A transition primarily shows relationship between the states it joins, specifying the task functions to be performed in a task list. Basically there are four kinds of relationships between two states, represented by transitions, including sequence, concurrency, choice, and synchronization [131].

*Goal Net* supports goal selection and action selection mechanism [120, 123]. *Goal Net* theory can be used to model the hierarchical goals in a complex system or process. As we discussed above, ASD itself is a complex process, so we use *Goal Net* method to model the typical *AUP* or *Scrum* process at high abstract level shown in Figure 8.

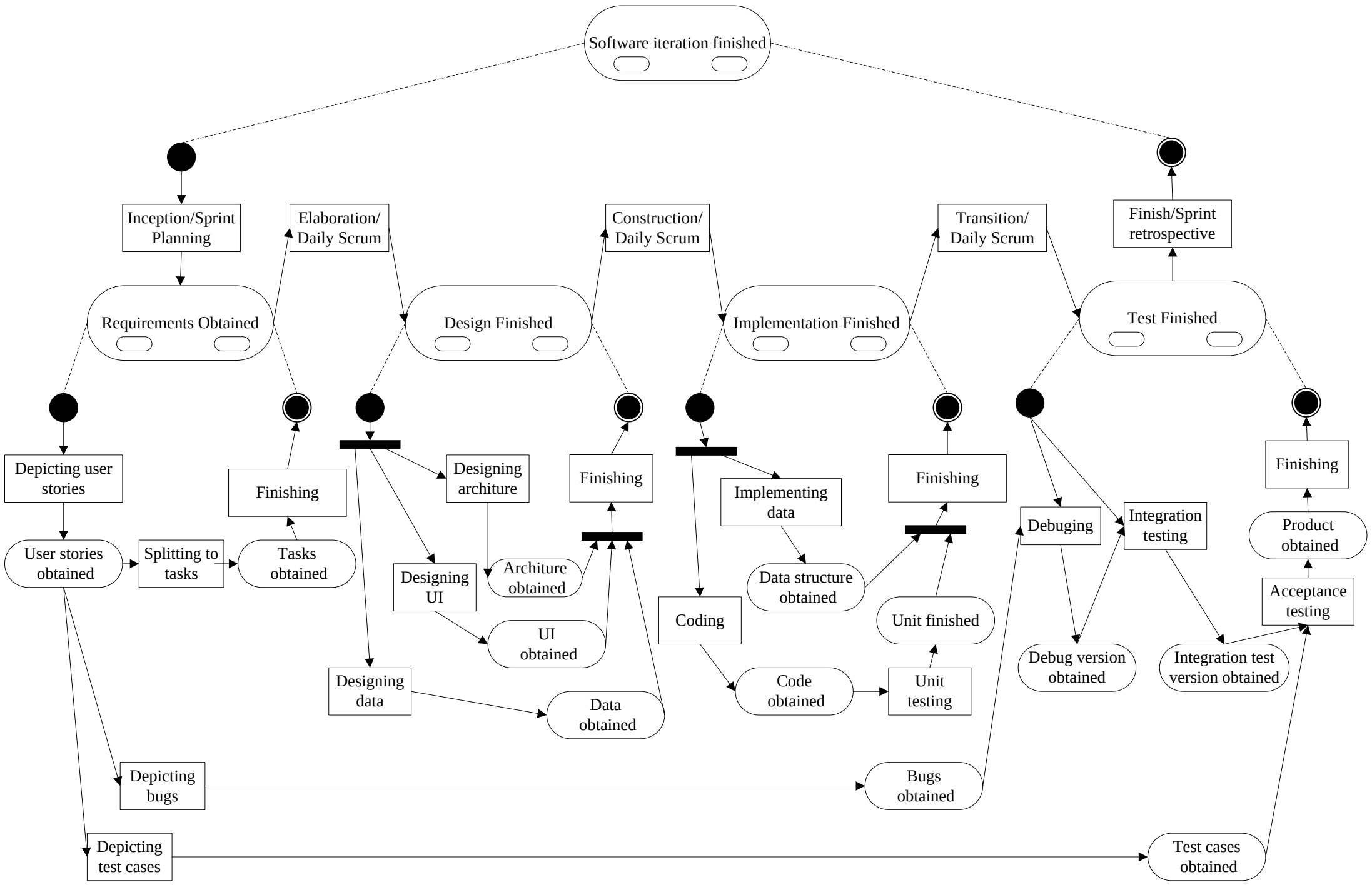

Figure 8. Goal Net model of a typical AUP/Scrum process

From the figure, we can see that the top goal for iteration is modeled as a composite state named '*Software iteration finished*'. To achieve the goal, we need to reach four sub-goals represented as four composite states, named '*Requirements Obtained*', '*Design Finished*', '*Implementation Finished*', and '*Test Finished*' sequentially. To achieve them, four transitions are required.

- *Inception*: the input transition for state of '*Requirements Obtained'*, which includes two atomic states: '*User stories obtained'* and '*Tasks obtained'*, and their three related transitions (tasks and conditions) shown in the figure. For *Scrum*, this transition can be executed in sprint planning activity.
- *Elaboration*: the input transition for state of '*Design Finished*' and output transition for state of '*Requirements Obtained*', which includes three concurrent atomic states, their corresponding input transitions and one synchronized atomic state shown in the figure. For *Scrum*, this transition can be executed in daily scrum activity.
- *Construction*: the input transition for state of '*Implementation Finished*' and output transition for state of '*Design Finished*', which includes three atomic states ('*Data structure obtained*' and '*Code obtained*' are concurrent, '*Code obtained*' and '*Unit test finished*' are sequential) and three corresponding input transitions. They are synchronized at a finished atomic state shown in the figure. For *Scrum*, this transition also can be executed in daily scrum activity.
- *Transition*: the input transition for state of '*Test Finished*' and output transition for state of '*Implementation Finished*', which includes three atomic states ('*Debug version obtained*', '*Integration test version obtained*' and '*Working software obtained*') and four corresponding input/output transitions. For *Scrum*, this transition also can be executed in daily scrum activity.

After these four goals are achieved, the *AUP* team can do some finishing work or *Scrum* team can so some sprint retrospective activity to end up the iteration.

There are two special atomic states shown at the bottom of Figure 8, '*Bugs obtained'* and '*Test cases obtained'*, can cut across their parent composite states. In agile process, team member can depict bugs or test cases at any time after user stories are obtained. Those bugs and test cases will be processed in transitions of '*Debugging and Acceptance testing'* respectively. This flexibility brings agility into the process.

*Goal Net* model has been applied to agent research field, as it provides a rich set of relationships and selection mechanism by providing a dynamic and highly autonomous agent problem-solving framework. Furthermore, a goal-oriented (GO) agent development methodology, namely GO methodology, based on *Goal Net* was also proposed in [114] by Shen et al. in 2005. GO methodology gives agent the ability to solve a complex problem by decomposing it into sub-goals. Sub-goals could be further decomposed until the hierarchical structure and the relationships of the goals are clearly defined. The temporal relationships and the transitions between the goals can be further identified. As a result, a *Goal Net* model can be constructed and serves as the brain of an agent or an autonomous system, which enables the agent or system to select the next goal to achieve selected goal, as well as to select the next action to pursue selected goal in a dynamic environment.

*Goal Net* also introduces an easy-to-use index card, which is called *Goal-Environment-Task* (GET) card [131]. GET card is an index card that is used to represent the environment and the task options for reaching the goal. The environment variables represent the environment situations during the goal pursuit. The three elements capture the essential dimensions of goal-oriented modeling based on *Goal Net*. GET card is an easy methodical approach in practice. We will equip it into our proposed method described in Chapter 3.

The research of *Goal Net* theory is still ongoing. As a modeling method, it's a novel way to present the overview goals structure of system. Its goal selection and action selection mechanism might also provide flexibility to the path selection and optimization for ASD process.

## 2.3 Task Allocation in Agile Software Development

Task allocation in agile software development is a challenging problem due to the need to efficiently utilize the team members' skills and capacity under changing environment and personal conditions for producing high quality software on time. Task allocation not only relies on information about internal properties of the tasks (such as priority, utility and effort level required), but also on human factors not depicted in product backlog (such as developers' mood, competency and the maximum effort per time step). A well-conceived task allocation strategy should reduce communication and coordination dependency between team members resulting in reduced delay and improved artifact quality.

From Table 2 and Table 4, we can see there are just a few research works in the field with primary focus on the problem of task allocation in ASD process. Nevertheless, the problem of task allocation in general software development process has been studied.

In 2006, Setamanit et al. proposed a hybrid computer simulation model of the software development process that is specifically architected to study alternative ways of configuring global software development projects. Their model has a hybrid system dynamics and discrete events, which includes phased-based, module-based, and follow-the-sun allocation strategies [132].

In 2009, Lamersdorf et al. conducted a series of researches on task allocation in global software development [133-135]. They introduced a model that aims to improve management processes in globally distributed projects by providing decision support for task allocation that systematically taking multiple criteria into account. The model uses existing approaches from distributed systems and statistical modeling. They also presented a customizable process for task allocation evaluation that is based on results from a systematic interview study with practitioners. During the process, the relevant criteria for evaluating task allocation alternatives are derived by applying the principles from goal-oriented measurement [136]. In addition, they integrated a risk model that is able to identify the possible risks for each assignment individually, an optimization model that uses Bayesian networks to suggest assignment alternatives with respect to multiple criteria, and an effort overhead model that is able to estimate the project effort for each assignment alternative, into the process for systematic evaluation and selection of task assignments in 2010 [137].

In 2012, Yilmaz and O'Connor introduced a market based mechanism to overcome task allocation issues in a software development process. They proposed a mechanism with a prescribed set of rules, where valuation is based on the behaviors of stakeholders (such as biding for a task) [138].

In practice, the strategy an agile team uses to sign up for work has significant implications for his work style and habits, and can ultimately impact the overall success of the iteration and its artifacts. Unfortunately, the agile community provides relatively little guidance on how this process should be carried out.

Existing task allocation approaches in ASD can be divided into three categories: 1) ad hoc task allocation, which allows any developer to sign up for any task he/she feels like working on that day; 2) dedicated developer/pair per story; or 3) user story swarming, in which the team swarms each user story in turn, seeing each one through to completion before starting the next. These approaches are mostly empirical, depending on the developers' intuitions and initiatives. There is a lack of an automated intelligent system that can produce task allocation plans that balance the considerations for quality and timeliness to improve the overall utility derived from an agile software development project.

## 2.4 Human and Social Aspects in Software Engineering

### *2.4.1 Human-Centered Software Engineering (HCSE) and Social Software Engineering (SSE)*

Human factors and social factors have a very strong impact to the success of software development and final system. The related researches on *SE* field, namely *Human-Centered Software Engineering* (HCSE) and *Social Software Engineering* (SSE), concern to the human and social aspects of software development process. One of the main observations in this field is that the concepts, principles, and technologies made for social software applications are applicable to software development itself as *SE* is inherently a social activity too. Accordingly, some methods and tools have been proposed to support different parts of HCSE/SSE, for instance, social system design or social requirements engineering.

A number of human factors or social factors have been subsequently mentioned by various researches in past decade.

In 2005, Miller et al. discussed that ethical analysis methods and related topics, which can inform a discussion about software development techniques when human values and ethical principles are considered, they suggested that all software engineers should have skill in some kind of ethical analysis, as well as another two human factors, utilitarian analysis and deontological analysis. The former helps a software engineer to think about consequences for developers, customers, users, and anyone else whose life may be affected by the software developed. The latter pushes a software developer into somewhat different emphases [5].

In 2005, Slaten et al. conducted a collective case study in a software engineering course at North Carolina State University to explore the effects of a collaborative pedagogy intervention on student perceptions. The pedagogy intervention was based upon the practices of *ASD* with a focus on pair programming. Six representative students in the course participated in the study. Their perspectives helped validate a social interaction model of student views. The findings showed that agile software methodologies contribute to more effective learning opportunities for computer science students, such as making them more confident and increasing interests to IT [139].

In 2006, Korkala et al. presented the empirical results from four different case studies to test the different communication and feedback methods. Three case studies had partially onsite customers and one had an onsite customer. The case studies used face-to-face communication to different extents along with email and telephone to manage customer-developer communication in the development iterations. Their results indicate that an increased reliance on less informative communication channels results in higher defect rates. These results suggest that the selection of communication methods, to be used in development iterations, should be a factor of considerable importance to agile organizations working with partially available customers. They also proposed some guidelines for selecting proper communication methods [140].

In 2006, Pichler et al. reported on the challenges and experiences gained during a three years multidisciplinary software development project in the insurance domain, focusing on the employed requirements process. Their lessons learned of applying an agile requirements process under the conditions of traditional processes at the customer's attitude; geographically distributed offices of the customer and development team; diverse interests of involved customers' departments, administrative and operational staff; limited availability of field workers etc. are provided as recommendations for other research [141].

In 2007, Whitworth et al. investigated the social natures that contribute to success of agile methodologies. They used qualitative grounded theory to explore socio-psychological experiences in agile teams, where agile teams were viewed as complex adaptive socio-technical systems. They found the end-goals and positive sources of motivation, such as pride, are very important for agile team. Their results support an understanding of how social identity and collective effort are supported by agile methods [6].

In 2008, Ahmadi et al. presented a survey of *SSE* relevant works from psychology, mathematics and computer science studies. They identified and discussed two main subcategories: the need to integrate results from social and psychological sciences in the software development lifecycle, and the need for engineering social networking services and collaborative tools. They also presented a set of mathematical methods that have been used for experimental validation of scientific contributions, which may be used as basic blocks for understanding the ideas proposed for social software engineering [142].

In 2008, Whitworth tried to explore the connection of agile software development and team cohesion by a qualitative study involving 22 participants in agile teams. He discussed participant experiences as seen through a socio-psychological aspect. It draws from social-identity theory and socio-psychological literature to explain, not only how, but why agile methodologies support team work and collective progress. Agile practices are shown to produce a socio-psychological environment of high performance, with many of the practical benefits of agile practices being supported and mediated by social and personal concerns [7].

In 2010, Esfahani et al. discussed some human factors in agile methods, such as improper role assignment, neglected team dependencies, and overlooked required skills, which have all been reported as reasons for failures during agile process. They advocated the use of goal-oriented modeling techniques to depict social aspects of agile methods. These social models can be used to identify the key factors that contribute to the success or failure of an agile method, thus providing guidance early during the introduction of the method in an organization [143].

Table 6 compares those methods, key points and applications of above research works during the past decade.

Table 5. Comparative study of HCSE/SSE research during the past decade

| Year | Author | Key Points | Methods &Human Factors |
|---|---|---|---|
| 2005 | Miller et al. [5] | Ethical analysis; Utilitarian Analysis; Deontological Analysis | human ethic; culture; |
| 2005 | Slatenet al. [139] | social interaction model of student views in agile process | confidence; interest |
| 2006 | Korkala et al. [140] | The selection of communication methods, to be used inside development iterations, should be a factor of considerable importance to agile | communication; feedback |
| 2006 | Pichleret al. [141] | Propose 10 recommendation to meet 7 social challenges | customer's attitude; geographically distributed teams; diverse interests; , |
| | | | administrative and operational people; limited availability |
| 2007 | Whitworth et al. [6] | found the end-goals and positive sources of motivation, such as pride, are very important | socio-psychological experiences; pride |
| 2008 | Ahmadiet al. [142] | The need to integrate results from social and psychological sciences in the software lifecycle; the need for engineering social networking services and collaborative tools. | social software engineering; social networking services; collaborative tools |
| 2008 | Whitworth[7] | Agile practices are shown to produce a socio-psychological environment of high performance, with many of the practical benefits of agile practices being supported and mediated by social and personal concerns | Social identity and in-group out-group bias, individual perceptions of security, efficacy, and control |
| 2010 | Esfahani et al.[143] | Current process modeling languages are not designed for describing or analyzing such human-related issues; goal-oriented modeling techniques | improper role assignment; neglected team dependencies; overlooked required skills |
| 2011 | Palacios et al. [144] | Affect Grid; Emotional assessment in Software Requirements Engineering | Emotions; Pleasure; Arousal |

All in all, those researchers on HCSE or SSE have noted that the importance of human factors to software development, such as goal, culture, confidence, pride and interest etc. However, few of them try to address another very important human factor: mood or mood swing, which is important to the team morale for ASD and is one of our research objectives.

### *2.4.2 User Centered Design (UCD)*

*User Centered Design (UCD)* is a design process focusing on user experience design, user interface design and usability evaluation. *Agile* emphasizes people, communication and the ability to adapt to change. Equipping *UCD* into ASD process might create a comprehensive user-centered software development methodology, and increase the chances to deliver a successful project. There are no inherent obstacles that could prevent such integration, but both of them need to change before it can succeed. Although there is no a unified *Agile-UCD* methodology has been established, there has been a growing interest in learning how to integrate these two proven approaches over the last decade.

In 2003, Kenia et al. did some research works intending to present three workflows for a new software development method, which besides focusing on cost and schedule, also includes some *HCI* aspects along the software development life cycle, such as usability, accessibility, acceptability requirements, guidelines application, model-based *User Interface (UI)* generation techniques, and evaluation techniques. They proposed a new method *UPi (Unified Process for Interactive systems)* based on *RUP*. The purpose of this integration is to develop interactive systems that are easy to learn and use, therefore, to help users in performing their daily tasks in an efficient manner [145].

In 2004, Beyer et al. proposed an agile user-centered method: *Rapid Contextual Design* that incorporates customer-centered techniques such as *CD* to provide additional solutions to the real problems recognized by agile methods. Their solution works in combination with agile methods' strengths resulting in a process that incorporates the customer goals, provides room for *UI* and user interaction design in agile process [146].

In 2005, Hodgetts discussed the coaching experiences of integrating sophisticated *User Experience Design (UED)* practices into the *Agile* process initiatives of several organizations, and integrating their *UED* best practices into the incremental, collaborative world of agile processes. In their paper, they thought agile processes are typically presented from the point of view of programmers, with the other disciplines often left feeling excluded and disenfranchised, such as *UED*. But in face the *UED* activities span the full lifecycle of software development, from early requirements analysis to construction and testing, with its work products forming key inputs and deliverables of many software development activities [147].

In 2006, Silva et al. proposed a streamlined approach to *HCI* design called extreme designing that follows on the principles of agile methods and is analogous to extreme programming, which brings together the advantages of sketching and prototyping as a communication tool, and of interaction modeling as a glue that binds together the sketches to allow designers to gain a more comprehensive view of and to reflection on the interactive artifact, thus promoting a more coherent and consistent set of design decisions [148].

In 2006, Chamberlain et al. reported a field study designed to investigate the use of agile methods alongside *UCD* in one particular organization. They compared the similarities and differences between *UCD* and *Agile* Development, and aimed to develop a framework for use by project teams wishing to integrate *UCD* practices with agile development. The study gave us five principles for integrating *UCD* and agile development, including *User Involvement, Collaboration and Culture, Prototyping, Project Lifecycle* and *Project Management* [149].

In 2007, Ferreira et al. thought that the integration of *UI* design into agile development is not well understood, as both agile development and *UI* design are iterative. While agile methods iterate on code with iterations lasting weeks, *UI* designs typically iterate only on the user interface using low technology prototypes with iterations lasting hours or days. Similarly, both agile development and *UI* design emphasize testing, but agile development involves automated code testing, while *UI* must done by expert inspectors or ideally potential end users. Then they proposed a qualitative grounded theory based on study of real agile projects involving significant *UI* design. The key results from their study are that agile iteration facilitates usability testing, as it allows software developers to incorporate results of those tests into subsequent iterations, therefore, it can significantly improve the quality of the relationship between *UI* designers and software developers[150]. They also interviewed interaction designers and other team members on two *XP* teams and reported on how they combined interaction design activities with *XP* [151].

In 2007, Williams et al. compared the *UCD* experiences associated with supporting non-agile projects and agile projects. Through their studies, we can see that agile and *UCD* methods are not at odds with each other. The iterative approach to agile team is a natural fit for *UCD*. They have successfully demonstrated how two *UCD* teammates on the agile team can aid in upfront and continual user input [152].

In 2008, Fox et al. conducted a study with participants that have previously combined Agile and *UCD* two methodologies. Their findings, combined with existing work show that the existing model used for Agile *UCD* integration can be broadened into a more common model. They described three different approaches taken by participants to achieve this integration, including Generalist, Specialist, and the Hybrid approach [153].

In 2008, Najafi et al. found that by incorporating *UED* in agile development, user research and testing can be utilized to prioritize features in the product backlog and to iteratively refine designs integrating *UED* and Agile to achieve better usability. Furthermore, processes can be accomplished with little or no impact on release schedules. Their two case studies demonstrated that the benefits of involving the User Experience team and applying *UED* practices to the agile development process more than offset the risks of potentially impacting product release dates. *UED* practices are iterative in nature and naturally complement the iterative nature of agile development. However, successful integration of the *User Experience* team requires full cooperation and collaboration with all cross-functional team members. Understanding users' expectation and goals helps to prioritize features in backlog. Consistent user testing and refinement of designs ensures that the product is developed to meet the needs and goals of its users [154].

In 2008, Singh proposed an agile methodology for promoting usability named *U-SCRUM*. He pointed out that the selected user stories in scrum process may not be good enough from the usability perspective, and also user stories of usability import may not be prioritized high enough. Therefore, given the fact that a product owner thinks in terms of the minimal marketable set of features in a just-in-time process, it is difficult for the development team to get a holistic view of the desired product or features. So they proposed *U-SCRUM* as a variant of the scrum methodology try to tackle this problem. Unlike typical scrum, where at best a team member is responsible for usability, *U-SCRUM* uses two product owners, one focused on usability and the other on the more conventional functions. Their preliminary result has showed that *U-SCRUM* improved usability than scrum [155].

In 2008, Ungar described their experience with the merger of *UCD* into agile development practice as manifest in a one day design studio. The design studio brings the domains of *UCD* and agile software development together in ways that benefit practitioners of both. They have observed these benefits in the course of daily work and seen the overall quality of designs improve as understanding of *UCD* and design best practices spreads among the teams and throughout the organization. So they encourage others attempting to practice *UCD* in an agile environment [156].

In 2009, Peixoto et al. discussed some limitations of agile methods and the results of using Scrum in a specific project. They found that good practices in development of *HCI* have been left aside in favor of reducing sprint duration. The behavior of listening to what customers say instead of watching what customers do for developers in agile process has a tendency to impact negatively the *HCI* usability. They also pointed the common characteristic between agile development and *HCI* design: the repetition aspect. They proposed a knowledge base representation of good practices in *HCI* design. A semantic network is used to represent main concepts in *HCI* design, which tried to use the conceptual modeling to implement an expert system to guide agile developers during *HCI* design [157].

In 2009, Wirfs-Brock suggested agile designers should to sharpen their communication and collaboration skills as well as their technical practices. They should value collaboration and collective understanding as much as good design and development practices. It’s a matter of attitude more than any specific technique or process [158].

In 2011, Salah proposed a *Software Process Improvement (SPI)* framework for *Agile* and *User Centered Design Integration (AUCDI)* by providing generic guidelines and practices for organizations aspiring to achieve *AUCDI*. For addressing *AUCDI*, the challenges include introducing systematicity and structure into *AUCDI*, assessing *AUCDI* processes, and accommodating project and organizational characteristics [159].

Table 7 compares those researches during the past decade, including their methods, key points and applications.

Table 6. Comparative study of *UCD-ASD* research during the past decade

| Year | Author | Methodology | Key Points | Application |
|---|---|---|---|---|
| 2003 | Kenia et al. [145] | UPi | based on RUP; includes some HCI aspects | Design and development stage of interactive systems |
| 2004 | Beyeret al. [146] | Rapid Contextual Design | incorporating customer-centered techniques such as CD | Design stage in agile process |
| 2005 | Hodgetts [147] | UED | sophisticated UED practices | agile process initiatives and incremental, collaborative world of agile processes |
| 2006 | Silva et al. [148] | Extreme Designing | Sketching; prototyping | Design activities for agile process |
| 2006 | Chamberlain et al. [149] | Case studies | User Involvement; Collaboration and Culture; Prototyping; Project Lifecycle; Project Management | Framework of Integrating Agile<br>Development and User-Centered Design |
| 2007 | Ferreira et al. [150, 151] | interview | qualitative, using grounded theory based on interviews | Design activities for agile process |
| 2007 | Williams et al. [152] | Case studies | UCD methods and principles can work well within an agile | UCD for agile process |
| | | | development process | |
| 2008 | Fox et al. [153] | Interview | a qualitative study using grounded theory approach | Agile UCD integration |
| 2008 | Najafi et al. [154] | Case studies | applying UED practices to the agile | UED practices for agile process |
| 2008 | Singh [155] | U-SCRUM | two product owners in team; one focused on usability | UCD for scrum team |
| 2008 | Ungar [156] | design studio | overall quality of designs is improved by Agile UCD | UCD for agile process |
| 2009 | Peixoto et al. [157] | semantic network | knowledge base HCI design | To guide agile developers during HCI design |
| 2009 | Wirfs-Brock[158] | attitude | communication and collaboration skills | agile attitude for UCD |
| 2011 | Salah[159] | SPI | Systematicity; structure; assess process; accommodating project | UCD for agile process |

All in all, from above works, we can see currently there are no systematical and clear principles or guidelines for practitioners to execute successful integration of *UCD* and *Agile*. In addition, as substantial differences exist between *Agile* and *UCD* approaches, the practitioners only can individually apply strategies, principles and methods in practice. For applying *UCD* to agile design, we need to answer some user-centered questions about users and their tasks and goals in advance, such as:

- Who are the users of the product?
- What are the users' tasks and goals?
- What are the users' experience levels with the product, and product like it?
- What functions do the users need from the product?
- What information might the users need, and in what form do they need it?
- How do users think the product should work?
- What are the extreme environments?
- Is the user multitasking?
- Does the interface utilize different inputs modes such as touching, spoken, gestures, or orientation? [160]

Then the answers will be used to make decisions for user centered design and the implementation in agile design process. Anyway, this survey gave us lot of new thoughts for this research.

### *2.4.3 Human Computation System (HCS)*

According to Edith and Luis's definition, "human computation is the idea of using human effort to perform tasks that computers cannot yet perform, usually in an enjoyable manner" [161]. In the survey of human computation systems made by Yuen et al. [162] in 2009, they claim that "human computation is a technique that makes use of human abilities for computation to solve problems. The human computation problems are the problems those computers are not good at solving but are trivial for humans", which usually apply to researches on social game, image search and *Computer Human Interaction* (CHI) etc., such as CAPTCHA, ESP Game, Verbosity and so on [163]. In this book, we investigate human computation research as we treat the ASD process as a human computation system because that ASD process uses human effort to perform tasks that computers are not good at solving, such as goals recognition and definition, task splitting and allocation, mood and morale tracking and monitoring etc. To assist agile teams better using computers to serve decision making in this human computation system we will propose and discuss a set of techniques as shown in Figure 9.

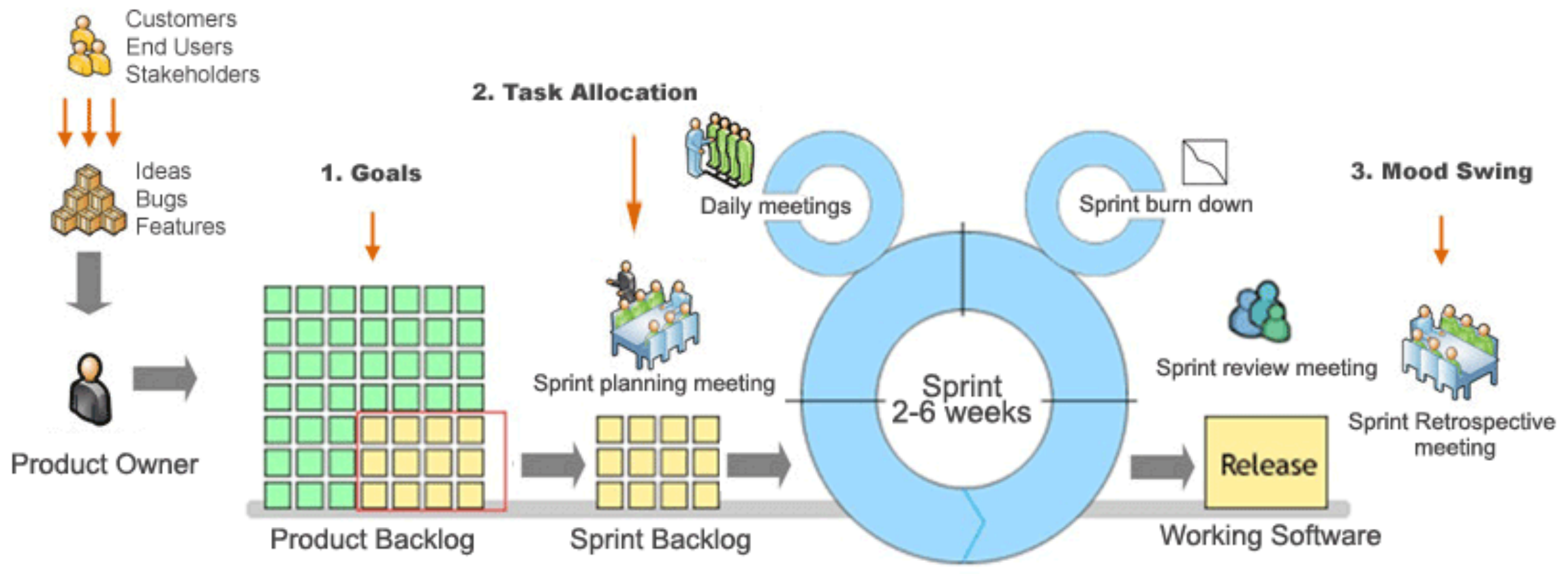

Figure 9. Scrum process as a human computation system

Firstly in Chapter 3 we will discuss requirements/goals management in ASD process. For a common ASD process, we propose a light weight Goal Net based method to model structured goal requirements according to existing user stories produced in the process. The approach has been applied in university level agile software engineering education. It was shown to result in significant improvement in the number and quality of user stories generated by students compared to the past approach.

Secondly in Chapter 4 we will discuss task allocation strategy in ASD process. We propose a *Simple Multi-Agent Real-Time* (SMART) task allocation approach for agile process based on distributed constraint optimization. Personal task allocation agents equipped with SMART helps individual developers to make situation-aware decisions on which incoming tasks to serve so as to reduce the risk of low quality task results while maintaining the timeliness of the overall software project.

Thirdly in Chapter 5 we will discuss the relationship between mood swing of developer and task execution in ASD process. We proposed a FCMs-based method to monitor mood swing of team members in agile process, which might help team leader to find right developer to implement tasks when some event occurred or some condition is triggered during ASD process.

Finally in Chapter 6 we will discuss more empirical insights about the task allocation, collaboration, and the team morale in Scrum agile process based on the data collected from our experiments, which help us to understand more about this human computation system, and then we can try to improve it through more focusing on human factors, such as using computer to make up for the shortcomings of humanity, or using computer to enhance human capacity etc.

## 2.5 Summary

In this chapter, we have reviewed some theories and methods related to our research. We did the literature reviews on many ASD methodologies, and the state-of-the-art on researches of *GORE, Goal Net, HCSE/ SSE, UCD, CHI,* and task allocation for ASD, or SE.

The comparisons of those research works have also been highlighted. They are foundations and starting point of our research work. From the literature review, we can see that based on current technologies and studies, there is no a systematical and full-fledged human-centered

ASD methodology to guide agile team to automatically manage goals, task allocation and mood swing activities during the development process.

# CHAPTER 3

# BUILDING GOAL-ORIENTED MODEL FOR ASD

Agile methodologies use user stories to capture software requirements. This often results in team members over emphasizing their self-understanding of these goals, without proper incorporation of goals come from other stakeholders and customers. Existing UML or other goal oriented modeling methods tend to be overly complex for non-technical stakeholders to properly express their goals and communicate them to the agile team [164]. In this chapter, we propose a light weight Goal Net based method to model goal requirements in agile software development process to address this problem. It can be used to decompose complex processes into phased goals, and model low level user stories to high level hierarchy goal structures. Our preliminary analysis and studies in educational software engineering contexts show that it can improve agile team's group awareness to project goals and, thus, improve team productivity and artifact quality. The proposed approach was evaluated in university level agile software engineering projects. It has achieved an improvement of over 50 percentage points in terms of the proportion of high quality user stories generated by students compared to the standard user story template used in *Scrum*.

## 3.1 Background

Instead of using requirement analysis and modeling activities in other plan-driven methodologies, ASD uses lean user story techniques to capture software requirements. However, there is more to requirements than user stories. For example, where is the business value in a user story? What are reasonable arguments for implementing a feature? Comparing to requirement, goal is more subjective but more important human factor in software development process. Goals may be formulated at different levels of abstraction, ranging from high-level, strategic concerns (such as “enhancing end-users’ satisfactions” or “providing ubiquitous payment service” for an online shopping system) to low-level, implementation concerns (such as “speeding up the display of product search results” or “supporting payment with *VISA* credit card in the system”). Goals can also cover different types of concerns: functional concerns associated with software features and non-functional concerns associated with quality of service (e.g. safety, security, accuracy, performance). The multidimensional hierarchy characteristics of goals bring complexity and difficulty for modeling goals in the software development process.

Existing research works have used traditional requirement modeling methods, such as User Case method, to model goals [165, 166]. However, for ASD, UML is too complex to apply and implement, especially when non-technical stakeholders such as customers are involved [164]. Agile team uses user story to depict requirements. A user story may include functions, features, enhancements, bugs, and so on. A user story is a short, simple sentence described from the perspective of the person who desires a new capability or expectation, usually a user of the system. User stories typically follow a simple template:

*As a* **<*role*>**, *I want to* **<*goal/desire*>** [*so that* **<*benefit*>**]

Elements in the angled brackets are to be specified and elements in the square brackets are optional. [167]

User story is often written on index card or sticky note, arranged on walls or tables to facilitate discussion. Therefore, the team can strongly shift the focus from writing features to talking about them. In fact, these discussions and communications are more important than whatever the text is written.

In a *Scrum* team, *Product Owner* (PO) is primarily responsible for user stories. But other team member also can contribute to them. In practice, many users write user stories. The first requirement may come from an end user. Others, such as the product owner, architect, scrum master, or business analyst etc., can update them.

User stories are often written in a non-technical manner from the perspective of an end user. This user story will be further defined. After fine tuning the stories to an extent that it can be put to review to the agile team, the entire agile team will work on these stories to understand it. Any technical constraints or limitations need to be noted down and presented to the customer. Finally, the user stories will be stored in the product backlog, and divided into small tasks for ASD team members to implement. The product backlog is a prioritized list of functionalities that will be developed into a software product or service.

One of the benefits for agile user stories is that they can be written at varying levels of detail. We can write user stories that cover large number of functionalities. These large user stories are generally known as “epics”. Here is an example epic for an online B2C marketplace services:

> *As a **customer**, I want to **pay via mobile phones** so that **I can buy goods on mobile phones quickly**.*

As an epic is generally too large for an agile team to complete in one iteration, it needs to be split into multiple smaller user stories first. The epic above can be split into many smaller user stories, for example:

*As a* ***VIP customer****, I want to* ***pay cash on delivery*** *so that* ***I can buy goods on mobile phones without paying immediately****.*

*As a* ***common customer****, I want to* ***be able to pay by credit card*** *so that* ***I can buy goods on mobile phones quickly****.*

Table 8-10 show examples of user stories at different levels of abstraction.

Table 7. An example of user story list

| ID | As a/an | I want to… | so that… |
|---|---|---|---|
| 1 | visitor | Easily search goods on mobile phones | I can find my favorite goods with no digital divide |
| 2 | visitor | Easily sort the search results | I can find my favorite goods according quickly |
| 3 | customer | Quickly pay via mobile phones | I can buy goods on mobile phones quickly |
| … | … | … | … |

Table 8. A detailed user story list extended from Table 8

| ID | As a/an | I want to… | so that… |
|---|---|---|---|
| 1 | visitor | Easily search goods on mobile phones | I can find my favorite goods with no digital divide |
| 1.1 | visitor | Search goods on mobile phones by voice input | I don't need to type |
| 1.2 | visitor | Search goods on mobile phones by clicking a category | I can find my favorite goods according to category what I'm choosing |
| 2 | visitor | Easily sort the search results | I can find my favorite goods quickly |
| 2.1 | visitor | Sort the result according to price | I can find my favorite goods at bargain prices |
| 2.2 | visitor | Sort the result according to location | I can find my favorite goods near me |
| 3 | customer | Quickly pay via mobile phones | I can buy goods on mobile phones quickly |
| 3.1 | VIP customer | Choose to pay cash on delivery | I can buy goods on mobile without paying first |
| 3.2 | common customer | Choose to pay by credit card | I can buy goods and pay on mobile quickly |
| … | … | … | … |

Table 9. A detailed user story list with tasks

| ID | As a/an | I want to… | so that… |
|---|---|---|---|
| 1 | visitor | Easily search goods on mobile phones | I can find my favorite goods with no digital divide |
| 1.1 | visitor | Search goods on mobile phones by voice input | I don't need to type |
| | Task 1.1.1 | Investigate voice input solutions for mobile phones | |
| | Task 1.1.2 | Design a new User Interface (UI) | |
| | Task 1.1.3 | Choose one solution and implement it on mobile phones | |
| | Task 1.1.4 | Integration and unit test | |
| 1.2 | visitor | Search goods on mobile phones by clicking a category | I can find my favorite goods according to category what I'm |
| | | | choosing |
| | Task 1.2.1 | Design a category tree | |
| | Task 1.2.2 | Design a new UI | |
| | Task 1.2.3 | Implement the function on mobile phones | |
| 2 | visitor | Easily sort the search results | I can find my favorite goods according to the results what I'm sorting |
| 2.1 | visitor | Sort the result according to price | I can find my favorite goods at bargain price |
| | Task 2.1.1 | Design a new UI | |
| | Task 2.1.2 | Adjust database structure to support this function | |
| | Task 2.1.3 | Implement the function on mobile phones | |
| 2.2 | visitor | Sort the result according to location | I can find my favorite goods near my place |
| | Task 2.2.1 | Design new UI | |
| | Task 2.2.2 | Adjust database structure to support this function | |
| | Task 2.2.3 | Implement the function on mobile phones | |
| 3 | customer | Quickly pay on mobile phones | I can buy goods on mobile phones quickly |
| 3.1 | VIP customer | Choose to pay by delivery | I can buy goods on mobile quickly without paying for now |
| | Task 3.1.1 | Adjust business flow to support this requirement | |
| | Task 3.1.2 | Design new UI | |
| | Task 3.1.3 | Implement the function | |
| 3.2 | common customer | Choose to pay by credit card | I can buy goods and pay on mobile phones quickly |
| | Task 3.2.1 | Design new UI | |
| | Task 3.2.2 | Adjust database structure to support this function | |
| | Task 3.2.3 | Implement the function on mobile phones | |
| … | … | … | … |

The split user stories then will be stored into backlog. In practice, there are three types of backlog used in agile process (e.g. *Scrum*) as follows:

- *Product Backlog*: A list of customer requirements for entire product, including user stories, bugs and features that need to be handled. Figure 9 shows an example of product backlog from a real project.

| | Item # | Description | Est | By |
|---|---|---|---|---|
| **Very High** | | | | |
| | 1 | **Finish database versioning** | 16 | KH |
| | 2 | **Get rid of unneeded shared Java in database** | 8 | KH |
| | - | **Add licensing** | - | - |
| | 3 | Concurrent user licensing | 16 | TG |
| | 4 | Demo / Eval licensing | 16 | TG |
| | | **Analysis Manager** | | |
| | 5 | File formats we support are out of date | 160 | TG |
| | 6 | Round-trip Analyses | 250 | MC |
| **High** | | | | |
| | - | **Enforce unique names** | - | - |
| | 7 | In main application | 24 | KH |
| | 8 | In import | 24 | AM |
| | - | **Admin Program** | - | - |
| | 9 | Delete users | 4 | JM |
| | - | **Analysis Manager** | - | - |
| | 10 | When items are removed from an analysis, they should show up again in the pick list in lower 1/2 of the analysis tab | 8 | TG |
| | - | **Query** | - | - |
| | 11 | Support for wildcards when searching | 16 | T&A |
| | 12 | Sorting of number attributes to handle negative numbers | 16 | T&A |
| | 13 | Horizontal scrolling | 12 | T&A |
| | - | **Population Genetics** | - | - |
| | 14 | Frequency Manager | 400 | T&M |
| | 15 | Query Tool | 400 | T&M |
| | 16 | Additional Editors (which ones) | 240 | T&M |
| | 17 | Study Variable Manager | 240 | T&M |
| | 18 | Haplotypes | 320 | T&M |
| | 19 | **Add icons for v1.1 or 2.0** | - | - |
| | - | **Pedigree Manager** | - | - |
| | 20 | Validate Derived kindred | 4 | KH |
| **Medium** | | | | |
| | - | **Explorer** | - | - |
| | 21 | Launch tab synchronization (only show queries/analyses for logged in users) | 8 | T&A |
| | 22 | Delete settings (?) | 4 | T&A |

Figure 10. An example of product backlog (source: mountain goat software)

- *Release Backlog*: A list of user stories, features and bugs that should be implemented in defined release. Figure 10 shows an example of release backlog from a real project.

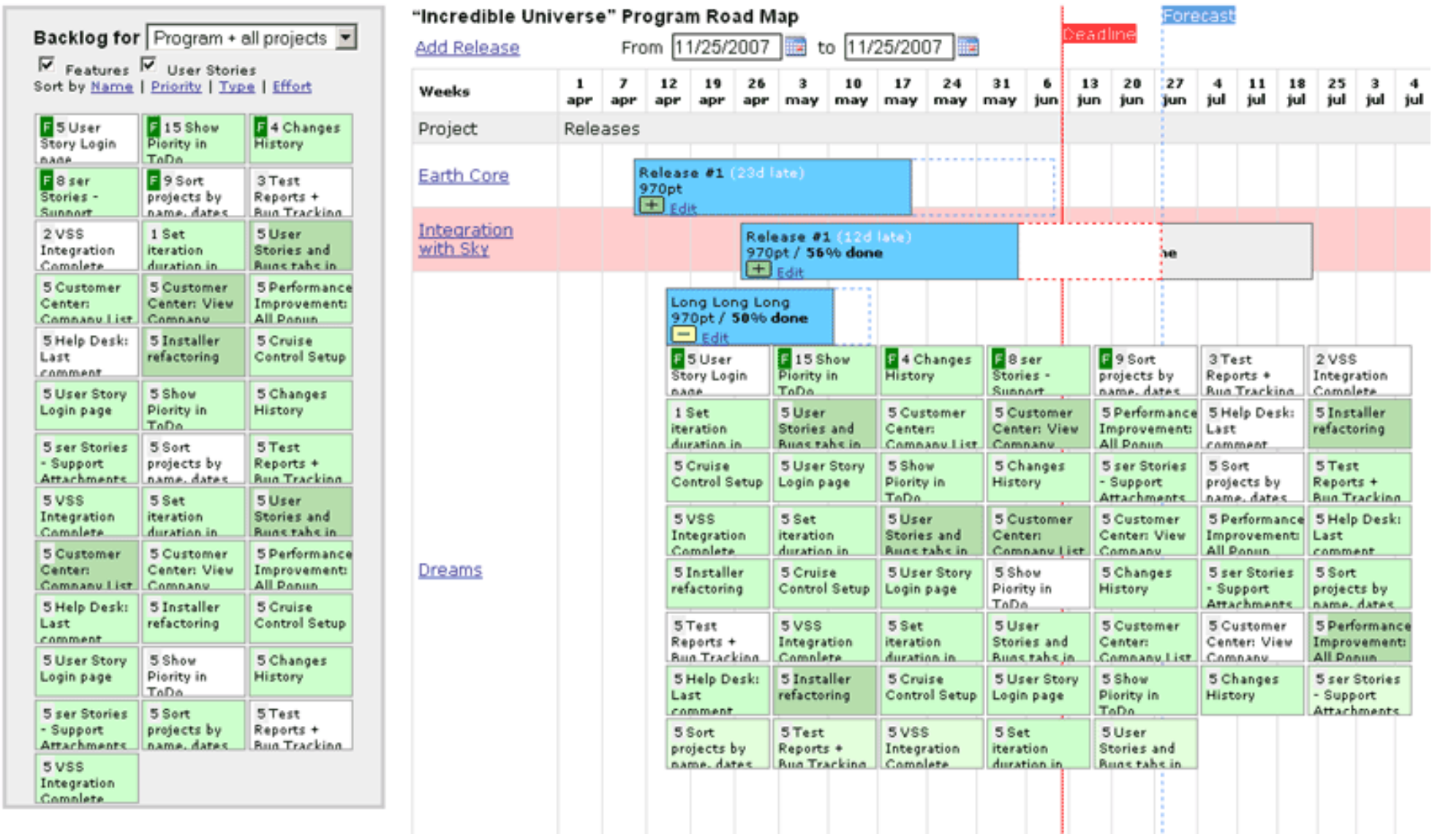

Figure 11. An example of release backlog (source: TargetProcess)

- *Iteration Backlog* (*Sprint Backlog* in *Scrum*): A list of user stories, features and bugs that should be implemented in defined iteration (e.g. one sprint in *Scrum*). Figure 11 shows an example of sprint backlog from a real project.

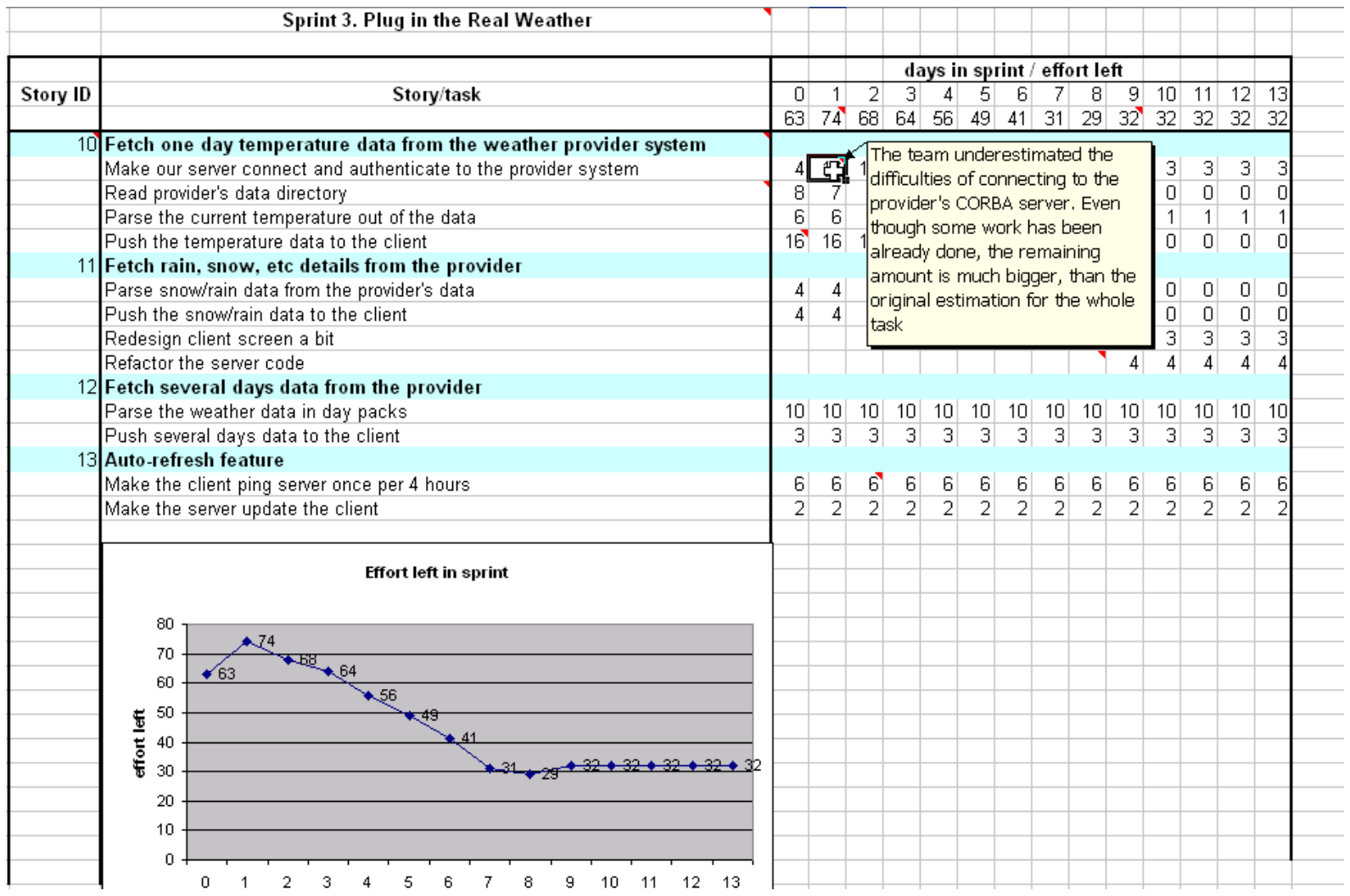

Figure 12. An example of sprint backlog (source: Agile Software Development Blog)

## 3.2 Problems

From the above examples, we can see that the *Agile*'s user story method has provided a way to present users' goals via an ***"I want to"*** clause in the template. To some extent the hierarchy of user stories also reflects the hierarchy of goals. However, in practice many agile teams, especially novice teams, often ignore this hierarchical structure between goals when the project enters the detailed iterative development process. We have conducted an eight week group-based agile software engineering project as part of an undergraduate course work in *Beihang University* from 01/04/2012 to 31/05/2012. We collected 118 user stories and they were split into 726 tasks by 17 teams with an average team size of 6 persons. By analyzing the user stories data, we discovered that about 4/5 of them ignored the hierarchical relationships, as most of them just described the developers' goals without considering users/other stakeholders' goals, which resulted in the grain size of their user stories is not good.

In addition, current approaches are based on nature language expressions, which can be ambiguous. This is especially true when requirements are not so clear for customers themselves. This situation occurs more often in the beginning of the software projects. If there is a well-defined modeling theory to support agile teams to model hierarchical goal structures, it will better facilitate the teams to understand the requirements. A graphical goal model can be more intuitive and effective for product owners to inspect and understand stakeholders' real goals too.

## 3.3 Method

Based on the *Goal Net* theory that we discussed in section 2.2.2, we propose a light weight method to enhance the ASD process. The proposed method consists of the following three steps:

1) *Defining High Level Goals:* to define stakeholders' high level goals by interviews;
2) *Identifying Middle Level Hidden Goals:* to identify different middle goals hidden in user stories;
3) *Modeling Hierarchical Goal Structure:* to model goal structure according to *Goal Net*. The improvement will not incur extra effort for the PO or other ASD team members.

## 3.4 Example

**Assumption**: an agile team is developing a mobile shopping app for iPhones. Before a new iteration/sprint, some elderly end users felt the working system was not easy to use. Therefore, they provided the product owner with new feedbacks. Based on these feedbacks, the PO has created some user stories, part of backlog is been listed in Table 8, 9 and 10. Then, the PO follows our proposed 3-step method to model the hierarchical goal structure.

**Input**: user story lists in Table 8, 9 and 10 (partially)

**Output**: a Goal Net model

**Modeling Steps:**

*Step 1. Defining high level goals by top-down approach*

After interviewing the elderly users, the PO knows that they actually want to enhance the user experience of the working system. The PO then clusters the user stories into four high-level goals: 1) improved user interface, 2) clearer navigation system, 3) friendly help system, and 4) simplified work flow. Therefore, the PO firstly designs an initial high-level Goal Net diagram shown in Figure 12:

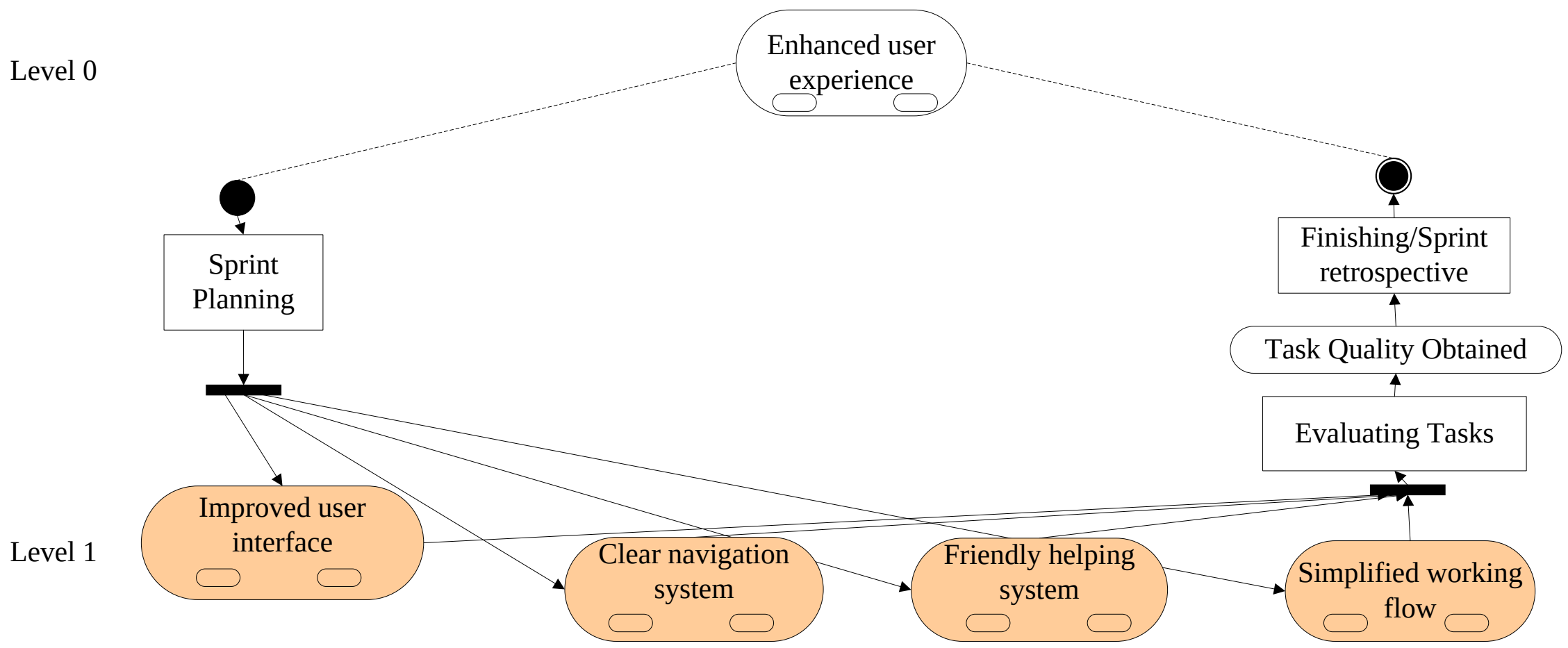

Figure 13. Initial Goal Net model without detailed goals/requirements

*Step 2. Identifying middle level hidden goals by bottom-up approach*

Based on the initial high level Goal Net model, the PO needs to find middle level hidden goals for each user story according to the ***"I want to"*** clause. For example, the hidden goal of user story 1.1 and 1.2 in Table 9 is "Easily search goods on mobile phones".

*Step 3. Modeling goal structure by Goal Net approach*

After the above two steps, the PO is able to build the full Goal Net diagram which is as shown in Figure 13. In this case, the full Goal Net model has four levels. In level 2, two goals of ***"Easily search goods on mobile phones"*** and ***"Easily sort the search results"*** are linked to their parent goal ***"Improved user interface"*** in level 1. The goal of ***"Quickly pay on mobile"*** is linked to its parent ***"Simplified work flow"*** in level 1. The goals depicted in the ***"I want to"*** clause of sub-user stories are linked to their respective parent goals in level 2.

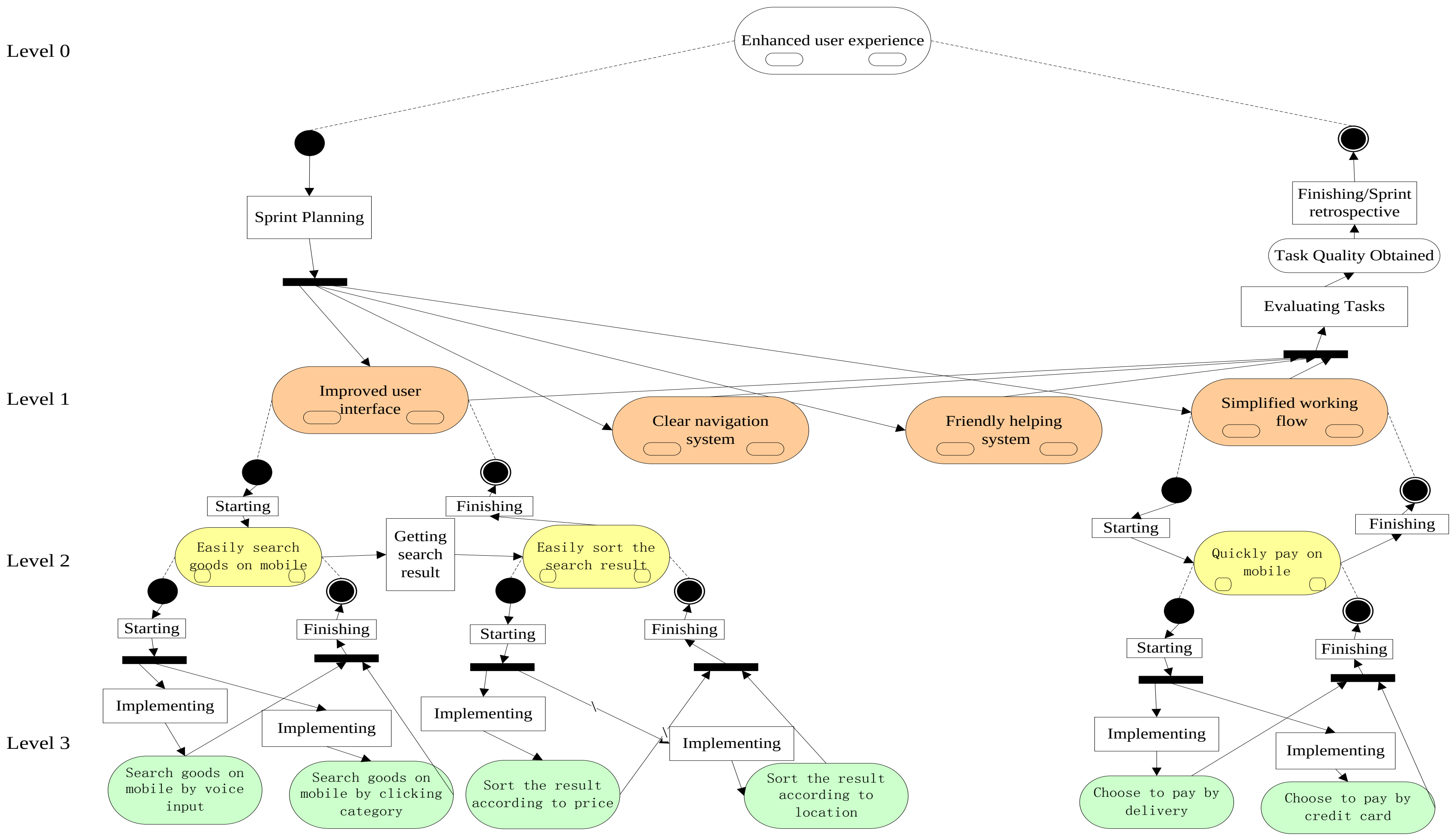

Figure 14. Final Goal Net model with detailed goals/requirements

Necessary sequence transitions, concurrency transitions, and synchronization transitions are included in this *Goal Net* model. For example, the transition between two sub-goals in level 2, ***"Easily search goods on mobile phones"*** and ***"Easily sort the search results"***, are sequence transitions. This is because only after executing ***"getting search result activities"*** can the implementations related to the goal of ***"Easily sort the search results"*** start. Three pairs of ***"Implementing activities"*** under level 2 are concurrency transitions, as after achieving two sub-goals, the process needs to be synchronized to reach their parent goal, so that these two ***"Implementing activities"*** can be executed by two or more developers concurrently.

The transition activity in the *Goal Net* represents a list of tasks for implementing the goals (user stories). For example, the ***"Implementing activity"*** for the goal ***"Search goods on mobile phones by voice input"*** can be depicted by the following *Goal-Environment-Task* (GET) card [131].

| **Goal 1.1**: Search goods on mobile phones by voice input | |
|---|---|
| **Environment Variables** | **Tasks** |
| 1. Man Power (location, role): David (LA, UI Designer), Michael (BJ, Developer), Grace (SG, Developer)<br>2. Device: iPhone5, iPhone 4s, HTC One<br>3. Third-part mobile voice input solutions<br>- Speech Input API for Android (Google)<br>- Xunfei Voice recognition API (USTC Xunfei) | 1.1.1 Investigate voice input solutions for mobile phones<br>1.1.2 Design new UI<br>1.1.3 Choose one solution and implement it on mobile phones<br>1.1.4 Integrate into the working system |

Figure 15. An example of GET card

The final *Goal Net* model can be used in subsequent iterations/sprints and can be updated with more user stories. It can help an entire ASD team review the user stories during iteration/sprint planning meetings and verify the working software retrospectively.

## 3.5 Evaluation

Since 2011, the *College of Software* at *Beihang University* has introduced Scrum into practical software engineering courses to train development and management skills for undergraduate students. These students were divided into teams with an average team size of 6 persons to carry out an 8-10 week group-based software development project. There are typically around 20 teams during each semester. All teams possess similar skill levels and backgrounds and adopt the Scrum process during the project.

At the end of each semester, the quality of the user stories created by the teams were graded by the course instructors with a score representing how well they reflect stakeholders' goals. Table 11 shows the comparison result between class 2011, 2012 and 2013.

Table 10. Comparison of teaching results of ASD course

| **Class** | **2011** | **2012** | **2013** |
|---|---|---|---|
| Using Goal Net to model user stories | No | No | Yes |
| Total number of student teams | 26 | 17 | 20 |
| Total number of user stories | 189 | 118 | 122 |
| The average number of user stories per team | 7.3 | 6.94 | 6.1 |
| Standard deviation of the number of user stories per team | 3.3 | 3.37 | 1.62 |
| The proportion of High Quality User Stories reflecting stakeholders' goals | 21% | 19% | 74% |

From Table 11, we can see that as we introduced proposed method and *Goal Net* model into the class 2013, the standard deviation of number of user stories per team decreased significantly and the proportion of high quality user stories increased significantly. Overall all teams understood more clearly for the stakeholders' goals and recorded them as correct-size user stories, but for class 2011 and 2012, they put too much emphasis on non-core or bad grain size goals. The proposed approach achieved an improvement of over 50 percentage points in terms of the proportion of high quality user stories generated by students compared to the standard user story template used in *Scrum*.

## 3.6 Summary

In this chapter, we firstly discuss the existing methods for requirements/goals management in ASD. Based on general ASD process, we proposed a light weight Goal Net based method to model structured goal requirements according to existing user stories produced in the agile process. The approach was applied in university level agile software engineering education. It was shown to result in significant improvement in the number and quality of user stories generated by students compared to the past approach.

By introducing *Goal Net* chart into ASD, the students can produce better user stories with good grain size and high quality, and then split them to tasks that will be distributed to team members or claimed by themselves. How to effectively allocate those tasks in agile team becomes another important question related to our human factors research in ASD. In next chapter, we will discuss more about it.

# CHAPTER 4

# TASK ALLOCATION STRATEGY FOR ASD

Current task allocation approaches for agile team are mostly empirical, depending on the developers' intuitions and initiatives. There is a lack of an automated computational system that can produce task allocation plans that balance the considerations for quality and timeliness to improve the overall utility derived from an agile software development project. In this chapter, we propose a *Simple Multi-Agent Real-Time* (SMART) task allocation approach for agile team based on distributed constraint optimization. Personal task allocation agents equipped with SMART helps individual developers to make situation-aware decisions on which incoming tasks to serve so as to reduce the risk of low quality task results while maintaining the timeliness of the overall software project. The analysis and simulation results show that it can achieve close to optimally efficient utilization of the developers' collective capacity and significantly outperform the prevailing practice in terms of both task quality and timeliness of completion.

## 4.1 Background

The work flow of traditional software development processes follows a push model. Tasks are assigned to individual team members by the project manager, and they stay pending in the team members' working queues until completed. This approach suffers from a major limitation - project managers often lack the ability to anticipate potential delays caused by each team member, which often results in inefficient utilization of the collective capacity of the team. In comparison, ASD advocates a pull model. Tasks are placed into a common queue and team members pull them from the front of the queue when they become available. For example, a typical Scrum team of assigning tasks in ASD is as follows: with the agile principle that teams are self-led, during the iteration each team member pulls out tasks they feel comfortable handling and assigns it to themselves. This pull model emphasizes on self-motivation of the team members compared to the push model of traditional software development processes, especially for a collocated agile team. In this way, it is expected to reduce local bottlenecks resulted from individual's unexpected absence or procrastination.

However, current task allocation approaches for agile are mostly empirical, depending on the developers intuitions, understandings and confidences. Recent studies [9, 10] have pointed out some important limitations of ASD reported by practitioners in the field. ASD is not necessarily a good choice for large scale software projects, especially for distributed agile with distrust [168]. ASD places heavy emphasis on the initiative of team members when distributing tasks, its success often depends on staffing the team with people who are highly competent, trusting each other, and have good interpersonal skills which takes time to build up and may not always be readily available in practice. The human and social factors that give small software development teams adaptability through ASD often hinder their performance in large software projects. In addition, practitioners have found it to be difficult for teams distributed over large geographic areas [10] to effectively adopt ASD. The discussion and coordination necessary for the team to agree on task allocations often need to be carried out in a face-to-face manner.

To efficiently utilize the capacity embedded in an agile team, task allocation plans should consider the trade-off between maintaining software quality (which tends to result in more tasks being allocated to highly competent team members) and delivering the software product on time (which needs tasks to be allocated to team members with less workload at the moment). Consistently making such decisions over the lifecycle of an agile software development manually is highly challenging for even experienced team managers.

## 4.2 Problems

### *4.2.1 Existing Problems*

Existing task allocation approaches in ASD can be divided into three categories: 1) ad hoc task allocation, which allows any developer to sign up for any task he/she feels like working on that day; 2) dedicated developer/pair per story; or 3) user story swarming, in which the team swarms each user story in turn, seeing each one through to completion before starting the next. These approaches are mostly empirical, depending on the developers' intuitions and initiatives. There is a lack of an automated computational system that can produce task allocation plans that balance the considerations for quality and timeliness to improve the overall utility derived from an agile software development project.

In practice, the strategy an agile team uses to sign up for work has significant implications for his work style and habits, and can ultimately impact the overall success of the iteration and its artifacts. Unfortunately, the agile community provides relatively little guidance on how this process should be carried out. For *Scrum* team, the idea behind daily stand-up meeting is to build common understanding of the situation and progress among team members. For example, when a new task arrives, the core questions of task allocation include whose skills are most suited to completing this task with high quality? And what effect will assigning this task to any given person have on the overall timeliness of the project? In essence, a good task allocation strategy should minimize both the risks of producing low quality work as well as failure to meet the deadlines at the same time. It is possible for effective solutions to be worked out when team members work together and are familiar with each other. However, for distributed or novice agile teams, it is difficult to make a timely decision that balance these considerations well.

#### *4.2.2 Preliminary Investigation*

We have conducted an eight week group-based software engineering project as part of an undergraduate course work in *Beihang University* from 01/04/2013 to 31/05/2013. In this round of data collection activity, we have:

- 122 students are divided into 20 teams with an average team size of 6 persons. During the course work project, they have no common venue to work together on a daily basis. Thus, all teams are distributed. Team members possess similar skill levels and backgrounds.
- All teams are required to adopt the Scrum to perform their software projects. In order to avoid bias and observe field case, they are not told to use specific task allocation strategy. Tasks are divided among the team members at the beginning of each one-week sprint based on internal discussions among each team.
- Each student may be assigned 0, 1, or multiple tasks during each sprint planning meeting. Each student reports his/her estimations of the difficulty of the task(s) assigned to or picked up by him/her and his/her confidence index of completing the task(s) with satisfactory quality in 10-point *Likert* scales [169, 170], and the expected number of days needed to complete the task(s). The team can use Planning Poker [171] technique to reach the consensus for task difficulty, confidence index and expected days.
- In the following sprint planning meeting, the timeliness of their completion relative to the expected time is evaluated by the members of each team together.

- At the end of the project, the performance of each team is graded by the course instructors with two scores representing the functional features of the software system developed and the time management of the team.

Finally, we collect a total of 726 tasks from 20 teams during whole period. According to the variance of number of tasks per capita, we can see that the strategies adopted by the teams can be generally grouped into three types: Type I - equality based group, their numbers of tasks per capita are almost same (variance less than 1); Type II - mixed strategy group, their variance of number of tasks per capita is between 1 to 5; Type III - competence based group, their numbers of tasks per capita very much depends on the agreed competence of the team members (variance greater than 5). Figure 15 shows the percentage of task delay for each team. From the data we can observe that the strategy an agile team uses to allocate tasks has significant implications for its task completion rate, accordingly results in significant turbulence for the team confidence index shown in Figure 16.

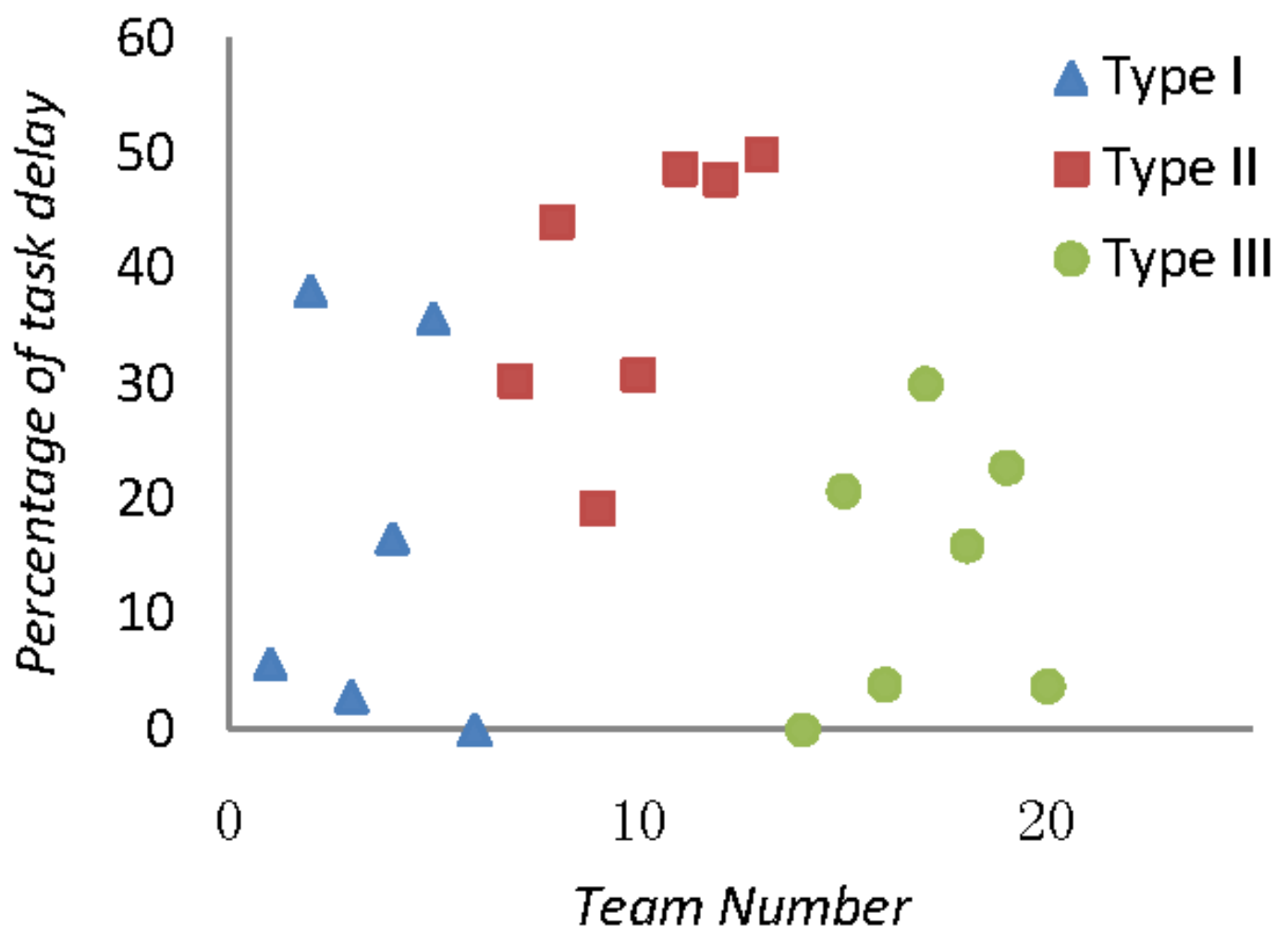

Figure 16. Distribution of percentage of task delay for each team

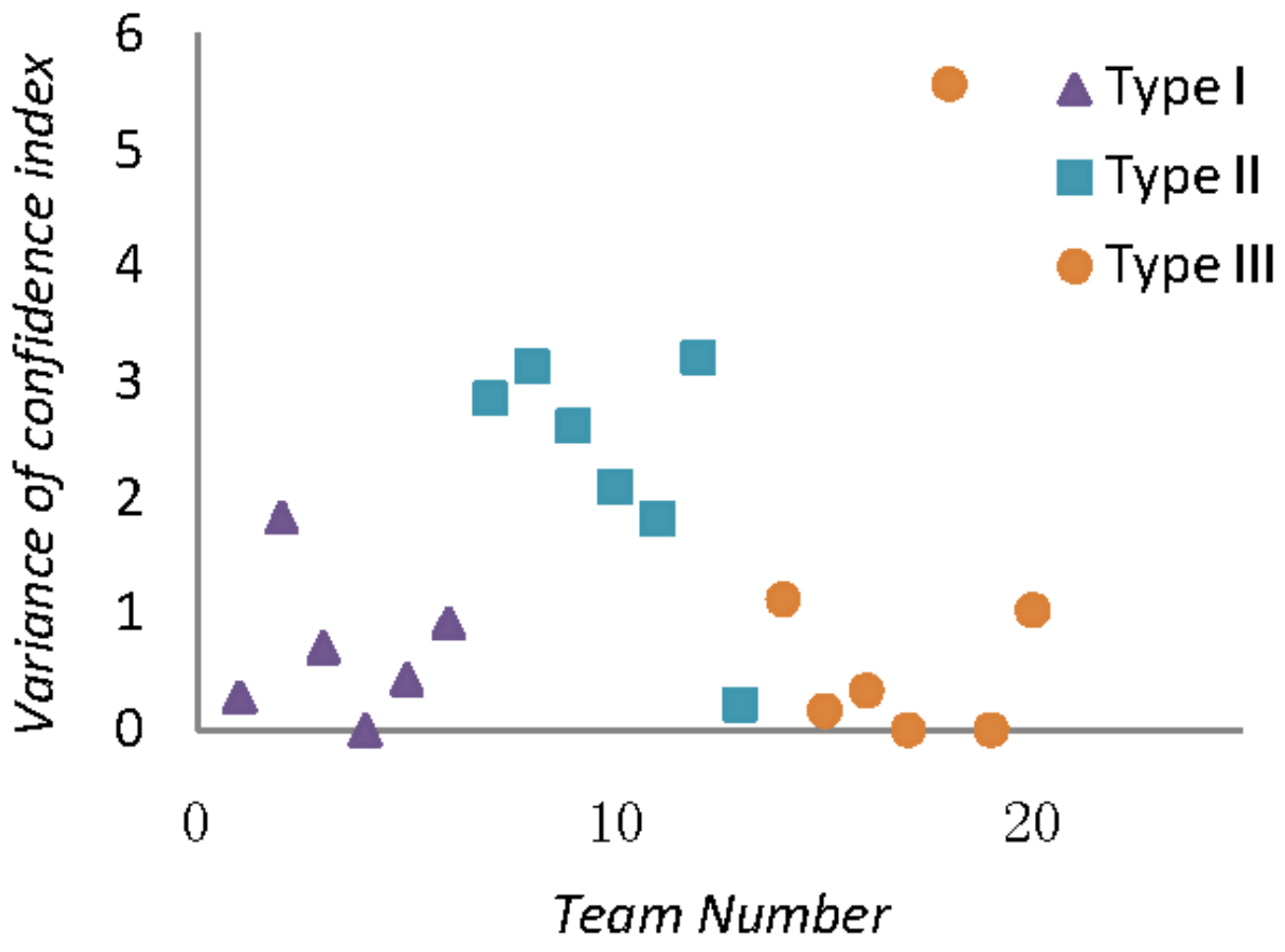

Figure 17. Distribution of variance of confidence index for each team

These two observations have empirically confirmed that the task allocation strategy of the ASD process can significantly affect the task execution progress and team morale. The resulting strategy adopted may not necessarily be suitable for the situation facing the distributed or novice agile team.

Although each team has an overall score for their final working software, it cannot reflect the relevance between task quality result and allocation strategy. So for the next round of data collection in 2014, we have conducted to collect quality data from three sources: 1) self-report, 2) peer evaluation and 3) product owner review.

## 4.3 Method and System Model

Task allocation and workload balancing not only rely on information about internal properties of the tasks (such as priority, utility, difficulty and effort level required etc.), but also on human factors not depicted in product backlog (such as developers' workload, competency, confidence and even psychology stress, etc.). To efficiently utilize the capacity embedded in an agile team, task allocation plans should consider the trade-off between maintaining software quality (which tends to result in more tasks being allocated to highly competent team members) and delivering the software product on time (which needs tasks to be allocated to team members with less workload at the moment).

In our research, we propose a *Simple Multi-Agent Real-Time* (SMART) task allocation approach to automate the process of task allocation in *Agile*. Personal software agents equipped with SMART interact with agile team members during the process of software development to monitor changes in their internal characteristics (e.g., competency in performing certain types of tasks, general effort spent on productive activities per unit time) as well as the external factors affecting their performance (e.g., amount of tasks currently pending in their personal task queues). The SMART agents then coordinate among themselves to pull new tasks on the team members' behalf in a situation-aware manner from the common task queue when they become available with the overall objective of maintaining the optimal quality and timeliness for the entire software development project.

Based on the principles of *Lyapunov* drift [172, 173], the SMART approach can achieve close to optimally achievable software project success rate subject to the limitations of the overall competency and capacity of a given agile team. Theoretical analysis has proven the performance bounds of the SMART approach which can produce solutions to the agile task allocation problem in linearithmic time. Through a highly dynamic simulation environment designed based on the observed team dynamics from 20 agile development teams in 2013 software engineering class of *Beihang University*, China, the SMART approach has been demonstrated to significantly outperform the existing agile task allocation approach under various team configurations and environment conditions.

### *4.3.1 Problem Formulation*

In this research, we aim to explore the multi-agent coordination principles to help ASD streamline the task distribution and management process. We formulate the challenges facing ASD as a *distributed constraint optimization problem* (DCOP) [174], and design efficient solutions to support the task allocation decision-making in agile teams. By letting software agents representing individual team member's interest with to coordinate their task acquisition decisions, the proposed research aims to strike a balance between maintaining the quality of the software product, keeping team members motivated, distributing tasks fairly among team members, and minimizing the delay experienced by the whole project while taking into account the current context facing each team member.

To achieve this goal, we first formalize the agile team management problem into a DCOP. The overall objective of ASD is to develop high quality software product according to the customer's requirement within the stipulated deadline. Through our analysis of projects developed with ASD in the past, the success of a project often hinges on how well they balance the quality aspect with the timely completion of the tasks. In order to produce high quality software, tasks should be completed by highly competent team members. However, over-concentrating tasks to these team members causes longer delays and may negatively affect their morale, thus, reducing their average work quality. To reduce delay, tasks should be divided fairly among the team members according to their individual context. By considering these two aspects together, the overall objective function for an *ASD* decision support system can be expressed as maximizing a ***quality-minus-delay*** expression for all team members.

The ***quality*** aspect depends on a wide range of factors innate to each team member. We summarize them into key variables represented by the 3-tuple $<M_i(t), C_i^{\tau}(t), E_i^{max}(t)>$. $M_i(t)$

denotes a team member *i*'s mood at time t, $C_i^\tau(t)$ represents *i*'s competency in performing task of type $\tau$ (e.g., C# programming), and $E_i^{max}(t)$ is the average total effort *i* can spend on performing tasks per unit time. The 3-tuple can be viewed as *i*'s internal context at a given time. $E_i^{max}(t)$ is assumed to be positively correlated with $M_i(t)$ subject to i's physical limitations.

To minimize the ***delay***, two conditions must be satisfied: 1) *i*'s pending task queue $Q_i^\tau(t)$ should not be allowed to grow indefinitely for all $\tau$ and t, and 2) $Q_i^\tau(t)$ should not be too short (i.e., i is idling). Based on i's internal context, we can derive its target workload for each type of task $q_i^\tau$ which positively correlates to $< M_i(t), C_i^\tau(t), E_i^{max}(t) >$. Thus, the objective of minimizing delay is equivalent to minimizing the drift of the collective workload in an agile team from its collective target workload. Therefore, the objective function now becomes maximizing the ***quality-minus-drift*** expression for all team members.

The properties of a task *j* in *ASD* are represented by the 3-tuple $<p^\tau, u^\tau, e^\tau>$. $p^\tau$ denotes the priority of this type of tasks and will affect their positions in the common task queue $Q(t)$ (such as product backlog for *Scrum*). $u^\tau$ represents the utility that can be derived by the team from successfully completing j on time. $e^\tau$ is the expected effort that needs to be expended to complete *j*. Since, tasks are often divided into small and more manageable pieces by the project manager in ASD, we can assume that $<p^\tau, u^\tau, e^\tau>$ for all the tasks belonging to the same type $\tau$ to be the same in the model. The ***utility*** that can be expected from letting i perform task j can at time t, thus, be expressed as:

$$utility(t) = u^\tau \cdot C_i^\tau(t) \cdot M_i(t) \tag{4.1}$$

The queuing dynamics for any pending task queue $Q_i^\tau(t)$ is:

$$Q_i^{\tau}(t) \leftarrow \max[Q_i^{\tau}(t-1) + \alpha_i^{\tau}(t-1) - \mu_i^{\tau}(t-1), 0] \quad (4.2)$$

Where $\alpha_i^{\tau}(t)$ denotes the number of new tasks of type $\tau$ admitted into $Q_i^{\tau}(t)$ at time t, $\mu_i^{\tau}(t-1)$ is the number of tasks from $Q_i^{\tau}(t-1)$ completed by i at time $t-1$. The $\max[\cdot,0]$ operator ensures that the size of the queue will never be negative. Therefore, the expected ***quality*** from a task acceptance decision can be expressed by:

$$quality(t) = \alpha_i^{\tau}(t) \cdot utility(t) \quad (4.3)$$

The ***drift*** is positively correlated to both $\alpha_i^{\tau}(t)$ and $\mu_i^{\tau}(t)$ (i.e., increase in either of them causes the drift to increase). Based on the principle of *Lyapunov* drift [172, 173], we have:

$$drift(t) = \alpha_i^{\tau}(t) \cdot \mu_i^{\tau}(t) \quad (4.4)$$

Eq. (4.4) can be trivially minimized by assigning both $\alpha_i^{\tau}(t)$ and $\mu_i^{\tau}(t)$ to 0 for all *i*, $\tau$ and *t* (i.e., all team members stay idle all the time). However, this is not a valid solution in practice. Through the above analysis, we can formalize the objective of managing an agile team as:

$$\text{Maximize: } \frac{1}{T}\sum_{t=0}^{T}\sum_{i,\tau}[\psi \cdot quality(t) - drift(t)]$$

$$= \frac{1}{T}\sum_{t=0}^{T}\sum_{i,\tau}\alpha_i^{\tau}(t)[\psi \cdot u^{\tau} \cdot C_i^{\tau}(t) \cdot M_i(t) - \mu_i^{\tau}(t)] \quad (4.5)$$

$$\text{Subject to: } \sum_{\tau}(\alpha_i^{\tau}(t) \cdot e^{\tau}) \le E_i^{\max}(t) \quad (4.6)$$

$$0 \le \alpha_i^{\tau}(t) \le \lambda^{\tau}(t) \text{ for all i, } \tau \text{ and t} \quad (4.7)$$

Where $\lambda^{\tau}(t)$ is the number of new tasks of type $\tau$ being added into the common queue Q(t) by the project manager at time t, and $\psi$ is a weight variable indicating the relative importance given to quality and drift while maximizing (4.5). With this formalization, we will develop *multi-agent system (MAS)* based approaches to find solutions for all $\alpha_i^\tau(t)$ whenever new tasks are proposed so as to maximize (4.5) subject to Constraints (4.6) and (4.7).

### *4.3.2 The SMART Approach*

In (4.5), $[\psi \cdot u^\tau \cdot C_i^\tau(t) \cdot M_i(t) - \mu_i^\tau(t)]$ is defined as the availability score of each task queue of i at t. A SMART agent helps an agile team member formulate a task request acceptance plan$\alpha_i^\tau(t)$ about how many new tasks of different types it shouldaccept at each time step based on his/her current situation which is represented by the 3-tuple $<M_i(t), C_i^\tau(t), E_i^{max}(t)>$.

In order to maximize (4.5), SMART proceeds as illustrated in algorithm 1.

---

**Algorithm 1** *SMART*

---

**Input**: $\psi \cdot u^\tau \cdot C_i^\tau(t) \cdot M_i(t) - \mu_i^\tau(t)$values for all$\tau$inan agile team member *i*, the incoming tasks $\lambda^\tau(t)$ for all $\tau$ at *i*, and $E_i^{max}(t)$.

$e_i(t) = E_i^{max}(t)$
**for** each $Q_i^\tau(t)$ in *i* in descending order of its $\alpha_i^\tau(t)$ **do**
**if** $\psi \cdot u^\tau \cdot C_i^\tau(t) \cdot M_i(t) - \mu_i^\tau(t) > 0$ **then**
**if** $\lambda_i^\tau(t) \cdot e^\tau \leq e_i(t)$ **then**
$\alpha_i^\tau(t) = \lambda_i^\tau(t)$
**else**
$\alpha_i^\tau(t) = \lfloor e_i(t)/e^\tau \rfloor$
**end if**
$e_i(t) \leftarrow e_i(t) - \alpha_i^\tau(t) \cdot e^\tau$
**else**
$\alpha_i^\tau(t) = 0$
**end if**
**end for**
Return ($\alpha_i^\tau(t)$ for each $Q_i^\tau(t)$ in *i* )

---

During one sprint cycle, the algorithm makes use of current task related information generated during sprint assessment phase (or in sprint planning meeting) and task completion information generated during previous sprint review phase/meeting, and characteristics of team members to produce a task allocation plan for allocating the sprint backlog to aid the team's decision. Some characteristics such as personal skills and competence can be obtained by self-reported survey and past performance. Current morale status such as confidence or interest to each task needs to be provided by developer in every sprint assessment phase, and current workload can be calculated by system automatically. Generally speaking, this design does not require much extra effort from the developers.

In essence, the higher the payoff per unit effort for a task $T^{\tau}$, the higher the quality of i in performing tasks of type c, and the more spare capacity i currently has in accommodating more requests for performing tasks of type c, the more likely $T^{\tau}$ will be accepted by the SMART agent on behalf of i. In the case where not all incoming requests are accepted, the SMART agent will inform other SMART agents or the agile team manager so that they can look for other alternatives.

## 4.4 Theoretical Analysis

In this section, we envision a situation where the suggestions made by SMART agents are fully complied by an agile team and analyze the impact on the size of the task queues and the overall quality of the software project.

In order to analyze the model, the first challenge is to quantify the level of congestion in a given agile team. We adopt the *Lyapunov* functions [173] to measure congestion. Based on this concept, we define the overall level of task queue congestion in a *MAS* at any *t* as

$$L(t) = \sum_{i,\tau} \left(Q_i^\tau(t)\right)^2 \tag{4.8}$$

A small value of $L(t)$ indicates that all $Q_i^\tau(t)$ are having a low level of congestion. $L(t)$ can be trivially minimized by making software agents reject all incoming tasks and, therefore, keeping all $Q_i^\tau(t) = 0$ for all the time. However, this is not a desirable mode of operation for any agile team. Instead, we want to limit the growth of the overall level of congestion while filling in spare capacities whenever they become available with new tasks (if there are enough new tasks from the team manager).

Assume there are positive constants $\psi$, $B$and $C$ such that the ***quality-minus-drift*** expression in (4.5) satisfies:

$$\psi \cdot quality(t) - drift(t) \geq \psi \cdot U^{opt} + B \sum_{i,\tau} Q_i^\tau(t) - C \tag{4.9}$$

where $U^{opt}$ is the total utility produced by the theoretical optimal solution for (4.5). Taking the expectations over the distribution of $Q(t)$ on both sides of (4.9), we have:

$$\psi \sum_{i,\tau} \mathbb{E}\{\alpha_i^\tau(t) \cdot utility(t)\} - \mathbb{E}\left\{L(Q(t+1) - L\big(Q(t)\big)\right\}$$

$$\geq \psi \cdot U^{opt} + B \sum_{i,\tau} \mathbb{E}\{Q_i^\tau(t)\} - C \tag{4.10}$$

which holds for all time steps t. Summing both sides over $t \in \{0, 1, \dots, T-1\}$, we have:

$$\psi \textstyle\sum_{t=0}^{T-1} \sum_{i,\tau} \mathbb{E}\{\alpha_i^\tau(t) \cdot utility(t)\} - \mathbb{E}\left\{L(Q(T) - L\big(Q(0)\big)\right\}$$

$$\geq \psi T U^{opt} + B \sum_{t=0}^{T-1} \mathbb{E}\{Q_i^\tau(t)\} - CT \tag{4.11}$$

Since $\alpha_i^{\tau}(\cdot) \cdot utility(\cdot) \geq 0$ and $L(\cdot) \geq 0$, and suppose the expected quality for completing $\alpha_i^{\tau}(t)$ tasks are bounded by

$$\sum_{i,\tau} \{\alpha_i^{\tau}(t) \cdot utility(t)\} \leq G_{max} \tag{4.12}$$

where $G_{max} = \sum_{i,\tau}\{\alpha_i^{\tau}(t) \cdot u^{\tau}\}$ can be achieved when all new tasks are completed successfully within their respective deadlines. By re-arranging the terms in (4.11) and dividing both sides by $TB$, the upper bound on the sizes of the task queues in the MAS is:

$$\begin{aligned} \frac{1}{T}\sum_{t=0}^{T-1}\sum_{i,\tau} \mathbb{E}\{Q_i^{\tau}(t)\} &\leq \frac{C + \psi G_{max} - \psi U^{opt}}{B} \\ &\leq \frac{C + \psi G_{max}}{B} \end{aligned} \tag{4.13}$$

Similarly, by rearranging (11) and dividing both sides by $T\psi$, the lower bound on the total quality produced by agents in the MAS is:

$$\begin{aligned} \frac{1}{T}\sum_{t=0}^{T-1}\sum_{i,\tau} \mathbb{E}\{\alpha_i^{\tau}(t) \cdot utility(t)\} &\geq U^{opt} + \frac{B}{T\psi}\sum_{t=0}^{T-1}\sum_{i,\tau} \mathbb{E}\{Q_i^{\tau}(t)\} - \frac{C}{\psi} \\ &\geq U^{opt} - \frac{C}{\psi} \end{aligned} \tag{4.14}$$

From above analysis, it can be deduced that if the condition in (4.9) can be fulfilled (which can be done through careful choice of the values for $\psi$, B and C), then, based on (4.13), a theoretical upper bound exists for all pending task queues for all software agents over the long run if the agents follow the recommendations made by SMART. This ensures that the task queue lengths will not keep on increasing and the software agents always can stop the growth of their task queues so their perceived quality of artifact can be maintained.

In addition, based on (4.14), the time averaged total utility achieved in a given agile team through SMART can approach that achieved by the theoretical optimal solution within $C/\psi$ in the long run. By increasing $\psi$, the utility produced by SMART can be made closer to the optimal social utility. However, increasing also causes the upper bound to the pending task queue lengths to rise according to (4.13), thereby increasing the expected time taken to complete a task. Due to the physical limitations of the agents in a realistic system, if the increase in the value of $\psi$ causes the expected completion time of tasks to start exceeding the stipulated deadlines, utility will start to decrease as the members' level of mood decreases. Setting the value of $\psi$ arbitrarily high will not make the utility produced by SMART be indefinitely close the optimal. Thus, the trade-off between quality and the timeliness in receiving tasks only exists within a limited range of the $\psi$ value. The actual range depends on the physical limitations of the agents in each given system. On the other hand, in agile teams adopting the existing task allocation approach, these upper and lower bounds cannot be guaranteed.

## 4.5 Experimental Evaluation

As the problem in this study is relatively new, there is no existing dataset that can be used to evaluate SMART. In addition, real project data are useful for designing a realistic experiment environment, but the behavior patterns of the developers are ad hoc for which we do not have ground truth. In order to comprehensively evaluate SMART under different circumstances, and to provide more flexible control of software agents' behavior, we implement it within a simulated *MAS* environment based on our system model and collected data from real teams. Our hypotheses in this section are:

- *Hypothesis 1*: The SMART approach can better mitigate the adverse effect of quality decline by using the SMART approach than using the existing task allocation approach for agile teams.
- *Hypothesis 2*: The total utility of an agile process can be improved through the use of the SMART approach.

### *4.5.1 Experiment Origin*

In our research, a simulated, highly dynamic multi-agent environment is designed based on agile team dynamics obtained from self-reported data of 20 agile software development teams (average number of members is 6) in *Beihang University*, *China* to study the *ASD* approach under various conditions. We collected and tracked some data concerning most of working variables that are shown in Table 12 during their performing agile processes, using a weekly report. The operational definitions of these variables are provided below.

Table 11. Working Variables Selected for Inclusion in *SMART*

| **Variable** | **Variable type** | **Developer variable** | **Task variable** |
|---|---|---|---|
| Developer’s mood | Individual | √ | |
| Developer’s competency | Individual | √ | |
| Developer’s average total effort | Individual | √ | |
| Developer’s task queue | Individual | √ | |
| Task priority | Process | | √ |
| Task type | Process | | √ |
| Task utility | Process | | √ |

| **Variable** | **Variable type** | **Developer variable** | **Task variable** |
|---|---|---|---|
| Task expected effort | Process | | √ |
| Task performed effort | Individual | √ | |

- Developer's mood (M): refers to a comprehensive emotions or feeling reported by an agile team member at any given time. The mood can include such feeling as motivated or bored, stressed or relaxed etc.
- Developer's competency (C) [175]: refers to the extent to which an individual team member is able to perform given type of tasks.
- Developer's average total effort (E): refers to the average total effort that a team member spends on performing tasks per unit time period.
- Developer's task queue (Q): refers to developer's pending task queue.
- Task priority (p): refers to the priority of a given type of tasks which will affect their positions in the common task queue.
- Task type (τ): refers to the type of task, e.g., C# programming, C++ programming, UI design etc.
- Task expected effort (e): refers to the expected effort that is needed in order to complete a task.
- Actual productive effort: refers to the actual effort that has been expended to complete task by a team member during a given period of time.

### *4.5.2 Experiment Design*

Based on the task processing behavior data collected from *Beihang University*, we design a multi-agent environment to simulate various team sizes and compositions as well as environment conditions to evaluate the SMART approach. The member agent population consists of between 20 to 160 agents exhibiting behavior patterns belonging to four different categories our experiments. They are labeled as:

1) HCA: highly competent developer agents who return high quality task results 90% of the time on average;

2) MCA: moderately competent developer agents who return high quality task results 70% of the time on average;

3) MIA: moderately incompetent developer agents who return high quality task results 30% of the time on average;

4) HIA: highly incompetent developer agents who return high quality task results 10% of the time on average.

By adjusting the number of different agents, we simulate agile teams with different developer composition. Table 13 – 15 are experiment settings for simulating small size agile teams.

Table 12. Small, Highly Incompetent Team, 20 people (S-I)

| **Developer Type** | **Number** | **Competence** | $\mathbf{E}^{\mathbf{max}}$ |
|---|---|---|---|
| HCA | 1 | 0.9 | 20 |
| MCA | 5 | 0.7 | 15 |
| MIA | 5 | 0.3 | 15 |
| HIA | 9 | 0.1 | 10 |

Table 13. Small, Mediumly Competent Team, 20 people (S-M)

| Developer Type | Number | Competence | $\mathbf{E}^{\mathbf{max}}$ |
|---|---|---|---|
| HCA | 5 | 0.9 | 20 |
| MCA | 5 | 0.7 | 15 |
| MIA | 5 | 0.3 | 15 |
| HIA | 5 | 0.1 | 10 |

Table 14. Small, Highly Competent Team, 20 people (S-C)

| Developer Type | Number | Competence | $\mathbf{E}^{\mathbf{max}}$ |
|---|---|---|---|
| HCA | 9 | 0.9 | 20 |
| MCA | 5 | 0.7 | 15 |
| MIA | 5 | 0.3 | 15 |
| HIA | 1 | 0.1 | 10 |

They will be simulated to perform 500 tasks in 100 days. The properties of tasks are shown in Table 16.

Table 15. The Properties of Tasks for Small Teams

| Task Type | Number | Utility | Effort |
|---|---|---|---|
| T1 | 100 | 10 | 10 |
| T2 | 100 | 8 | 8 |
| T3 | 100 | 5 | 5 |
| T4 | 100 | 3 | 3 |
| T5 | 100 | 1 | 1 |

Table 17 – 19 are experiment settings for simulating medium size agile teams.

Table 16. Medium, Highly Incompetent Team, 50 people (M-I)

| **Developer Type** | **Number** | **Competence** | $\mathbf{E^{max}}$ |
|---|---|---|---|
| HCA | 2 | 0.9 | 20 |
| MCA | 13 | 0.7 | 15 |
| MIA | 13 | 0.3 | 15 |
| HIA | 22 | 0.1 | 10 |

Table 17. Medium, Mediumly Competent Team, 50 people (M-M)

| **Developer Type** | **Number** | **Competence** | $\mathbf{E^{max}}$ |
|---|---|---|---|
| HCA | 12 | 0.9 | 20 |
| MCA | 13 | 0.7 | 15 |
| MIA | 13 | 0.3 | 15 |
| HIA | 12 | 0.1 | 10 |

Table 18. Medium, Highly Competent Team, 50 people (M-C)

| **Developer Type** | **Number** | **Competence** | $\mathbf{E^{max}}$ |
|---|---|---|---|
| HCA | 22 | 0.9 | 20 |
| MCA | 13 | 0.7 | 15 |
| MIA | 13 | 0.3 | 15 |
| HIA | 2 | 0.1 | 10 |

They will be simulated to perform 1500 tasks in 100 days. The properties of tasks are shown in Table 20.

Table 19. The Properties of Tasks for Medium Teams

| **Task Type** | **Number** | **Utility** | **Effort** |
|---|---|---|---|
| T1 | 300 | 10 | 10 |
| T2 | 300 | 8 | 8 |
| T3 | 300 | 5 | 5 |
| T4 | 300 | 3 | 3 |
| T5 | 300 | 1 | 1 |

Table 21–23 are experiment settings for simulating large size agile teams.

Table 20. Large, Highly Incompetent Team, 160 people (L-I)

| **Developer Type** | **Number** | **Competence** | $\mathbf{E^{max}}$ |
|---|---|---|---|
| HCA | 10 | 0.9 | 20 |
| MCA | 40 | 0.7 | 15 |
| MIA | 40 | 0.3 | 15 |
| HIA | 70 | 0.1 | 10 |

Table 21. Large, Mediumly Competent Team, 160 people (L-M)

| Developer Type | Number | Competence | $E^{max}$ |
|---|---|---|---|
| HCA | 5 | 0.9 | 20 |
| MCA | 5 | 0.7 | 15 |
| MIA | 5 | 0.3 | 15 |
| HIA | 5 | 0.1 | 10 |

Table 22. Large, Highly Competent Team, 160 people (L-C)

| Developer Type | Number | Competence | $E^{max}$ |
|---|---|---|---|
| HCA | 9 | 0.9 | 20 |
| MCA | 5 | 0.7 | 15 |
| MIA | 5 | 0.3 | 15 |
| HIA | 1 | 0.1 | 10 |

They will be simulated to perform 5000 tasks in 100 days. The properties of tasks are shown in Table 24.

Table 23. The Properties of Tasks for Large Teams

| Task Type | Number | Utility | Effort |
|---|---|---|---|
| T1 | 1000 | 10 | 10 |
| T2 | 1000 | 8 | 8 |
| T3 | 1000 | 5 | 5 |
| T4 | 1000 | 3 | 3 |
| T5 | 1000 | 1 | 1 |

Two simulated multi-agent simulations are run in parallel. In one of them, the developer agents adopt the general competence based competence based *Accept-When-Requested* (AWR) approach for handling incoming task requests. In the other, the developer agents adopt the proposed SMART approach for handling incoming task requests. The results from these two sets of experiments are labeled as AWR and SMART respectively in the following figures. If no developer agent is willing to accept a task request under SMART in a particular time step, the task will be put into the following time steps to be picked up by other developer agent. Each simulation is repeated 10 times to reduce the effect of random variations in the system.

### *4.5.3 Analysis of results*

***Hypothesis 1***

Figure17 to 25 compare the proportion of task allocated to agents for agile teams of various sizes and compositions over 100 time steps under both AWR and SMART. It can be seen that under AWR, the agent's allocated task proportion fluctuates significantly. The sequence of event is: during the development process, agents with high competence will be allocated a large number of tasks. The influx of tasks resulted in long backlog in their task queues. On the other hand, many other agents are in idle status, especially when the team size becomes very large. The SMART approach avoids this issue and keep members' task queue in an average level while still assigning more tasks to competent worker without overworking them.

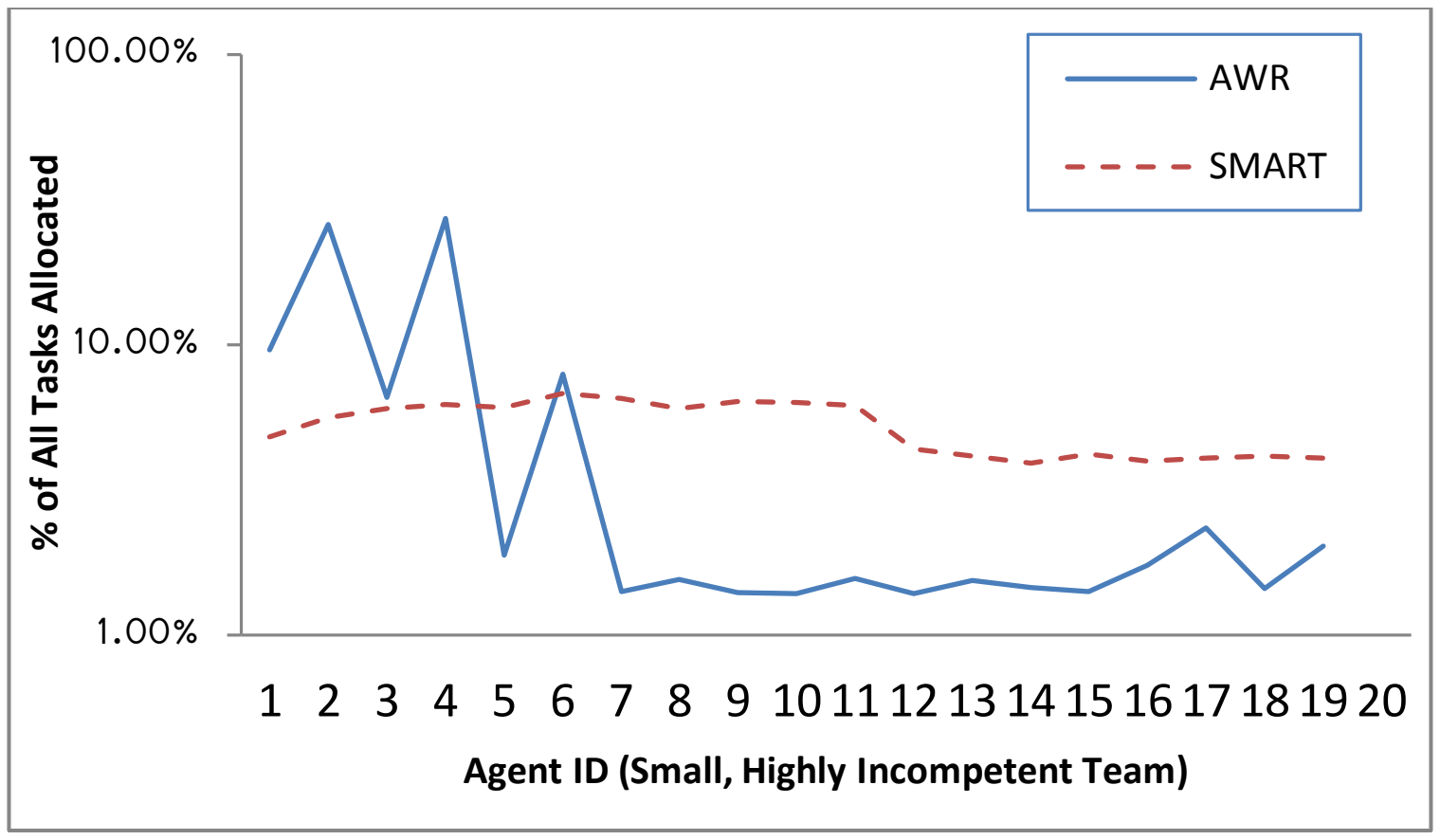

Figure 18. Task allocation proportion for S-I team

Figure 18 – 25 show the similar situation for other teams.

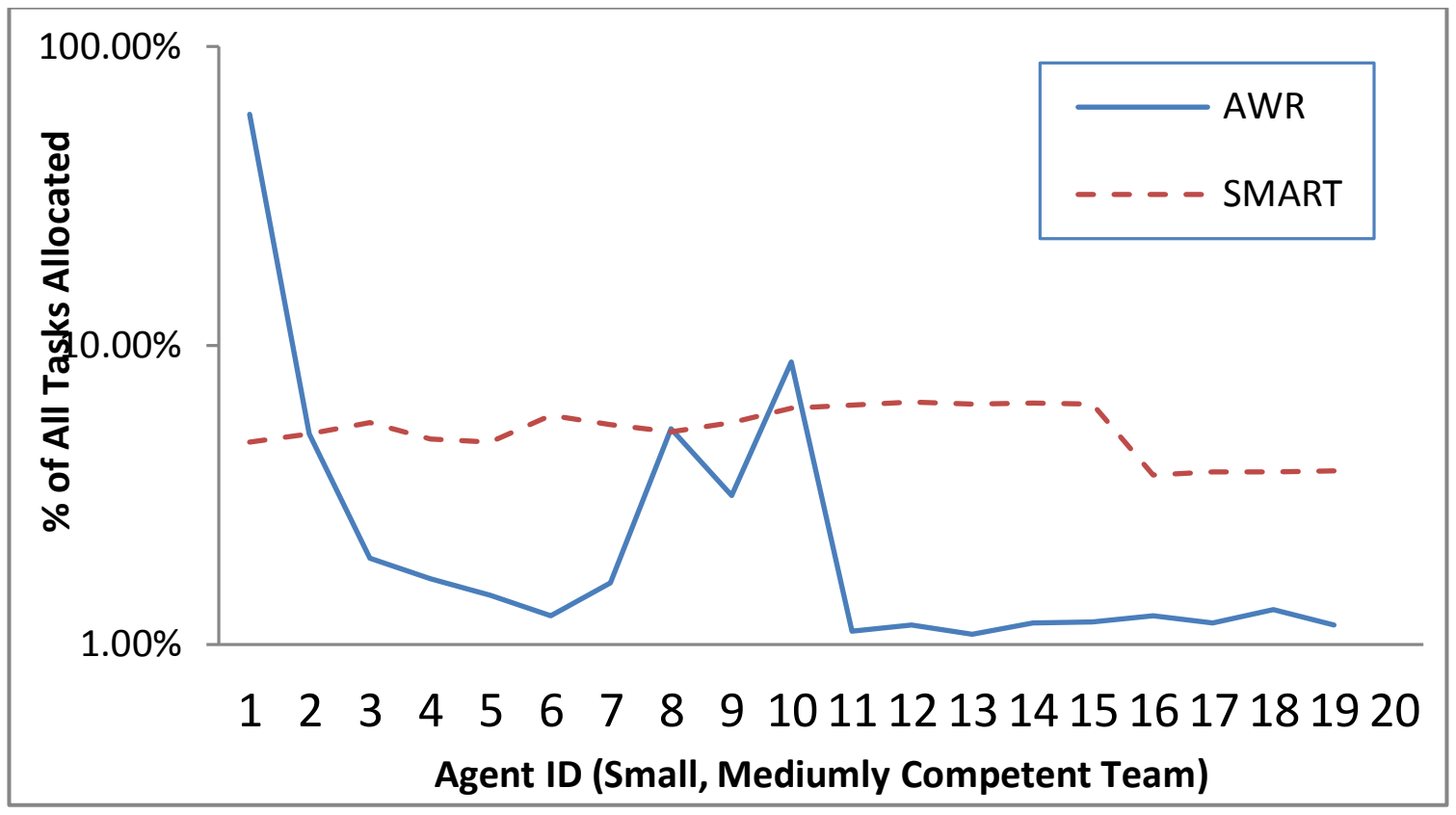

Figure 19. Task allocation proportion for S-M team

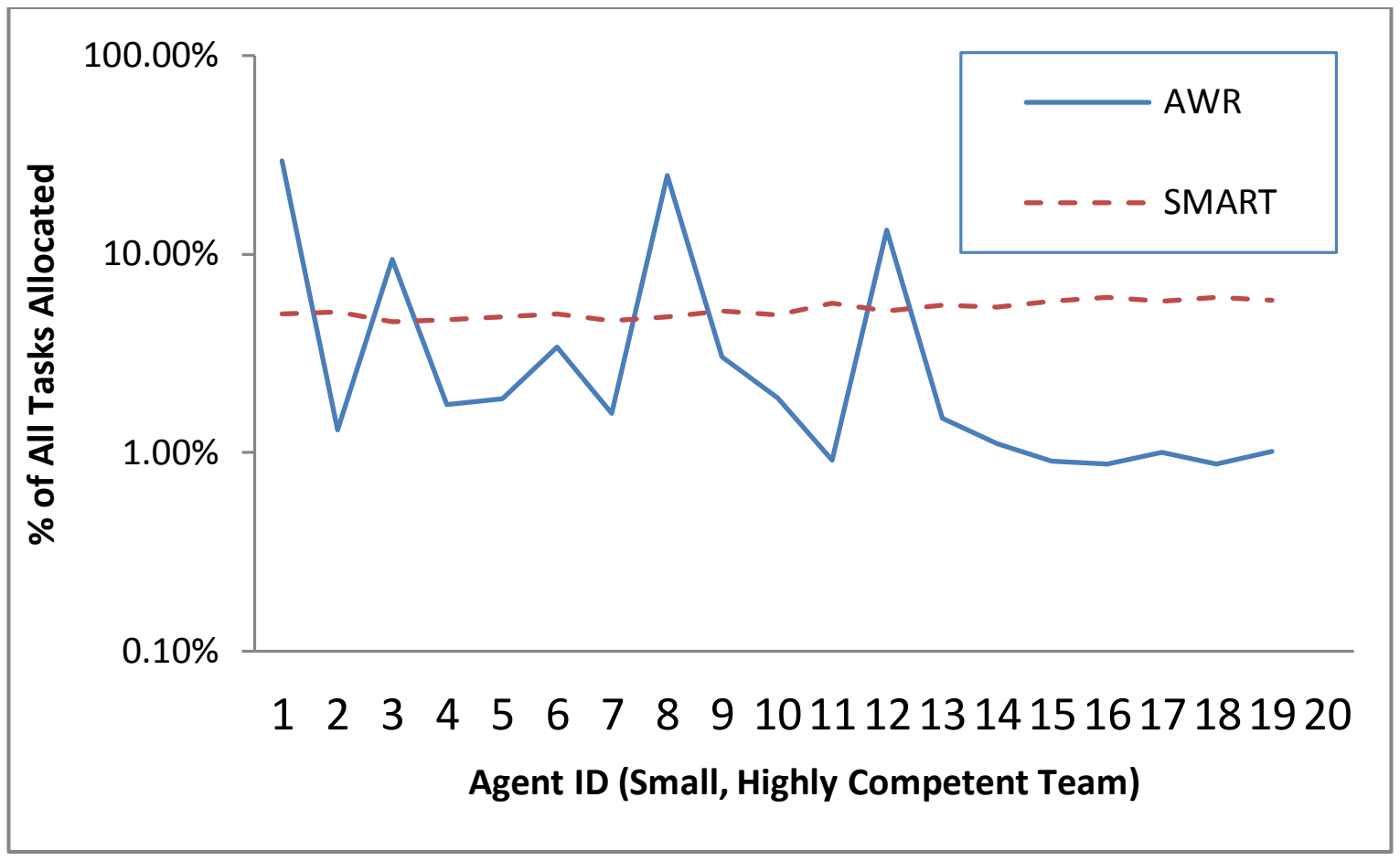

Figure 20. Task allocation proportion for S-C team

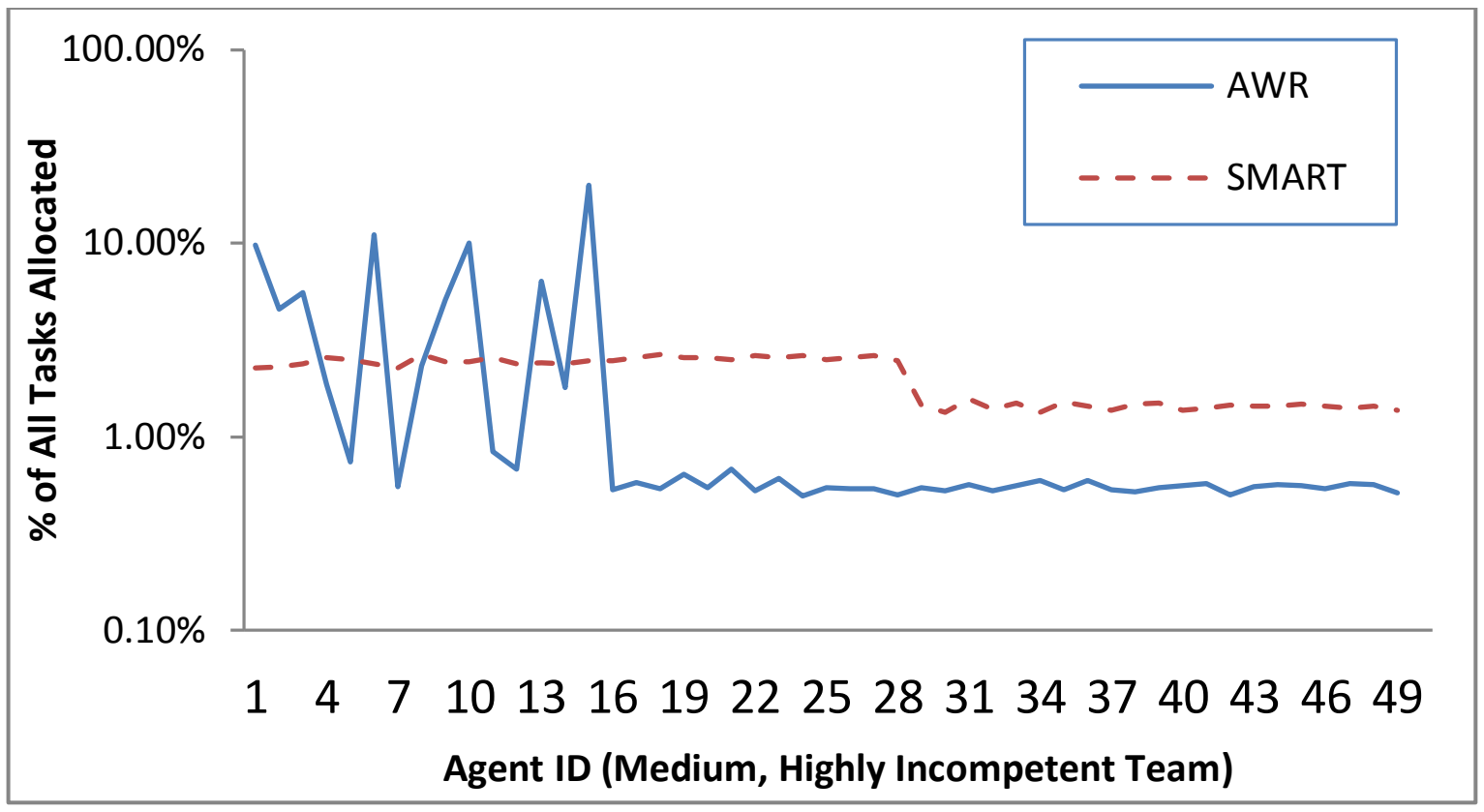

Figure 21. Task allocation proportion for M-I team

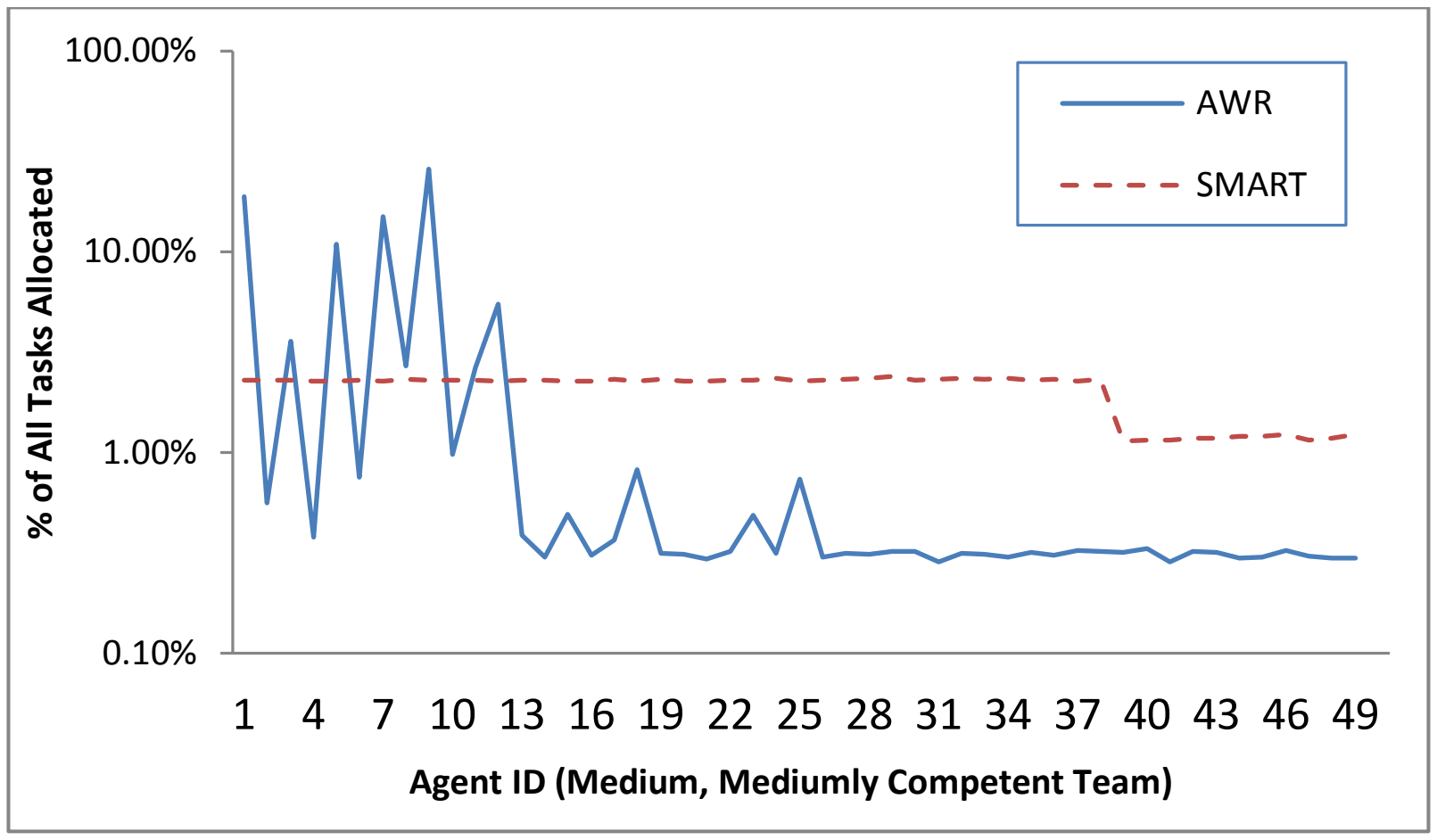

Figure 22. Task allocation proportion for M-M team

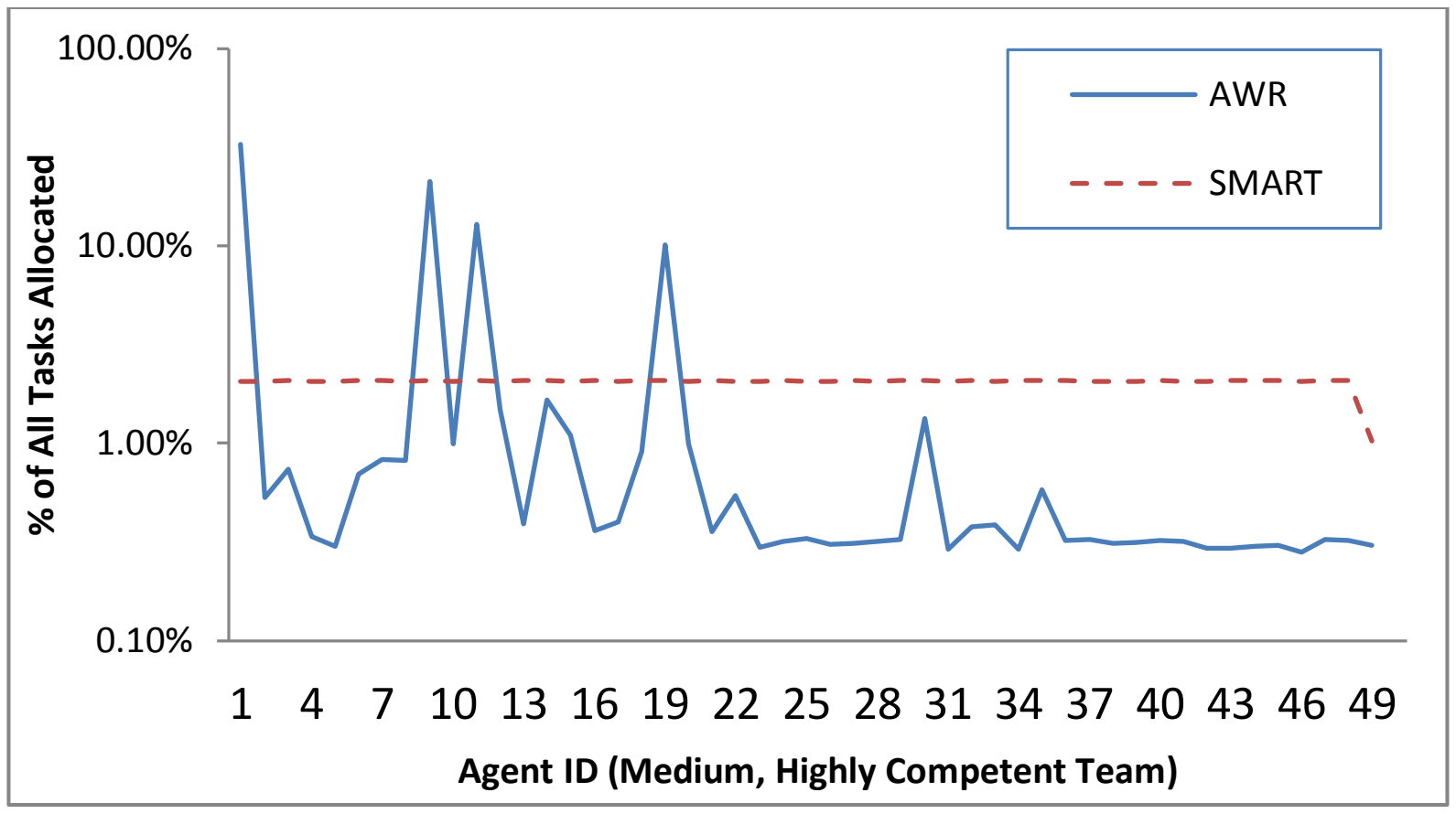

Figure 23. Task allocation proportion for M-C team

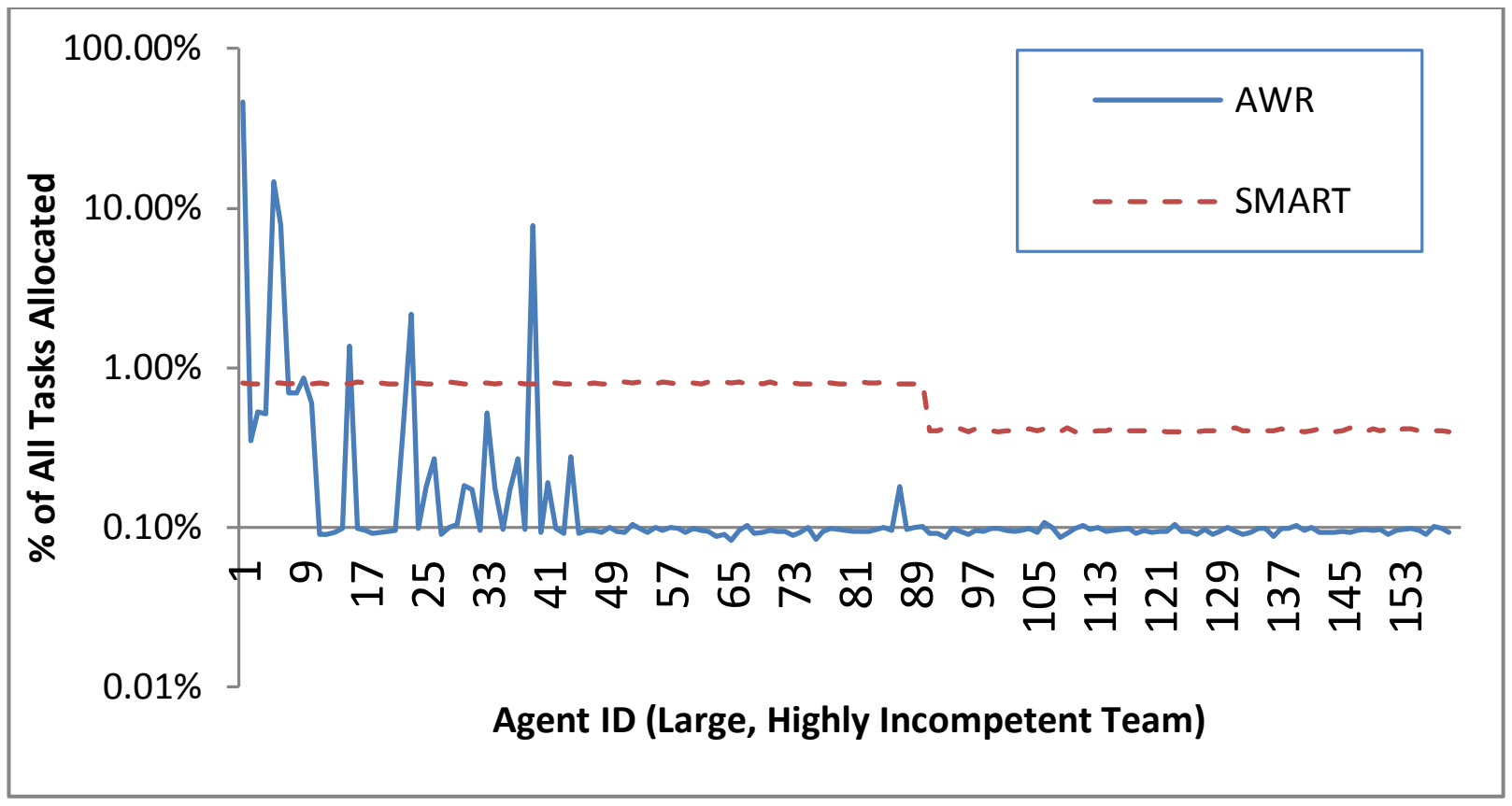

Figure 24. Task allocation proportion for L-I team

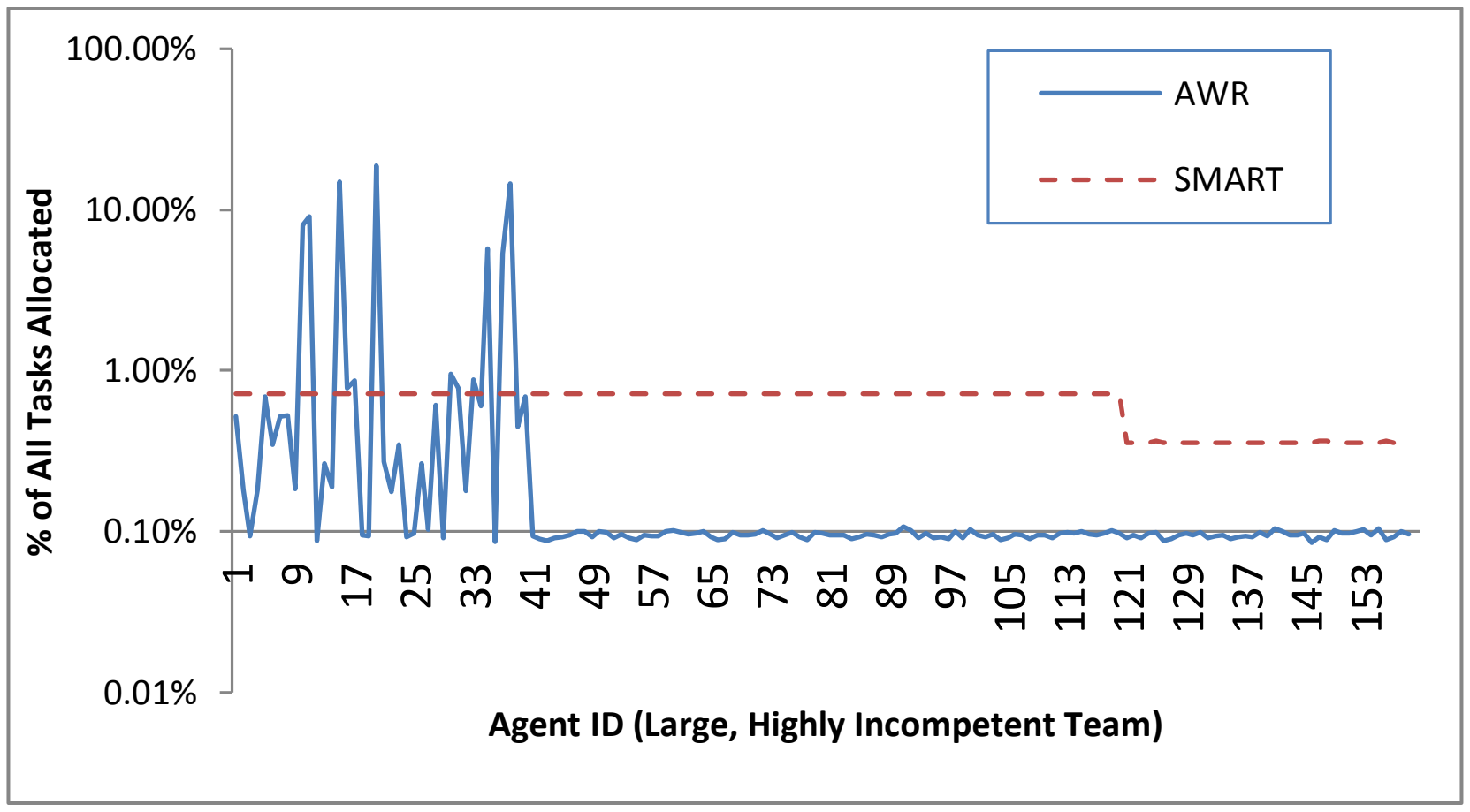

Figure 25. Task allocation proportion for L-M team

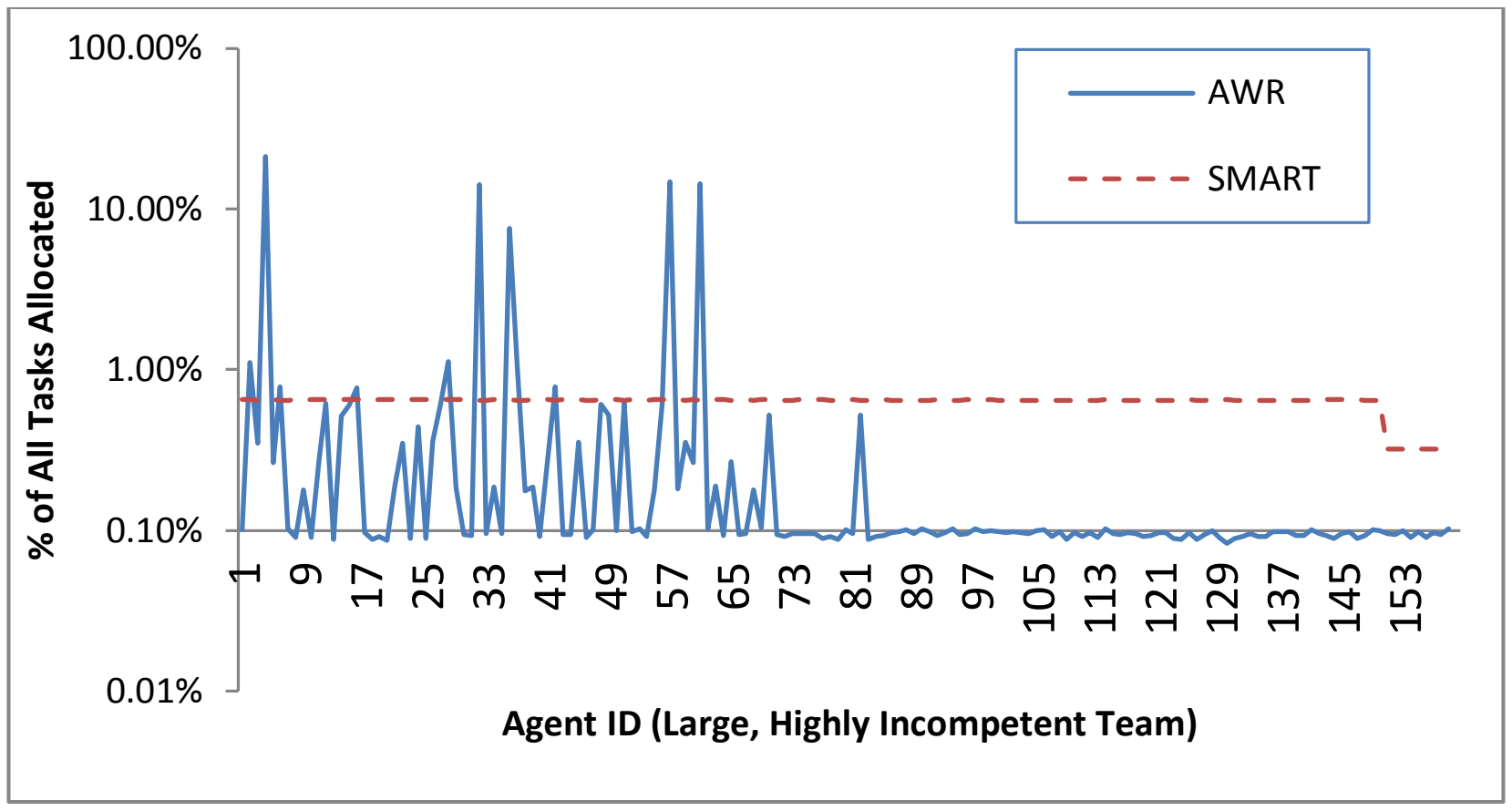

Figure 26. Task allocation proportion for L-C team

***Hypothesis 2***

Figure 26 shows the global utility achieved by given different agent populations under AWR and SMART. By reducing the adverse effect of uneven task allocation and the corresponding inefficient utilization of team resources caused by the existing approach through efficient utilization of developer agents' capacities, the agile team equipped SMART task allocation approach consistently achieved significantly higher utility than the compared approach for all team sizes and compositions. During the process, SMART will always try the best way to make sure that: 1) high utility tasks will be performed by high competent members; and, 2) tasks assigned to team members will not overwhelm them based on their past observed performance. In contrast, the compared approach can only achieve very low utility as no guarantee can be provided in terms of either task quality or timeliness of completion.

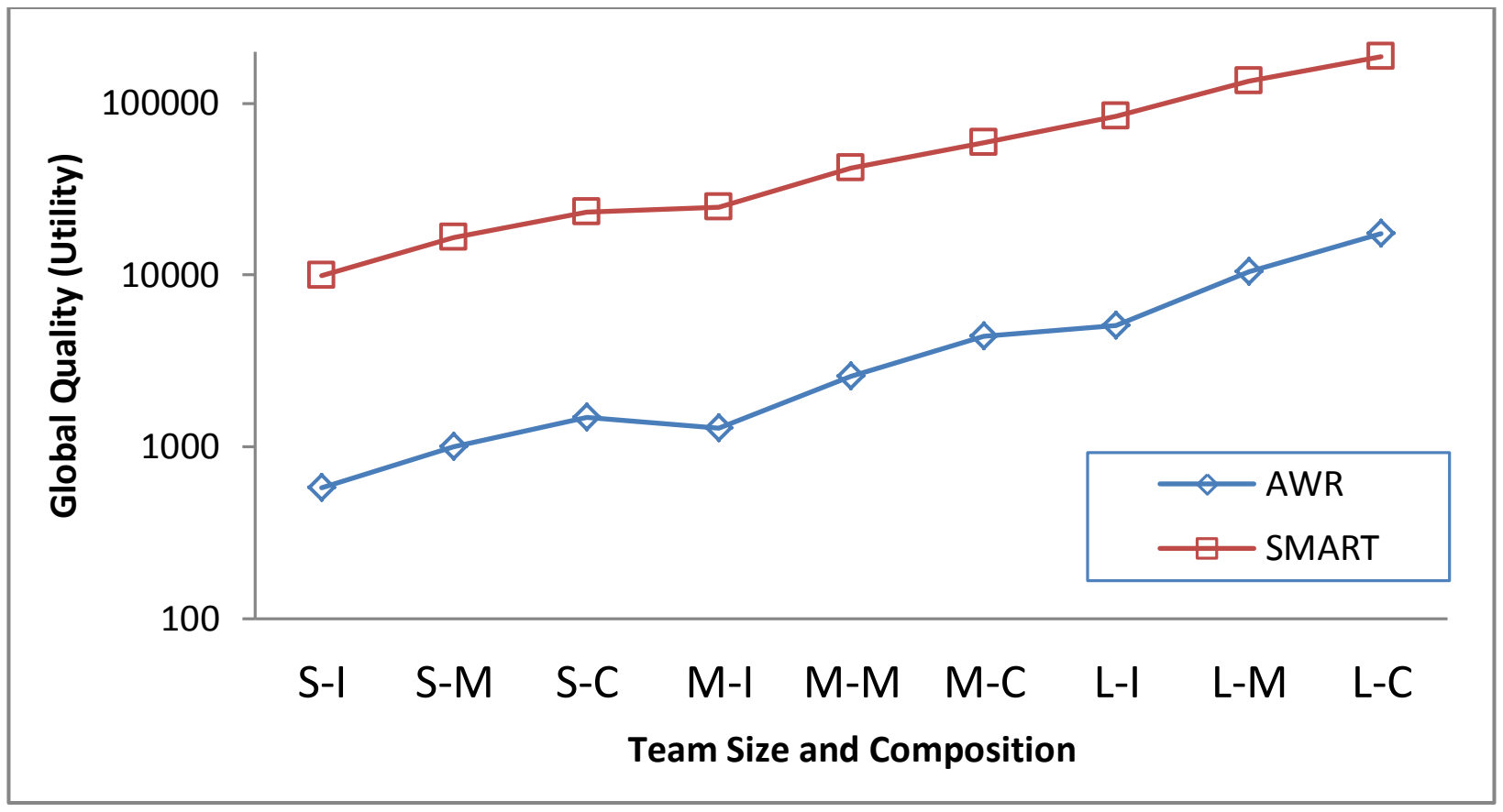

Figure 27. Comparisons of global utility for AWR and SMART

## 4.6 Summary

In this chapter, we explore the use of multi-agent coordination principles to help ASD automatically streamline the tasks distribution and management process to more efficiently utilize the collective capacity of an agile team. Motivated by observations from 20 agile teams consisting of university software engineering students, we propose a *Simple Multi-Agent Real-Time* (SMART) task allocation approach for agile process based on distributed constraint optimization. Personal task allocation agents equipped with SMART helps individual developers to make situation-aware decisions on which incoming tasks to serve so as to reduce the risk of low quality task results while maintaining the timeliness of the overall software project.

To the best of our knowledge, it is the first real-time approach designed to help resource constrained agile team members determine how to react to incoming task requests to protect their quality of artifacts through minimizing the delay experienced by members. Through theoretical analysis, we prove the performance bounds of the SMART approach. The results show that it can achieve close to optimally efficient utilization of the developers' collective capacity and significantly outperform the prevailing practice in terms of both task quality and timeliness of completion.

# CHAPTER 5

# AFFECTIVE MODEL FOR ASD

Understanding emotion status of developers in agile team is essential for the success of software project, especially for a distributed team. If the enhanced agile methodology with emotion prediction can help team leader to monitor developers' mood swing and current team morale to avoid the effect of mood swing to task execution result, it will be wonderful features for ASD. Unfortunately, existing agile or other methodologies seldom consider this challenge.

For analyzing the people's emotion, we firstly need to investigate the different roles in an agile team and some research fields related to this work.

## 5.1 Background

### *5.1.1 Current Roles in ASD*

Generally in an agile team, there are several roles. Roles are not positions, any given person takes on one or more roles can switch roles over time, and any given role may be assigned to zero or more people at any given point in one project.

Figure 27 shows the overview structure of an agile team. The core agile team includes the team of developers who lead by team leader, working closely with a product owner to build high-quality working software during the iterative and incremental process. Sometimes an architecture owner is also involved. The supporting casts including technical experts, domain experts and independent testers etc.

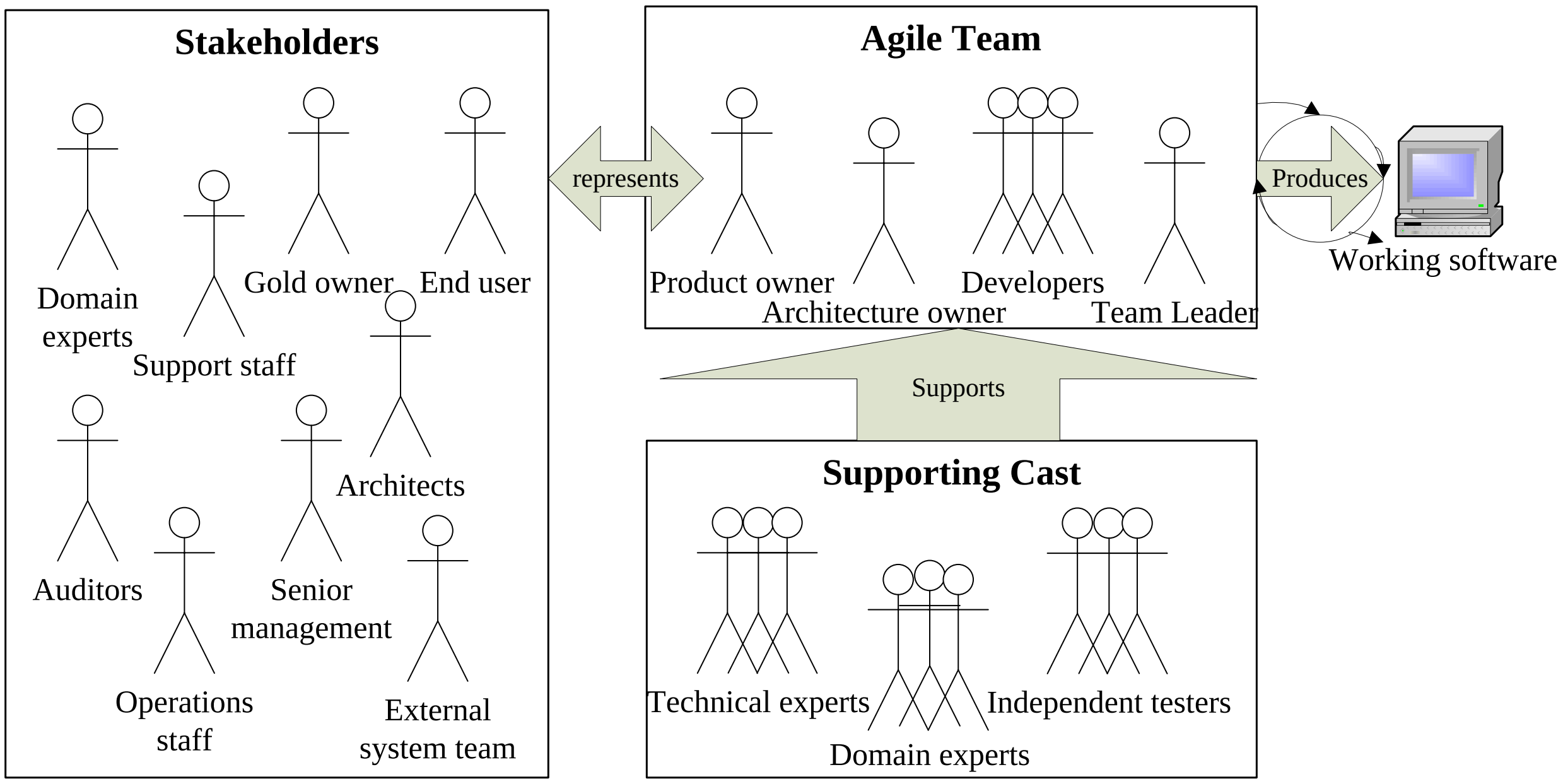

Figure 28. Organization structure of a typical agile project

From the Figure 27 we can see there are the following roles in a general agile team:

**Core roles in agile team:**

- *Team leader* – The person whose role is responsible for facilitating the team, obtaining resources for it, and protecting it from problems. This role encompasses the soft skills of project management but not the technical ones such as planning and scheduling, activities which are better left to the team as a whole. Team leader is accountable for removing impediments to the ability of the agile team to deliver the iteration (in *XP*) or sprint (in Scrum) goal/deliverables. In Scrum, this role is called *Scrum Master* and requires (or strongly recommended) certification, whereas, *XP* or other methods define the role of coach quite informally and the role may float between members of the team. Team leader acts as a buffer between the team and any distracting influences. He/she ensures that the agile process is used as intended. This role is the enforcer of the rules of *Agile*, often chairs key meetings, and challenges the team to improve. The role has also been referred to as a servant-leader to reinforce these dual perspectives [176].

- *Developer* – The team members, including designer, tester, and programmer, whose role is responsible for the creation and delivery of a system. Their activities include analyzing, designing, modeling, programming, testing, and maintaining etc.

- *Product owner* – The product owner, called on-site customer in *XP* and active stakeholder in *AM*, represents the stakeholders. This is the one person whose role is responsible for a team (or sub-team for large projects). He/she is also responsible for the prioritized work item list (the product backlog in *Scrum*), for making decisions in a timely manner, and for providing information in a timely manner.

**Additional roles at scale:**

- *Architecture owner* – The person whose role is responsible for facilitating architectural decisions on a sub-team and is part of the architecture owner team which is responsible for overall architectural direction of the project.

- *Stakeholder* – A stakeholder is anyone who is a direct user, indirect user, manager of users, senior manager, operations staff member, the gold owner who funds the project, support IT staff member, auditors, program manager, developers working on other systems that integrate or interact with the one under development, or maintenance professionals potentially affected by the development and/or deployment of a software project.

**Supporting roles at scale:**

- *Technical expert* – Sometimes the agile team needs the help of technical experts, such as build masters to set up build scripts or a DBA to help design and manage database. Technical experts are brought in on an as-needed, temporary basis, to help the team overcome a difficult problem and to transfer their skills to one or more developers on the team.

- *Domain expert* – As you can see in Figure 27, the PO represents a wide range of stakeholder, not just end users, and in practice it isn't reasonable to expect them to be experts at every specific domain. As a result, PO will sometimes bring in domain experts to work with the team.

- *Independent tester* – Effective agile teams sometimes need an independent test team working in parallel that validates their work throughout the lifecycle. This is an optional role, typically adopted only on very complex or big projects.

### *5.1.2 Existing Approach and Problems*

Because of some stereotypical belief and characters, team members especially programmers aren't very keen on talking about their emotions. But team members are all human beings with human emotions that will significant affect the project progress and final software production. During the development process, their emotion states can be affected by many aspects and events, such as the expectation of goals, delay of tasks, quality of artifacts etc., and further affect back to the process and result. In real projects, some practitioners have noted this and try to use a *Happiness Chart* shown in Figure 7 to monitor team members' daily mood and encourage emotional openness among whole team [11].

According what we discussed above, agile team consists of different peoples and peoples have emotions that highly influence their activities, including those activities in software development process. The team leader knows that the emotion status of members will affect their tasks and artifacts, but human's emotion and its transition are too complex to observe and control, so there are no magical ways to solve this problems because the complexity is essential to the software development processes [2].

In 2011, Palacios et al. have tried to use the affect grid psychological tool created by psychologist Russell to characterize emotions in software requirement engineering. Their results revealed that emotions are key issues in software development activities, and the importance of emotions management in SDLC is obvious as software development is a human capital intensive activity. Knowing the emotional state of the development team helps the project manager to build a good environment or create an award for team to avoid the effects of "bad" emotions [144]. However, current method is too psychological to agile process without a modeling and computing tool.

## 5.2 Related Works

### *5.2.1 Affective Computing*

Affective computing is one of new areas of artificial intelligence today, which is an interdisciplinary field spanning computer sciences, psychology, and cognitive science and might be applied to the ASD process improvement. The concept of "affective" was originally a topic of study in the psychology and cognitive sciences. The modern branch of computer science originated with Rosalind Picard's paper in 1995 on affective computing [177, 178].

Emotions in human communication process are very important for perception, decision-making, interaction, and intelligence. Therefore, investigating affective computing model to ASD is essential to reduce the risk of the developers' mood swing to the development process.

### *5.2.2 A Computable Affective Model: OCC*

Ortony, A., Clore, G., Collins, A. proposed the OCC emotional model in 1988. They think that emotions are the results of the following three types of subjective appraisals [179]:

- The appraisal of the pleasantness of the consequences of events with respect to the agent's goals. Those emotions that we call goal-based or event-driven emotions will be stimulated by valenced reaction to consequences of events and appraised by agent's internal goals, including happy-for, resentment, gloating pity, hope fear, satisfaction, fears-confirmed, relief, disappointment, joy and distress;
- The appraisal of the approval of the actions of the agent self or another agent with respect to a set of behavior standards. Those emotions that we call standard-based or agent-driven emotions will be stimulated by valenced reaction to actions of agents and appraised by agent's internal standards, including pride, shame, admiration and reproach;
- The appraisal of the liking of objects with respect to the attitudes of the agent. Those emotions that we call attitude-based or object-driven emotions will be stimulated by valenced reaction to aspects of objects and appraised by agent's internal attitudes, including love and hate.

Generally speaking, in our understanding of OCC model, emotions are reacted by both internal factors and external factors. Internal factors include goals, preferences and attitudes, whereas external factors include events, conditions and other objects in environment. OCC model also considers another four compound emotions which result from both consequences of events and actions of agents, they are: gratification, remorse, gratitude and anger.

The original OCC model, with its 22 different types of emotions, looks probably too much fine grained. In practice, some researchers think the OCC model should be simplified to match the abilities of the character [180]. A simplified version of this theory was presented in 2003 by Ortony, where he considered only two different categories of emotional reactions: positive and negative [181].

Based on OCC theory, Conati et al. presented a probabilistic model that assesses student emotional reaction during a game in 2002. Their model can predict a player's emotional state by assessing the player's appraisal of interaction with the game, in light of the player's goals and personality [182]. In 2009, Zhang et al. developed an emotional agent for serious game DINO, which is designed based on Goal Net model. Their agent's emotions are modeled by the OCC model and incorporated into FCM inference [183].

### *5.2.3 FCMs research and applications*

According to OCC theory, we know that the developers' mood or emotion is led to and affected by specific events, conditions or other situations in the process. So we want to build an affective model for agile team members to simulate interactions between development events and human emotions. During our surveying, we found FCMs' modeling ability and computational ability might be used to build causal relationship model for ASD process, which bring agile team the ability of viewing some state transitions and changes during process. So here we will firstly introduce the state of the art of FCMs research firstly.

FCMs are fuzzy-graph structures for representing causal reasoning, which is developed from the concept of *Cognitive Maps* (CMs). CMs were initially introduced by *Robert Axelrod* in 1976. CMs were used to model a system with some concepts and cause-effect relationships, those relationships can be divided into three types: positive, negative, or neutral. CMs can be drawn as a simple directed graph, the nodes corresponding to relevant concepts or variables in the given domain, and the directed edges denote the mutual relationships between two concepts or variables. Type of relationship is denoted by a sign that is associated with the edges. Positive sign means the positive type of relationship with promoting effect, which describes a situation in which the start concept causes promoting effect on the end one. This means that increasing in start concept's value will lead to increase in end one's value. Analogically, negative sign means negative type with inhibiting effect, which expresses the situation in which the start concept causes inhibitory effect on the end on, i.e. increasing in start concept's value leads to decreasing in end one's value. No connection between two concepts means that concepts have no relationship with each other, which expresses that they are independent.

CMs have two main drawbacks: 1) they do not allow feedback, which significantly limits its usefulness; 2) the insufficient representation of relationships. Therefore, using of cognitive maps to model complex systems is infeasible, such as the software development environment. For tackling them, the *Fuzzy Cognitive Maps* (FCMs) were proposed as an extension. Comparing to CMs, two significant enhancements introduced in the FCMs are that:

1) **Causal relationships between concept nodes are fuzzified.** This character enriches the description of link by numerical value instead of only using positive, negative and no signs. It allows using varying degrees to denote different causal influence.
2) **It is a dynamic model to express dynamic system.** FCMs can evolve with time, and involve feedback mechanisms in the model. Specifically, the effect of changing the value of one concept node in the model may change the values of other concept nodes, and the change could loop back to the original changing node.

The strength of relationship between two nodes takes on any value in the [-1, 1] range. Value -1 represents full negative effect, whereas +1 means full positive causal effect. Zero denotes no causal effect. Other values denote different fuzzy levels of causal effect. The knowledge of system relationships can be described by a matrix, called connection matrix. Each cell of this matrix stores a value of corresponding relationship. Commonly used convention is to place start nodes in rows and end nodes in columns.

For example, considering the system with N concept nodes, we have $N \times N$ matrix representing the FCM, the elements in the matrix are the causal link strengths. Any one state of the system can be determined by one state vector, which specifies current values of all system concept nodes. The FCM iteratively updates the state of the system. In one iteration process, the value of each node is calculated based on the current values of every node by exerting influence on it through its causal link. After multiplying these values by edge weight between the two nodes, which represents the strength of the relationship between the nodes, the sum of these products is taken as the input to a transformation function, which is used to reduce unbounded inputs to a certain range.

The value of each node in any iteration is computed from values of nodes in preceding state, using the following equation:

$$N_j(k+1) = f\left(\sum_{i=1}^{n} e_{ij} \cdot N_i(k)\right)$$

where $N_i(k)$ is the value of $i^{th}$ node in the $k^{th}$ iteration (system state), $e_{ij}$ is edge weight (relationship strength) between nodes $N_i$ and $N_j$, $k$ is the corresponding iteration, n is the number of concepts, and $f$ is the transformation function.

Three types of transformation function which are commonly used are *Binary*, *Trivalent* and *Sigmoid* function [184].

*(1) Bivalent*

$$f(n) = \begin{cases} 0, & n \le 0 \\ 1, & n > 0 \end{cases}$$

*(2) Trivalent*

$$f(n) = \begin{cases} -1, & n \le -0.5 \\ 0, & -0.5 < n < 0.5 \\ 1, & n \ge 0.5 \end{cases}$$

*(3) Sigmoid*

$$f(n) = \frac{1}{1 + e^{-cn}}$$

*Or other sigmoid functions* [184]

The final results of a simulation performed with FCM strongly depend on the transformation function. If we use function which results in binary values, the simulation of a FCM system leads to either fixed state pattern of node values, which is called hidden pattern or fixed-point attractor, or a cycling between several states, which is known as the limit cycle. If we use a continuous-output transformation function, the simulation may result in a different outcome. The system may continue to produce different state vector values for successive cycles. In this case, this unstable situation is called chaotic attractor [185].

FCMs theory has been applied in many application areas more than two decades, Table 25 lists some research contributions about its application, which shows that FCMs are successfully applied to modeling system or environment with complex causal relationship in many areas, including software project management [185], software quality risk analysis [186] and agile software development [187].

Table 24. FCMs are used to model complex system or environment in many application areas

| Year | Author | Application area |
|---|---|---|
| 1989 | Gotoh et al. [188] | plant control |
| 1991 | Styblinski et al. [189] | analysis of electrical circuits |
| 1991 | Taber [190] | disease diagnosis |
| 1992 | Kosko [191] | political affairs |
| 1993 | Dickerson et al. [192] | modeling of virtual worlds |
| 1994 | Dickerson et al. [193] | modeling of virtual worlds |
| 1995 | Pelaezet al. [194] | analysis of failure modes effects |
| 1996 | Ndousseet al. [195] | fault management in distributed network environment |
| 1997 | Kardaras et al. [196] | modeling and analysis of business performance indicators |
| 1998 | Stylios et al. [197] | modeling of supervisory systems |
| 1998 | Banini et al. [198] | factors affecting slurry rheology |
| 1999 | Stylios et al. [199] | large manufacturing systems |
| **2004** | **Stach et al. [185]** | **software project management** |
| 2006 | Niskanen [200] | prisoner's dilemma |
| 2007 | Niskanen [201] | business planning models |
| 2008 | Froelich et al. [202] | stock market modeling and forecasting |
| 2009 | Papageorgiou et al. [203] | cotton yield management in precision |
| 2010 | Song et al. [204] | prediction of time series |
| **2010** | **Cao et al. [187]** | **Dynamics in Agile Software Development** |
| **2011** | **Bhatia et al. [186]** | **software quality risk analysis** |
| 2012 | Papageorgiou et al. [205] | prediction of pulmonary infections |

## 5.3 Method

As FCMs theory and method can be used to build causal relationship model for complex system or process. So we will use FCMs to build the affective model for agile team, and then use their computational functionality to simulate the influence between development events and human moods.

Our proposed method assumes that everyone has a relatively stable emotion quotient (EQ) status. External events or activities can affect developers' emotion or mood, so events (including its results) and developers' emotions will be counted into this model. Our method includes the following four steps:

- Step 1: Design a general EQ FCMs model for all team members, the concept nodes will include emotions, external events/activities and their results. There are two ways to determine the weights between concepts, one is self-reported by members, and another is learned from members' past working data.
- Step 2: Execute FCMs to reach to the equilibrium state. The state can be treated as a stable personalized EQ level of members.
- Step 3: When a new event/activity comes out or a new event/activity result occurs, to change the concept value and execute FCMs from previous equilibrium point to new equilibrium state. Especially pay more attention on the changes at early stage when emotions and results value sharply shock, and the time it come back to the new equilibrium point. In this step, different people will have different performance.

- Step 4: Show a comparison score list to help team leader to predict who will be the most suitable candidate to handle the new event/activity.

## 5.4 Simulation and Analysis

In this case, *Michael* and *Grace* are working in an agile team as developer. For simplifying the model, we consider their complex emotions as a comprehensive mood factor.

### *5.4.1 Simulation Scenario I*

***Step 1: Building FCMs Model***

According common software development experience and ignoring other factors, we assume the causal relationships between developer's mood status, progress of task execution and quality of task execution during development process shown in FCMs of figure 28.

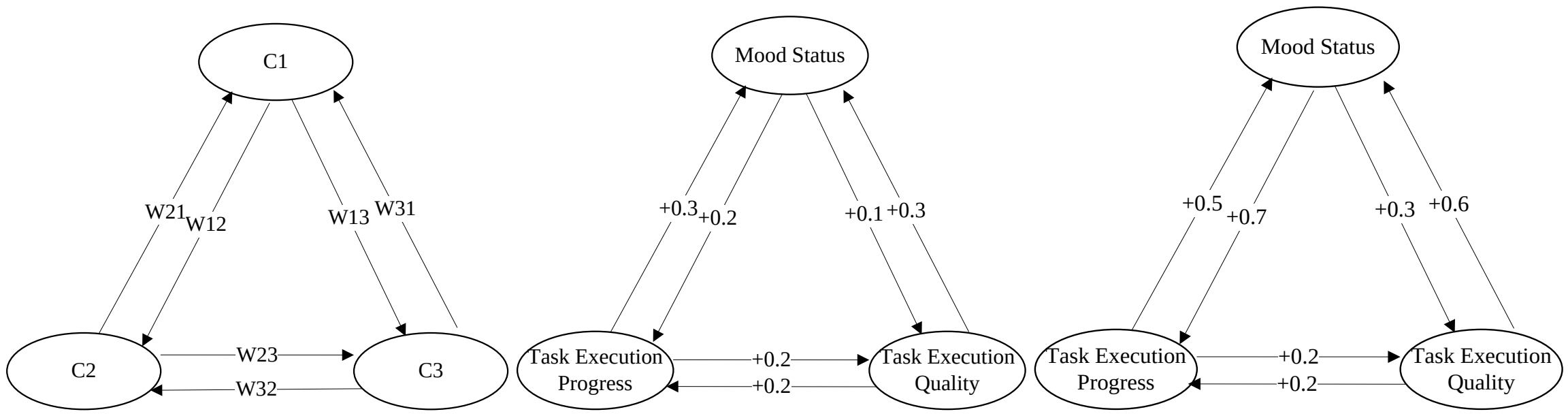

Figure 29. FCM-based first model of team member's mood-task execution causal relationships (left: general model; middle: developer *Michael*'s model; right: developer *Grace*'s model)

The first general model presents relations that are essential during software development process, which consists of three concept nodes as follows:

- C1: Worker's *Mood Status* (called *Mood* for short), which reflects a comprehensive emotions or feeling to a task, event or event result. Those emotions might include interesting or boring, like or dislike, stressed or relaxed etc., which only affect to or is affected by the result of task execution. Its value can be gained by affective button technique [206] and then be transferred into a fuzzy number range from 0 to 1.
- C2: *Task Execution Progress* (called *Progress* for short), which can be interpreted as a relative comparison concept for task execution progress. Its value can be calculated by the ratio of actual task completed time to estimated task completed time, and then be transferred into a fuzzy number range from 0 to 1.
- C3: *Task Execution Quality* (called *Quality* for short), which can be interpreted as a relative comparison concept for artifact quality after task executing. Its value can be gained by questionnaire and then be transferred into a fuzzy number range from 0 to 1.

The casual relationships between nodes, which are represented by directed edges with a fuzzy number of weight acquired from human experiences by questionnaires. Table 26 shows a sample questionnaire in our scenario study.

Table 25. A sample questionnaire for determining the weights of FCM

| |
|---|
| Thank you for attending this survey as a developer. Please help us improve our team intelligence by completing this survey.<br>The entire survey will take approximately 1minutes to complete.<br>----------------------------------------------------------------------------------------------------------<br>Hi, my dear friend, please choose your feeling when you meet the following situation during development.<br>1. When you work get bad quality, will you feel sad?<br>☐ Not at all ☐ A little ☐ Moderately ☐ Mostly ☐ Completely |

2. When you work is delayed, will you feel bad?
   ☐ Not at all ☐ A little ☐ Moderately ☐ Mostly ☐ Completely

3. When you're feeling down today, do you think that will affect your work quality?
   ☐ Not at all ☐ A little ☐ Moderately ☐ Mostly ☐ Completely

4. When you're feeling down today, do you think that will affect your work progress?
   ☐ Not at all ☐ A little ☐ Moderately ☐ Mostly ☐ Completely

5. If give you more time, do you think that your works will be better than now?
   ☐ Not at all ☐ A little ☐ Moderately ☐ Mostly ☐ Completely

6. You fulfill task very well, do you think which will help you get good progress for next task?
   ☐ Not at all ☐ A little ☐ Moderately ☐ Mostly ☐ Completely

From the FCM model, we can see the mood status of developer who picks up the task has positive effect on the task execution progress (0.2 directed edge from C1 to C2 for *Michael*; and 0.7 directed edge from C1 to C2 for *Grace*), which means the progress of task execution will be good when developer's mood status is at high level. Meanwhile, the task execution progress also positively influence the mood status of developer (0.3 directed edge from C2 to C1 for *Michael*; and 0.5 directed edge from C2 to C1 for *Grace*), which means the developer might feel depress when task delayed or feel happy when progress is good. This obeys common sense in real projects.

Furthermore, the developer mood status positively influences the task execution quality (+0.1 directed edge from C1 to C3 for *Michael*; and +0.3 directed edge from C1 to C3 for *Grace*). On the other hand, increase in quality can lead to higher mood status for developer, e.g. he/she can be excited to the good evaluation result of software (+0.3 directed edge from C3 to C1 for *Michael*; and +0.6 directed edge from C3 to C1 for *Grace*). Generally, good task execution progress also leads to high quality (+0.2 directed edge from C2 to C3 for *Michael*; and +0.2 directed edge from C2 to C3 for *Grace*), as well as high quality with few bugs will make the progress better (+0.2 directed edge from C3 to C2 for *Michael*; and +0.2 directed edge from C3 to C2 for *Grace*). The strength of the relationships was established by developer's personal experience, and reflects common perception of the strength of these relationships.

***Step 2: Run FCMs Simulation***

The FCM-based model for *Michael* can be presented in a matrix form:

$$W = \begin{bmatrix} 0 & 0.2 & 0.1 \\ 0.3 & 0 & 0.2 \\ 0.3 & 0.2 & 0 \end{bmatrix}$$

And the matrix form for *Grace's* FCM model:

$$W = \begin{bmatrix} 0 & 0.7 & 0.3 \\ 0.5 & 0 & 0.2 \\ 0.6 & 0.2 & 0 \end{bmatrix}$$

Next, the developed model was simulated. The starting vector is denoted Iteration #0. Each state vector consists of three numbers, which correspond to conceptual nodes as follows: *Mood* (C1), *Progress* (C2), and *Quality* (C3). The experiment makes possible to examine the mutual relationships among these elements. The simulation begins with the start state vector Iteration #0 = (0.5, 0, 0) for both developers, which represents a situation that beginning of task when *Emotion* concept is active and set at middle value 0.5, and other concepts are inactive, i.e. their values are set as zero, which indicates that the workers did not start to work yet. As the simulation continues successive values of nodes show trends which occur with the progressing time. By analyzing states of nodes in model computational process, some observations and consequents can be learned and analyzed.

The experiment was carried out using the logistic sigmoid function as a threshold function, which is a continuous-output transformation function and thus provides true fuzzy conceptual node states. We mainly want to see the comparative results between two developers, so the constant c can be set to a common number 5. Rounding to six significant digits, during the simulation the first 20 states are achieved in Table 27:

Table 26. Results of simulation of first model (left: *Michael*; right: *Grace*)

| Iteration# | Mood | Progress | Quality |
|---|---|---|---|
| **0** | 0.5 | 0 | 0 |
| **1** | 0.5 | 0.851953 | 0.679179 |
| **2** | 0.984744 | 0.919025 | 0.832291 |
| **3** | 0.991792 | 0.986331 | 0.916533 |
| **4** | 0.994597 | 0.987725 | 0.9223 |
| **5** | 0.994708 | 0.987912 | 0.922701 |
| **6** | 0.994717 | 0.987922 | 0.922726 |
| **7** | 0.994717 | 0.987922 | 0.922728 |
| **8** | 0.994717 | 0.987922 | 0.922728 |
| **9** | 0.994717 | 0.987922 | 0.922728 |
| **10** | 0.994717 | 0.987922 | 0.922728 |
| **11** | 0.994717 | 0.987922 | 0.922728 |
| **12** | 0.994717 | 0.987922 | 0.922728 |
| **13** | 0.994717 | 0.987922 | 0.922728 |
| **14** | 0.994717 | 0.987922 | 0.922728 |
| **15** | 0.994717 | 0.987922 | 0.922728 |
| **16** | 0.994717 | 0.987922 | 0.922728 |
| **17** | 0.994717 | 0.987922 | 0.922728 |
| **18** | 0.994717 | 0.987922 | 0.922728 |
| **19** | 0.994717 | 0.987922 | 0.922728 |

| Iteration# | Mood | Progress | Quality |
|---|---|---|---|
| **0** | 0.5 | 0 | 0 |
| **1** | 0.5 | 0.622459 | 0.562177 |
| **2** | 0.85532 | 0.743106 | 0.705257 |
| **3** | 0.897757 | 0.826436 | 0.763284 |
| **4** | 0.915644 | 0.840378 | 0.781651 |
| **5** | 0.919313 | 0.845181 | 0.785532 |
| **6** | 0.920274 | 0.846166 | 0.786648 |
| **7** | 0.920505 | 0.846437 | 0.786894 |
| **8** | 0.920561 | 0.846499 | 0.786959 |
| **9** | 0.920575 | 0.846514 | 0.786974 |
| **10** | 0.920578 | 0.846518 | 0.786977 |
| **11** | 0.920579 | 0.846519 | 0.786978 |
| **12** | 0.920579 | 0.846519 | 0.786979 |
| **13** | 0.920579 | 0.846519 | 0.786979 |
| **14** | 0.92058 | 0.846519 | 0.786979 |
| **15** | 0.92058 | 0.846519 | 0.786979 |
| **16** | 0.92058 | 0.846519 | 0.786979 |
| **17** | 0.92058 | 0.846519 | 0.786979 |
| **18** | 0.92058 | 0.846519 | 0.786979 |
| **19** | 0.92058 | 0.846519 | 0.786979 |

For *Michael*, the model steadily reaches equilibrium at state of

Iteration #14= (0.920579509, 0.846519301, 0.786978702)

And for *Grace*, the model steadily reaches equilibrium at state of

Iteration #7= (0.994717128, 0.987922232, 0.92272765)

Figure 29 shows the result of *Michael's* first model in a plot, and figure 30 shows the result of *Grace's* first model in a plot.

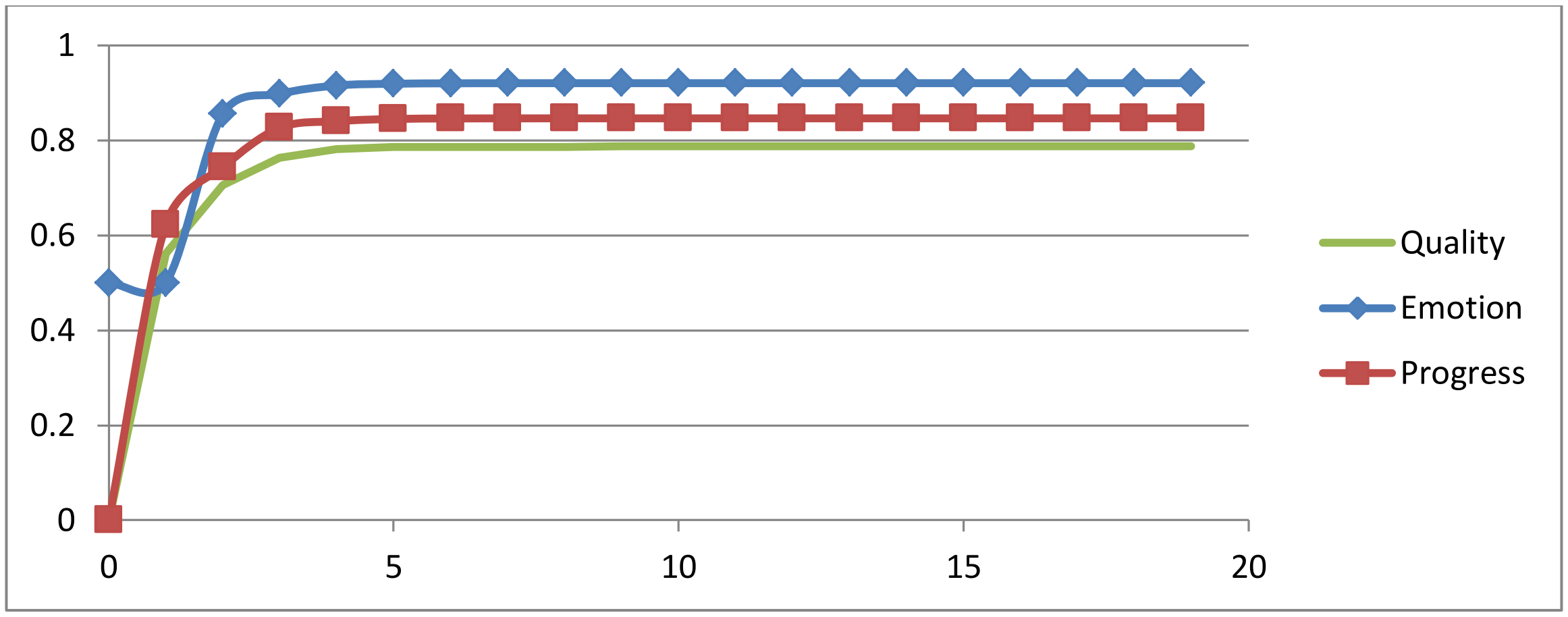

Figure 30. Result of *Michael*'s first model

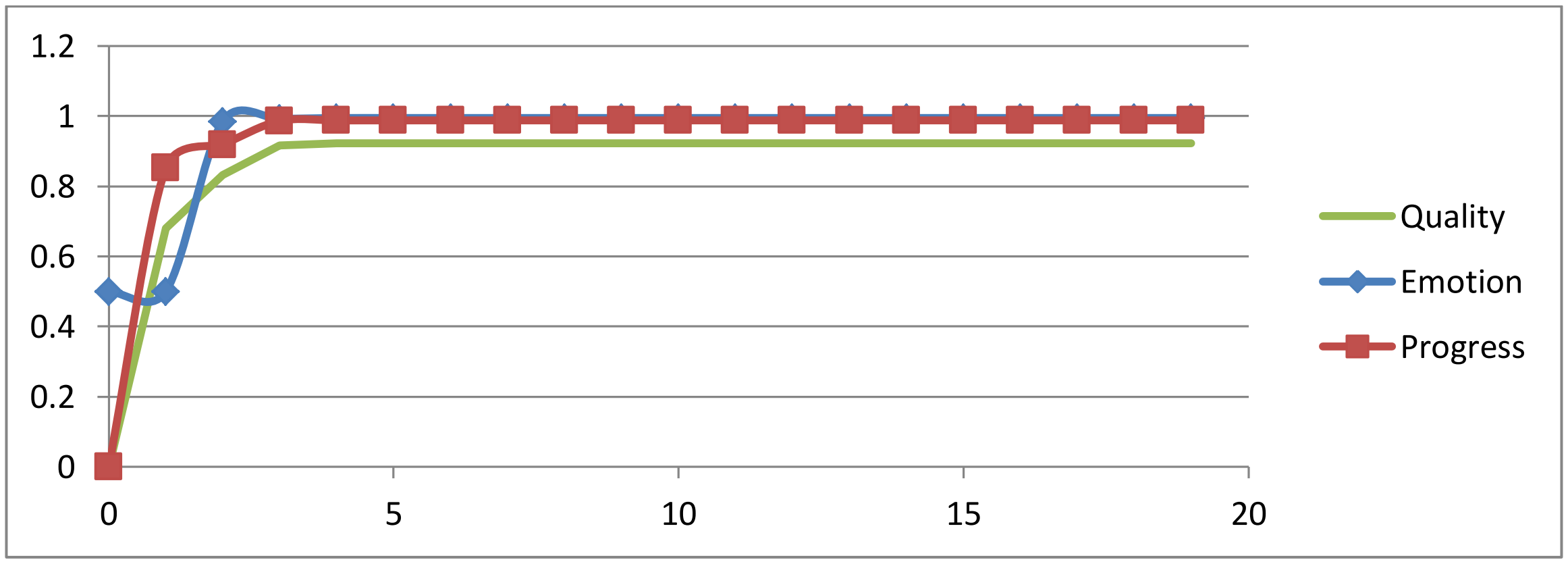

Figure 31. Result of *Grace*'s first model

***Step 3: Analysis and Comparison:***

This model describes state changes trend in a task development activities. It characterizes a situation when a worker starts to do a task. Successive states of the modeled system show changes in considered nodes, which represent workers' emotional status and task execution aspects. Values of the nodes describe trends, i.e. qualitative changes evolving with time, of the considered aspects. The final state achieved by the model shows the state where equilibrium between the considered concepts, which shows the relative task execution result which is achieved for the modeled system.

The first iteration (#1) shows that situation when tasks have been started to execute by worker at the beginning phase. With the time changes, the workers' emotional status and quality increase rapidly. At iteration #1, the workers need some time to know well the new task and environment, so their progress are not too fast and emotional states are not so high. After this period, the emotional status increases quickly and then stays at the equilibrium, especially for emotional *Grace*. For *Michael*, after the quality reaching the highest at iteration #12 and the progress reaching the highest at iteration #11, the rates increase slightly and stay at the equilibrium at iteration #14. For *Grace*, before the quality reaching the highest at iteration #7, the rates and the progress both reach the highest at iteration #6 and then stay at the equilibrium. This result shows the emotional person need less time to reach the stable state, and their final state performs better than others. The plot shows the emotional *Grace* have relative better performance than *Michael*, which might because that the emotional developers have strong motivation to chase the high quality and good progress for tasks.

So according to the first model, *Grace* will be better person for picking up those tasks. And then let's consider a new factor in next FCMs model of simulation scenario II.

### *5.4.2 Simulation Scenario II*

#### *Step 1: Building Model*

Considering one more human factor, when developer picks up a new task, the difficulty of new task will affect his/her mood status and the progress of task execution. We assume the causal relationships between the difficulty of task, developer's mood status, progress of task execution and quality of task execution shown in FCMs of figure 31.

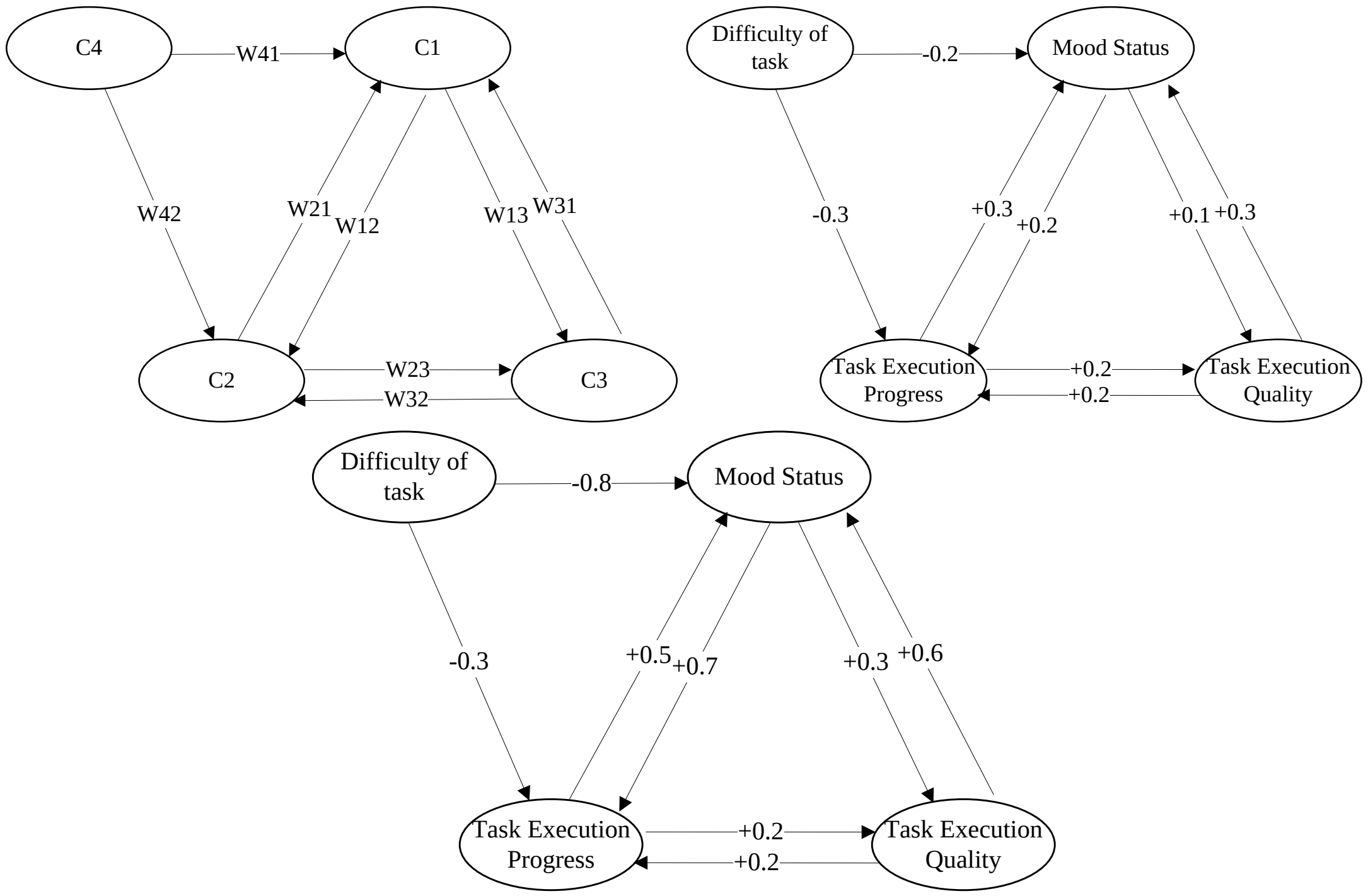

Figure 32. FCM-based first model of team member's mood-task execution causal relationships (up-left: general model; up-right: an example for developer *Michael*; down: an example for developer *Grace*)

The new concept node is:

- C4: *Difficulty of Task* (called *Difficulty* for short), which reflects a generally recognized difficulty factor for task. The value can be determined by whole team or a group of supporting experts, and then will be transferred to a fuzzy number range from 0 to 1 for computation in *FCM*s.

Generally, the difficulty of task has negative effect on developer's mood status (-0.2 directed edge from C4 to C1 for *Michael*; and -0.8 directed edge from C4 to C1 for *Grace*), which means that a task is more difficult, the developer will be more stressful. And the difficulty of task negatively influence the progress of task execution (-0.3 directed edge from C4 to C2 for Michael; and -0.3 directed edge from C4 to C2 for Grace), which reflects more difficult tasks need more time to execute.

The causal relationships between other three nodes are same to the first model.

***Step 2: Run FCMs Simulation***

The FCM-based model for *Michael* can be presented in a matrix form:

$$W = \begin{bmatrix} 0 & 0.2 & 0.1 & 0 \\ 0.3 & 0 & 0.2 & 0 \\ 0.3 & 0.2 & 0 & 0 \\ -0.2 & -0.3 & 0 & 0 \end{bmatrix}$$

And the matrix form for *Grace's* FCM model:

$$W = \begin{bmatrix} 0 & 0.7 & 0.3 & 0 \\ 0.5 & 0 & 0.2 & 0 \\ 0.6 & 0.2 & 0 & 0 \\ -0.8 & -0.3 & 0 & 0 \end{bmatrix}$$

Then, the developed model was simulated again. Each state vector consists of four numbers, which correspond to conceptual nodes as follows: *Mood* (C1), *Progress* (C2), *Quality* (C3) and *Difficulty* (C4). The simulation begins with start state vector Iteration #0 = (0.920579509, 0.846519301, 0.786978702, 1) for *Michael* and Iteration #0 = (0.994717128, 0.987922232, 0.92272765, 1) for *Grace*, which represent a situation when a top difficult task is picked up by them and set at value 1, and other concepts start from the equilibrium state with first model.

The simulation was also carried out using the same logistic sigmoid function as a threshold function. Rounding to six significant digits, during the simulation the first 20 states of *Michael's* are achieved in Table 28, *Grace's* results are achieved in Table 29:

Table 27. Results of simulation of second model (*Michael*)

| Iteration# | Emotion | Progress | Quality | Difficulty |
|---|---|---|---|---|
| 0 | 0.92058 | 0.846519 | 0.786979 | 1 |
| 1 | 0.810036 | 0.551704 | 0.786979 | 0.5 |
| 2 | 0.818768 | 0.699941 | 0.722465 | 0.5 |
| 3 | 0.836663 | 0.688096 | 0.752003 | 0.5 |
| 4 | 0.840258 | 0.698184 | 0.751463 | 0.5 |
| 5 | 0.842171 | 0.698827 | 0.753676 | 0.5 |
| 6 | 0.84274 | 0.699695 | 0.753973 | 0.5 |
| 7 | 0.842971 | 0.699877 | 0.754186 | 0.5 |
| 8 | 0.843049 | 0.69997 | 0.754241 | 0.5 |
| 9 | 0.843079 | 0.699999 | 0.754266 | 0.5 |
| 10 | 0.843089 | 0.70001 | 0.754274 | 0.5 |
| 11 | 0.843093 | 0.700014 | 0.754277 | 0.5 |
| 12 | 0.843095 | 0.700015 | 0.754278 | 0.5 |
| 13 | 0.843095 | 0.700016 | 0.754278 | 0.5 |
| 14 | 0.843095 | 0.700016 | 0.754279 | 0.5 |
| 15 | 0.843095 | 0.700016 | 0.754279 | 0.5 |
| 16 | 0.843095 | 0.700016 | 0.754279 | 0.5 |
| 17 | 0.843095 | 0.700016 | 0.754279 | 0.5 |
| 18 | 0.843095 | 0.700016 | 0.754279 | 0.5 |
| 19 | 0.843095 | 0.700016 | 0.754279 | 0.5 |

Table 28. Results of simulation of second model (*Grace*)

| Iteration# | Emotion | Progress | Quality | Difficulty |
|---|---|---|---|---|
| 0 | 0.994717 | 0.987922 | 0.922728 | 1 |
| 1 | 0.775214 | 0.948056 | 0.922728 | 0.5 |
| 2 | 0.958446 | 0.947149 | 0.891956 | 0.5 |
| 3 | 0.954511 | 0.970589 | 0.915662 | 0.5 |
| 4 | 0.959823 | 0.970871 | 0.917007 | 0.5 |
| 5 | 0.960006 | 0.971429 | 0.917632 | 0.5 |
| 6 | 0.960131 | 0.971464 | 0.917695 | 0.5 |
| 7 | 0.960142 | 0.971478 | 0.917712 | 0.5 |
| 8 | 0.960145 | 0.97148 | 0.917714 | 0.5 |
| 9 | 0.960145 | 0.97148 | 0.917715 | 0.5 |
| 10 | 0.960145 | 0.97148 | 0.917715 | 0.5 |
| 11 | 0.960145 | 0.97148 | 0.917715 | 0.5 |
| 12 | 0.960145 | 0.97148 | 0.917715 | 0.5 |
| 13 | 0.960145 | 0.97148 | 0.917715 | 0.5 |
| 14 | 0.960145 | 0.97148 | 0.917715 | 0.5 |
| 15 | 0.960145 | 0.97148 | 0.917715 | 0.5 |
| 16 | 0.960145 | 0.97148 | 0.917715 | 0.5 |
| 17 | 0.960145 | 0.97148 | 0.917715 | 0.5 |
| 18 | 0.960145 | 0.97148 | 0.917715 | 0.5 |
| 19 | 0.960145 | 0.97148 | 0.917715 | 0.5 |

For *Michael*, the model steadily reaches equilibrium at state of

Iteration #14= (0.843095, 0.700016, 0.754279, 0.5)

And for *Grace*, the model steadily reaches equilibrium at state of

Iteration #9= (0.960145, 0.97148, 0.917715, 0.5)

Figure 32 shows the result of *Michael's* second model in a plot, and figure 33 shows the result of *Grace's* second model in a plot.

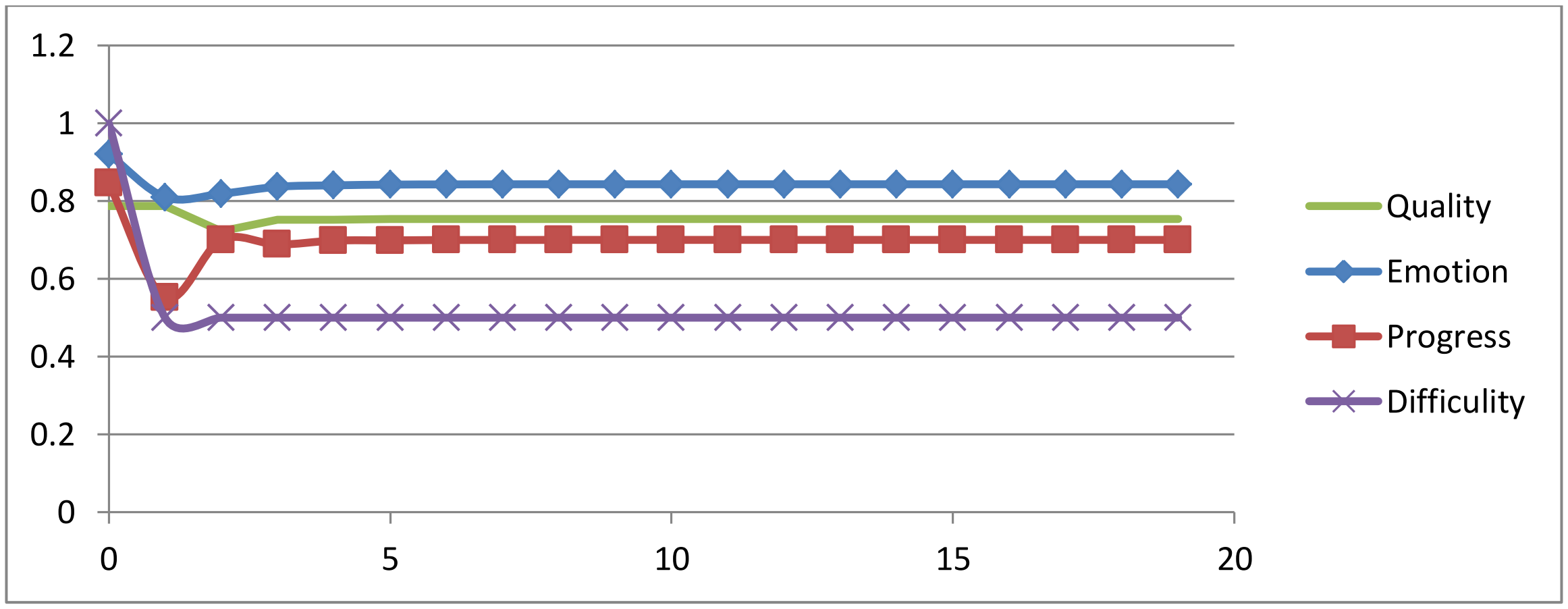

Figure 33. Result of Michael's second model

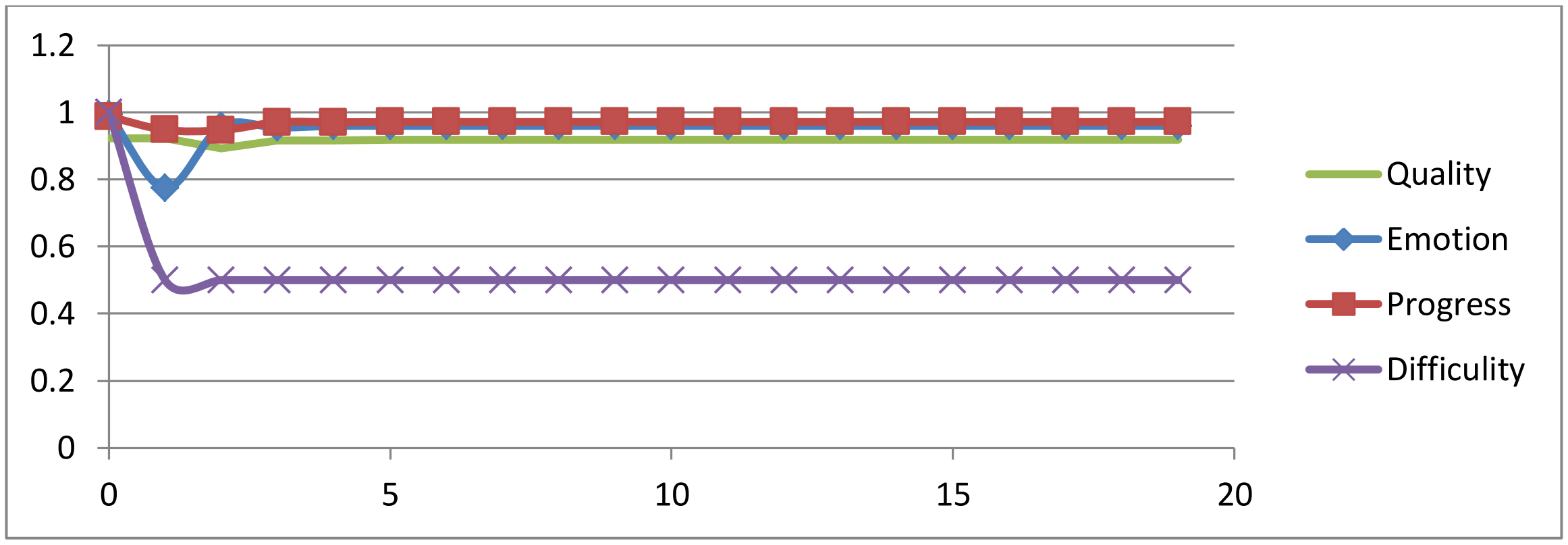

Figure 34. Result of Grace's second model

***Step3: Analysis and Comparison***

This model describes state changes trend in a new task execution. The simulation shows that upon adding new concept into the model, task execution progress and quality are different with the first model for both two developers. From the result, we can see that emotional *Grace's* mood status is more turbulent than *Michael's* at beginning, which affect her progress and quality of task execution until they reach the equilibrium.

Focusing our attention on quality and progress node, both final states of *Michael's* are lower than *Grace's*. So *Grace* is still better person then *Michael* to execute those tasks from the result.

### *5.4.3 Other Simulations*

Several additional simulations were also performed with the second model. This time, the initial state vector was changed. Several experiments, which illustrate how the initial value of the *Difficulty of Task* node impacts on the progress, were performed. Initial values of the other nodes remained unchanged, which allows observing and comparison of influence of varying initial values of the new *Difficulty of Task* node on the system behavior. The obtained results are summarized in Figures 34-37.

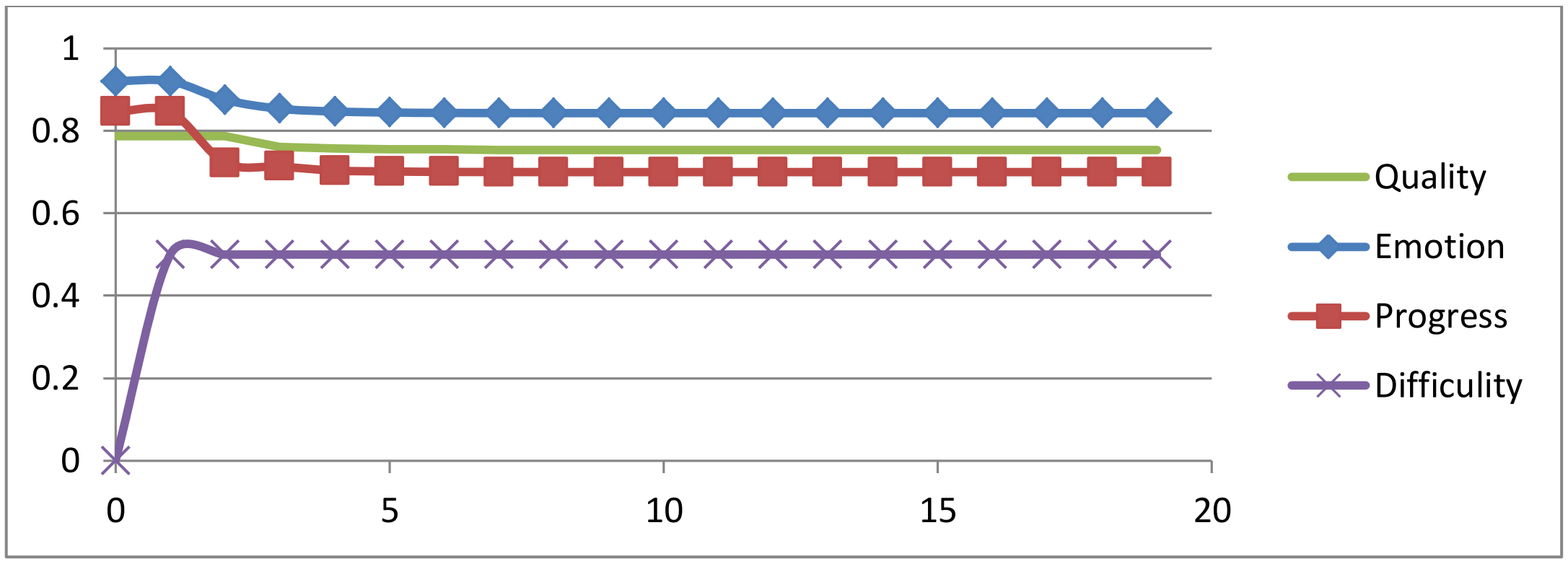

Figure 35. Result of *Michael*'s second model (initial value of Difficulty is 0)

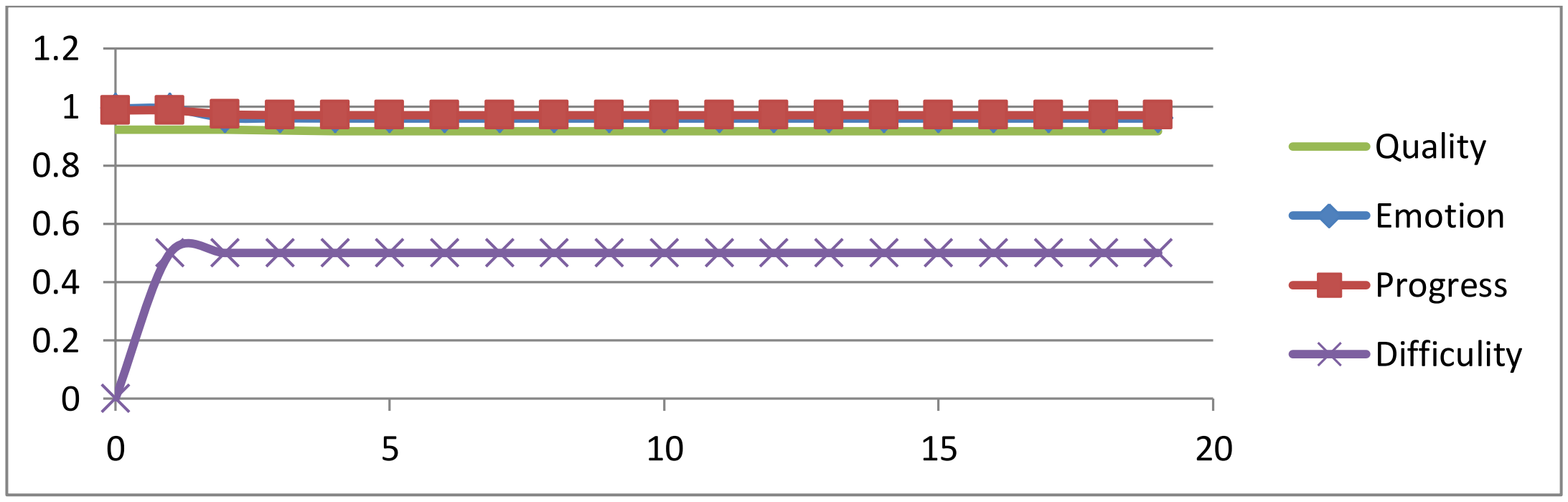

Figure 36. Result of *Grace*'s second model (initial value of Difficulty is 0)

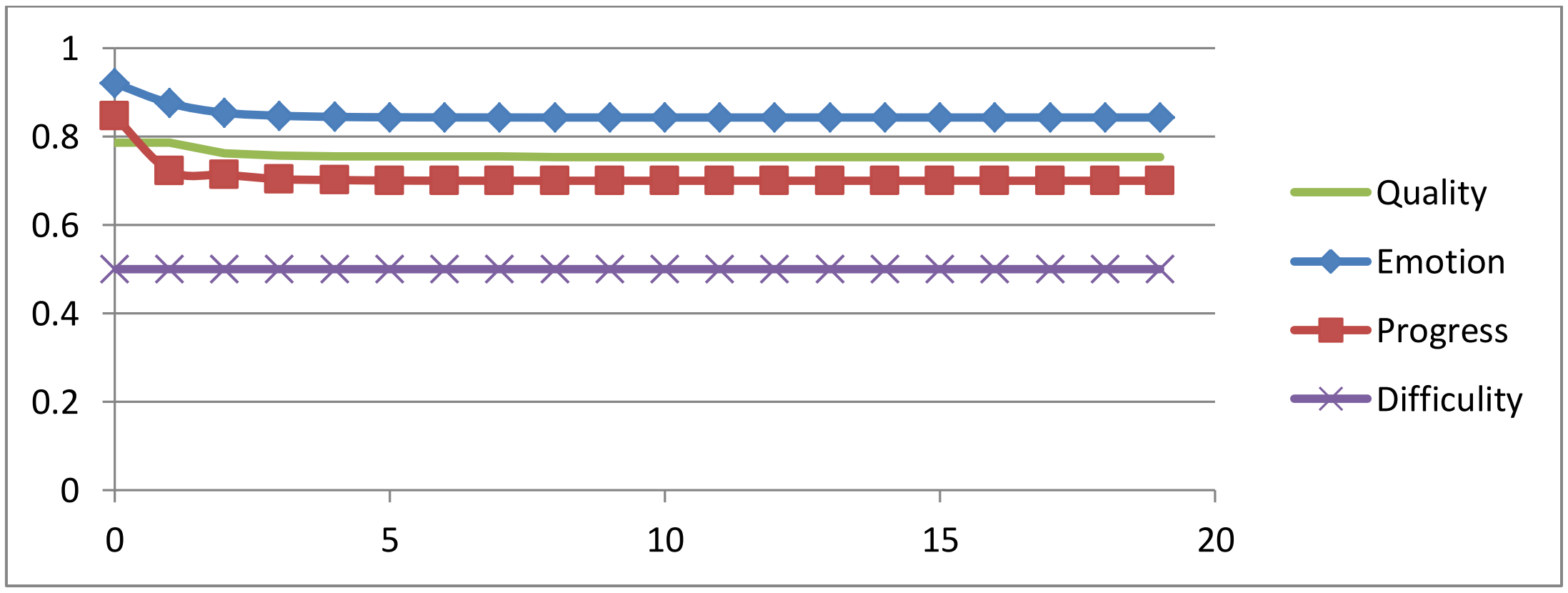

Figure 37. Result of *Michael*'s second model (initial value of Difficulty is 0.5)

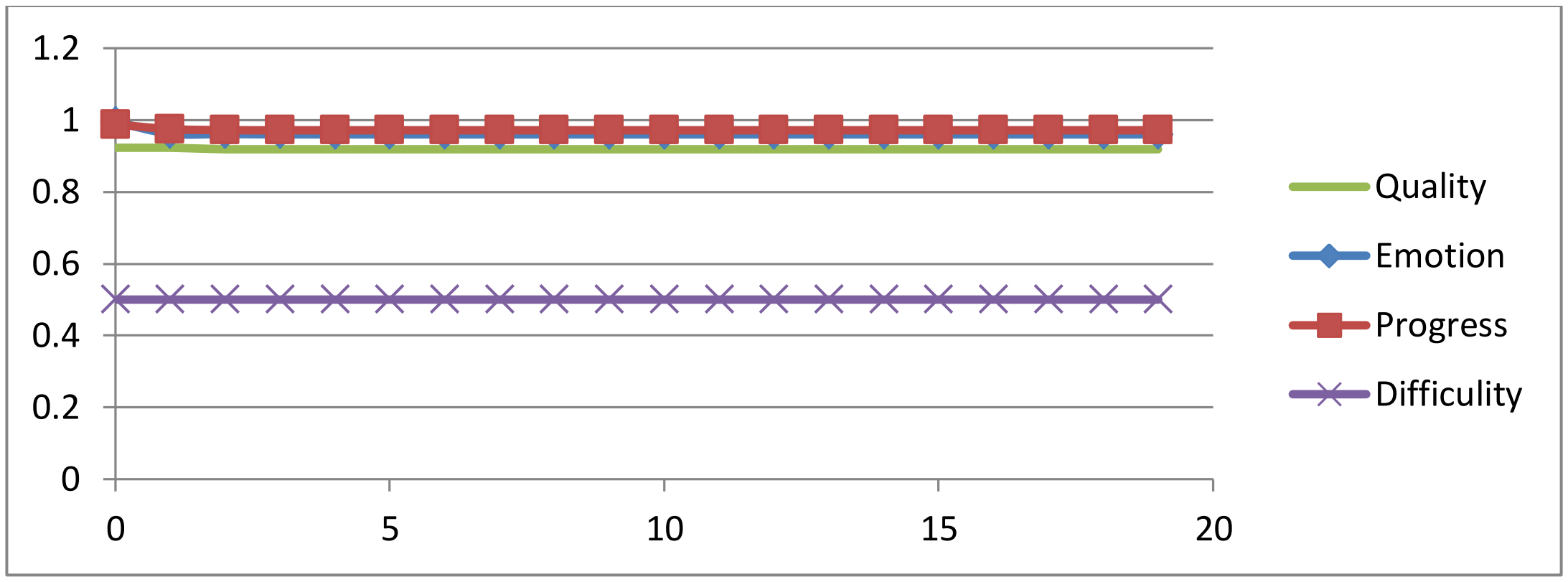

Figure 38. Result of *Grace*'s second model (initial value of Difficulty is 0.5)

The simulation results show that obviously *Grace* is the better candidate for executing those tasks. With considering more human factors into the FCM model, we can find more interesting observations hiding in the complex development process.

## 5.5 Summary

Human's emotion or mood might be the most difficult thing that can be modeled and calculated by computer. But actually developers' emotion and the mood swing affect the software development process and software production.

In this chapter, we analyzed current agile methodologies from aspects of peoples and roles in agile team, as well as emotion factors in development process. And then we proposed a FCMs-based method to monitor mood changes of team members in agile process, which might help team leader to find right developer to implement tasks when some event occurred or some condition is triggered during ASD process.

Our method has been applied on some educational game studies [207, 208]. For applying this method on ASD process, this study is at very preliminary stage and needs to be further verified in the future's experiments. Although it is a real challenge, exploring the relationships between personal emotions or team morale and the development activities will be our further main research direction. In next chapter, we will also discuss some more empirical insights into the process data collected from the 2014 student teams.

# CHAPTER 6

# EMPIRICAL INSIGHTS INTO SCRUM NOVICES

Through our four years experiments on students' *Scrum* based ASD process, we have more deeply understanding and thinking on the human factors of agile methodology. In this chapter, we present empirical findings regarding the aspects of task allocation decision-making, collaboration, and team morale related to the *Scrum* ASD process which have not yet been well studied by existing research. We draw our findings from a 12 week long course work project in 2014 involving 125 undergraduate software engineering students from a renowned university working in 21 *Scrum* teams. Instead of the traditional survey/interview based methods, which suffer from limitations in scale and level of details, we obtain fine grained data through logging students' activities in our online agile project management (APM) platform - HASE. During this study, the platform logged over 10,000 ASD activities. Deviating from existing preconceptions, our results suggest negative correlations between collaboration and team performance as well as team morale for those student or novice agile team.

## 6.1 Background

Over the past decade, many companies have fundamentally changed the way they tackle challenges in the software engineering. In place of the traditional plan-driven approaches, the ASD approaches are becoming increasingly widely adopted. ASD embodies a new way of thinking and working. It alters the philosophy and practice governing how companies collaborate with their customers, how companies coordinate development, and how engineers develop software. ASD values a set of core principles such as iterative shipments of working software increments, early attention into software quality by all developers, close customer collaboration for fast feedback, and a focus on collaboration within a software development team.

In recent years, there has been a growing interest to understand how well software developers adapt to ASD [209-212]. Nevertheless, to date, research in this field remains limited and more empirical studies are constantly sought after by researchers. According to the *8th annual State of Agile Survey* in 2013 [84], *Scrum* and its variants are the most popular ASD methodologies. However, most of the existing empirical research focused on studying the XP methodology instead of the Scrum methodology [23]. This dichotomy between research effort and industry practice results in ASD training providers lacking important insight into the *Scrum* ASD process for effectively training prospective software engineers.

In our experiment of 2014, we attempt to bridge this research gap with a 12 week field study involving 125 second year undergraduate software engineering students from March to June 2014. The students were new to the *Scrum* methodology and self-organized into 21 *Scrum* teams of 5 to 7 persons each. This study provides a unique opportunity to study the impact of the ASD methodology used in *Scrum* teams. Students in this study carried out activities at various stages of the *Scrum* methodology in our online agile project management (APM) tool - the HASE platform. The activities for each team member supported by HASE mainly occur during the sprint planning and sprint review/retrospective phases. They include proposing tasks, estimating the priority, difficulty and time required for each task, deciding how to allocate tasks, collaboration information, reviewing the timeliness and quality of completed tasks, and providing feedback about each team member’s mood at different points in time during a sprint. During the study, students logged 10,779 ASD activities in the HASE platform.

In our previous round of study in 2013 [213], this new form of ASD activity data has been shown to produce insights into the ASD process traditional survey/interview based approaches are unable to achieve. The results from this study offer additional insights into ASD task allocation decision-making, collaboration, and team morale which, to the best of our knowledge, have not been reported by published research study before. Specifically, the results point towards strong positive correlations between a student's technical productivity and the amount of workload allocated to him/her. In addition, contrary to popular preconceptions, collaboration among student team members who are new to *Scrum* has shown negative correlation with team performance and team morale.

## 6.2 Related Work

In [211], the authors present the results of a systematic literature review concerning agile pair programming effectiveness. The paper analyzed compatibility human factors, such as the feel-good, personality, and skill level factors, and their effect on pair programming effectiveness as a pedagogical tool in *Computer Science* and *Software Engineering* education. Four metrics were used in the analysis: 1) academic performance, 2) technical productivity, 3) program/design quality and 4) learning satisfaction. The general findings are that pair programming is more effective in terms of technical productivity, learning satisfaction and academic performance, while not significantly different in terms of program quality as compared to solo programming. While our study also looks into similar metrics, our data are in the form of ASD activity logs. In addition, our study focuses on the *Scrum* methodology instead of XP.

A number of studies about the *Scrum* ASD processes start to emerge in recent years. In [214], the authors investigate decisions related to software designs. They employ content analysis and explanation building to extract qualitative and quantitative results from interviews with 25 software designers. The study finds that the structure of the design problem determines the designers' choice between rational and naturalistic decision making.

The study in [215] focuses on decision-making by Scrum teams. It examines decisions made across the four stages of the sprint cycle: sprint planning, sprint execution, sprint review and sprint retrospective. The authors employ the technique of focus group study with 43 *Agile* developers and managers to determine what decisions are made at different points of the sprint cycle. In another publication by the same research group [216], interviews with an additional 18 professional Agile practitioners from one global consulting organization, one multinational communications company, two multinational software development companies, and one large museum organization are analyze to identify six key obstacles facing decision making in Agile teams.

Nevertheless, the techniques used by existing studies mainly involve interviews and surveys. This limited the scale of the study as well as the level of details of the collected data. As a result, the form of findings from such studies tends to be qualitative in nature. For instance, in [215], some obstacles facing agile teams during sprint decision-making can "people are unwilling to commit to a decision", "lack of ownership" and "lack of empowerment". There is a lack of quantitative results indicating the extent of each obstacle facing agile team members with different competence levels.

## 6.3 Study Design

In this section, we present our research approach, the metrics that have been measured, and key characteristics of the student population involved in the study.

### *6.3.1 Research Approach*

Our goal is to investigate the aspects of decision-making, collaboration, and team morale in the *Scrum* ASD process as practiced by student developers who are new to *Scrum* in the natural settings where these activities occur. Therefore, students in this study perform *Scrum* ASD activities in our HASE online APM platform. The platform provides five main features to support agile project management which cover the sprint planning and sprint review/retrospective phases:

1) *Registration*: In order to build user profiles, HASE requires registrants to specify their self-assessed competence levels in different areas of expertise such as familiarity with specific programming languages, system de-sign methodologies, and user interface (UI) design tools, etc.
2) *Team and Role Management*: HASE supports the creation of teams, the selection of product owners and stakeholders into the teams, and the assignment of different roles within a team (e.g., programmers and UI designers).
3) *Task Management*: Task information including task description, skills required for the task, and the person who proposed each task is displayed for all team members to view. The difficulty value of each task, is recorded using an 11-point *Likert* scale [169, 170] (with 0 denoting “extremely easy” and 10 denoting “extremely hard”). Each team member can input his/her estimated difficulty value for each task into the HASE platform. The HASE platform then uses the average difficulty value for the task ($D_{\tau}$).

The students were asked to take into account the technical challenge as well as the amount of effort required when judging the difficulty of a task. The priority value of each task $\tau$, is also recorded using an 11-point Likert scale (with 0 denoting "extremely low priority" and 10 denoting "extremely high priority"). Each team member can input his/her estimated priority value for each task into the HASE platform. The HASE platform then uses the average priority value for the task ($\mathrm{Prio}_{\tau}$).

4) *Sprint Planning*: HASE records the teams' decisions on which tasks are assigned to which team member during each sprint. Once assigned, the status of the task becomes "Assigned". The assignee $i$ inputs his/her confidence value ($\mathrm{Conf}_{\tau}^{i}$) for each task $\tau$ on an 11-point Likert scale (with 0 denoting "not confident at all" and 10 denoting "extremely confident"). Each team member also inputs the estimated required time to complete each task (in number of days). The HASE platform uses the average estimated time required to generate the deadline for the task ($\mathrm{T}_{\tau}^{est}$). Apart from a primary assignee, multiple students can collaboratively work on a task. The collaborator information for each task is also recorded by HASE.

5) *Sprint Review/Retrospective*: Once a task is completed, the assignee changes its status in the HASE platform to "Completed". This action will trigger HASE to record the actual number of days ($\mathrm{T}_{\tau}^{act}$) used to complete this task. HASE also provides functions for team members to peer review the quality ($\mathrm{Qual}_{\tau}$) of each completed task $\tau$. The quality of a completed task is recorded in the platform using a 11-point Likert scale with 0 representing ("extremely low quality") and 10 representing ("extremely high quality"). The average quality rating for each task is used by HASE as the final quality rating for that task.

6) *Team Morale Monitoring*: During the sprint planning meeting, team members can report their current mood values into the HASE platform. A person i's mood at the beginning of a sprint $t$ ($\mathrm{m}_i^{begin}(t)$) is represented on a 5-point Likert scale with 1 representing "very low" and 5 representing "very high". During the sprint review/retrospective meeting, each task assignee i can report his/her mood after completing a task at the end of sprint $t$ ($\mathrm{m}_i^{end}(t)$) using the same 5-point *Likert* scale.

The input data to the HASE platform required from ASD teams are as a result of students' activities following the Scrum methodology. In this way, users of HASE can behave as if they are using any APM tool without expending additional effort to help with data collection. Thus, the data collection process remains unobtrusive to the participants.

A total of 125 second year undergraduate software engineering students from the *Beihang University, China* were involved in this study. The students need to form into *Scrum* teams of 5 to 7 persons each to complete a team-based software engineering project over a 12 week period of time. As this is part of the students' course work, the curriculum dictates that the students must decide among themselves how to form into teams. Eventually, the students formed into 21 teams. Each team then proposes a software engineering project for the course instructor to approve. The projects are mediated by the instructor so that they are of comparable scale and complexity across all teams. Some examples of the proposed software projects in this study are "An android system for interest-based music recommendation", "A mobile health information app for the elderly", "A mobile app for monitoring user mobility pattern", etc. The teams then adopt the Scrum process to develop their projects. Each sprint lasts a week. During the sprint planning meeting, team members propose the tasks that need to be completed over this sprint. A total of 893 tasks have been proposed by all teams during this study. Students perform all Scrum ASD activities using the HASE platform.

### *6.3.2 Metrics*

As we mentioned in section 1.3, in this study, we adopt the exploratory data analysis (EDA) approach to analyze the data collected. We use the following metrics to facilitate our analysis:

1) *Technical Productivity* ($\mu_i$): it refers to the average amount of workload a student i can complete during a sprint. In this study, we use the task difficulty value as an indicator of the workload of a task as the task difficulty values reported by students denote both the technical challenge and the amount of effort required to complete the task.

2) *Competence* ($\mathrm{Comp}_i$): it refers to the probability a student $i$ can complete a task assigned to him/her with satisfactory quality before the stipulated deadline. In this work, the outcome of a task needs to achieve an average quality rating higher than 5/10 in order to be considered as having satisfactory quality. This metric is similar to a student's reputation. Thus, we adopt a reputation computation model - the Beta Reputation model [217] - which is widely used in the fields of artificial intelligence and network communications [218, 219]. It is calculated as follows:

$$\mathrm{Comp}_i = \frac{\alpha_i + 1}{(\alpha_i + 1) + (\beta_i + 1)} \in (0,1) \tag{6.1}$$

where $\alpha_i$ and $\beta_i$ are calculated as:

$$\alpha_i = \sum_{\tau \in \emptyset(\mathrm{i})} 1_{[\mathrm{T}_\tau^{act} - \mathrm{T}_\tau^{est} \leq 0 \ and \ \mathrm{Qual}_\tau > 5]} \mathrm{D}_\tau \tag{6.2}$$

$$\beta_i = \sum_{\tau \in \emptyset(\mathrm{i})} 1_{[\mathrm{T}_\tau^{act} - \mathrm{T}_\tau^{est} > 0 \ or \ \mathrm{Qual}_\tau \leq 5]} \mathrm{D}_\tau \tag{6.3}$$

The function $1_{[\text{condition}]}$ in Eq. (6.2) and Eq. (6.3) equals to 1 if "condition" is true. Otherwise, $1_{[\text{condition}]}$ equals to 0. $\emptyset(i)$ denotes the set of tasks i has previously worked on until the current point in time. The "+1" terms in the numerator and denominator of Eq. (6.1) are *Laplace smoothing terms* [220] which ensure that if i has no previous track record, $\text{Comp}_i$ evaluates to 0.5 indicating maximum uncertainty about i's performance.

3) *Workload* ($w_i\ (t)$): it refers to the actual amount of workload assigned to a student *i* during sprint t.
4) *Final Score* ($s_i^f$): it refers to the final score a student *i* achieves for this course. It ranges from 0 to 100 marks.
5) *Team Score* ($s_j$ ): it refers to the score given to a team *j* by the course instructor based on the assessment of the software produced by the team at the end of the course work project. It ranges from 0 to 30 marks.
6) *Collaborators/Task* ($c_\tau$ ): it refers to the number of students working on a same task.
7) *Team Morale* (Begin) ($M_j^{begin}(t)$): it refers to the average of the mood values reported by members of team j during the sprint planning meeting of sprint *t*.
8) *Team Morale* (End) ($M_j^{end}(t)$): it refers to the average of the mood values reported by members of team *j* during the sprint review/retrospective meeting of sprint *t*.

### *6.3.3 Subject Characteristics*

Before the commencement of their course work projects, students involved in our study did not have any experience practicing ASD methodologies. Nevertheless, they have received standard training in software engineering concepts in their first year of undergraduate study. The distribution of the students' competence and technical productivity is shown in Figure 38.

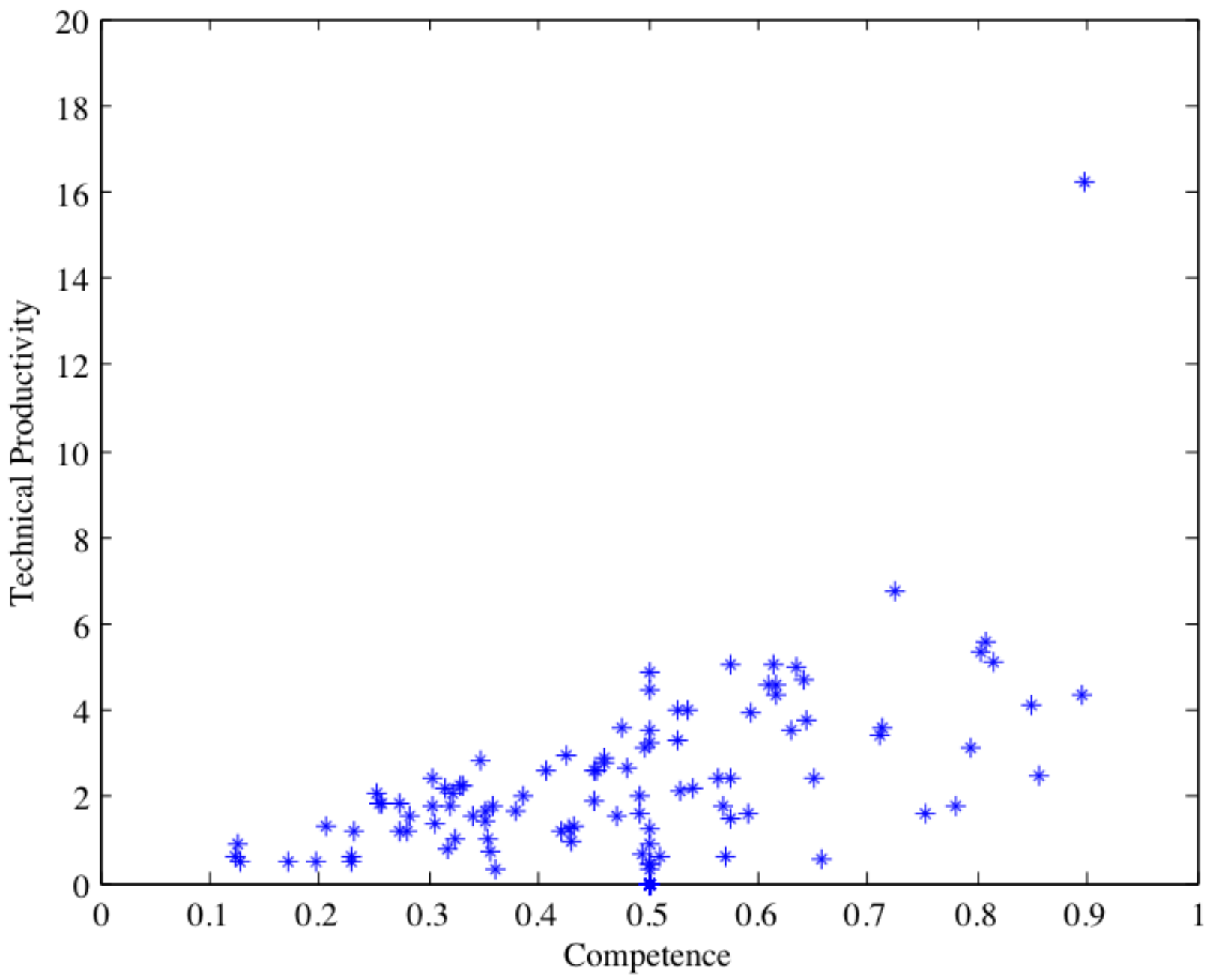

Figure 39. The distribution of students' capabilities in the study

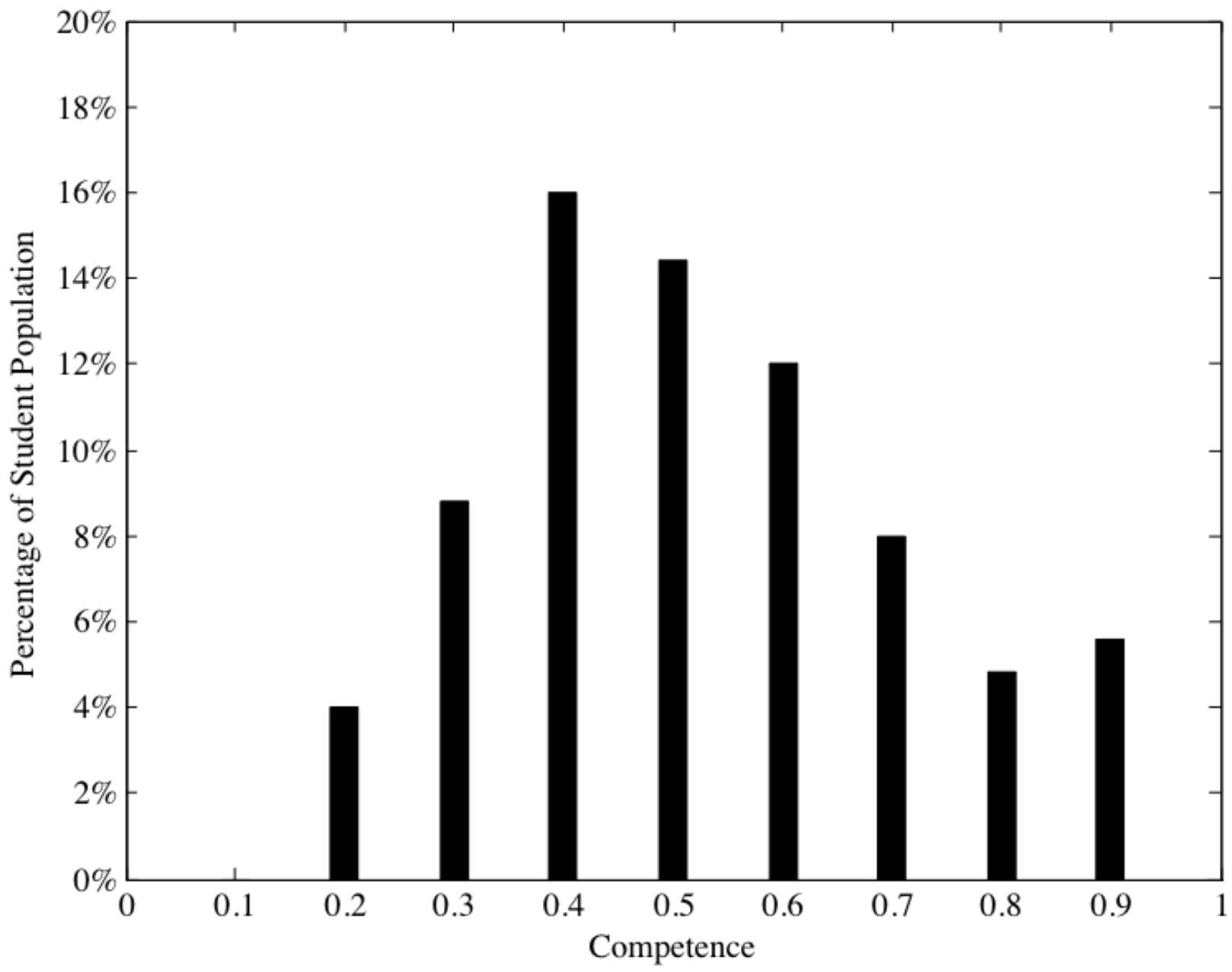

Figure 40. The distribution of students' competence

The *Pearson Correlation Coefficient* (PCC) [221] between students' competence and their respective technical productivity values is ($r = 0.7443$, $p < 0.01$), indicating a statistically significant positive correlation. This means students involved in our study who often complete the tasks assigned to them on time also tend to do so with high quality, and vice versa.

As illustrated in Figure 39, the percentages of student population showing various competence levels roughly follow a bell curve centered around 0.4 to 0.5. The distribution of the students' final scores, which include their examination test scores together with their course work project scores, is shown in Figure 40.

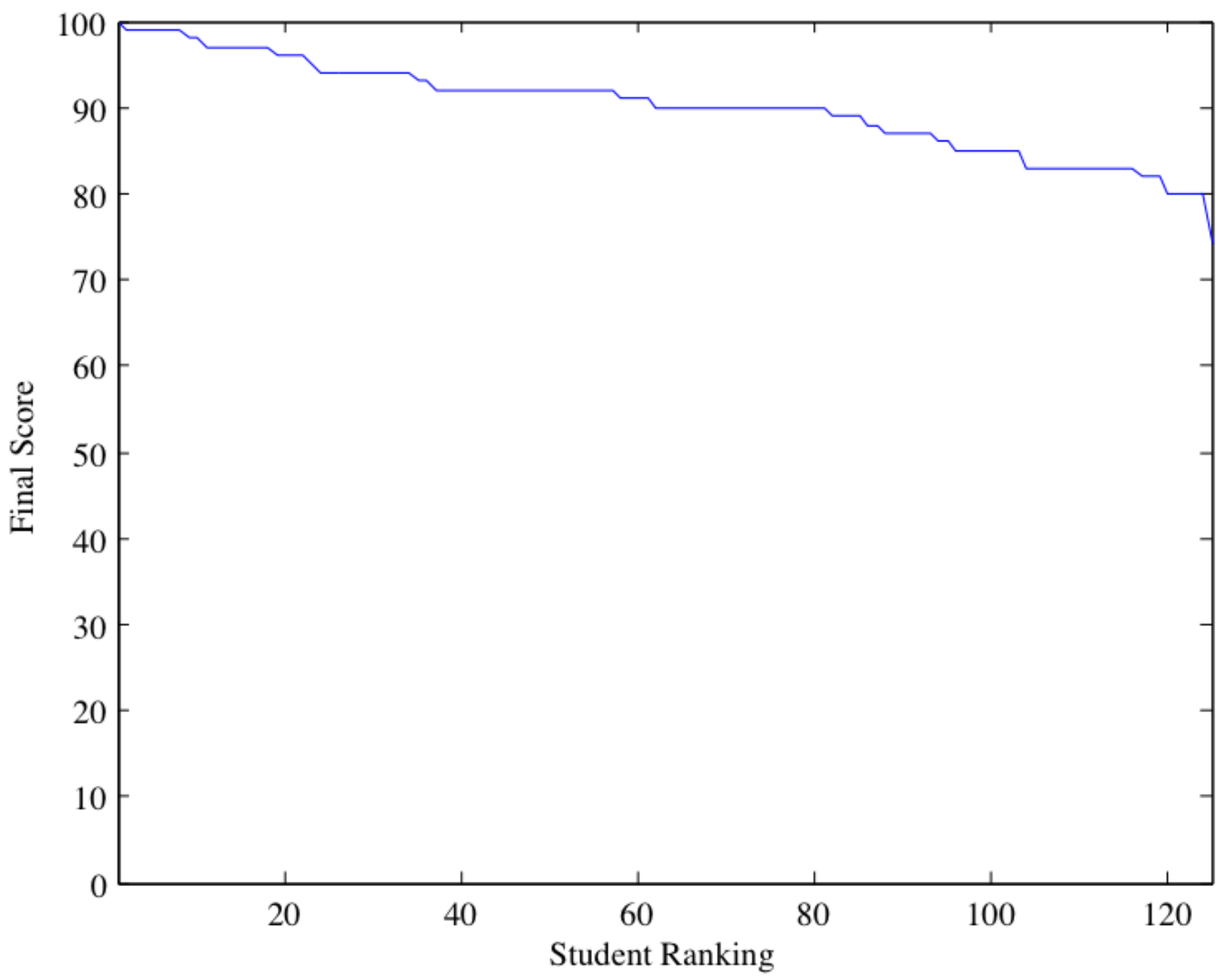

Figure 41. The distribution of students' final scores

The final scores range from 74 to 100 marks with the majority of students scoring between 80 and 90 marks. As can be observed in Figure 41, it turns out that the distribution of the teams' competence and technical productivity is similar to the distribution of individual students' competence and technical productivity in Figure 38. As the students decide among themselves on which team they wish to join, we have no control of team characteristics in this study.

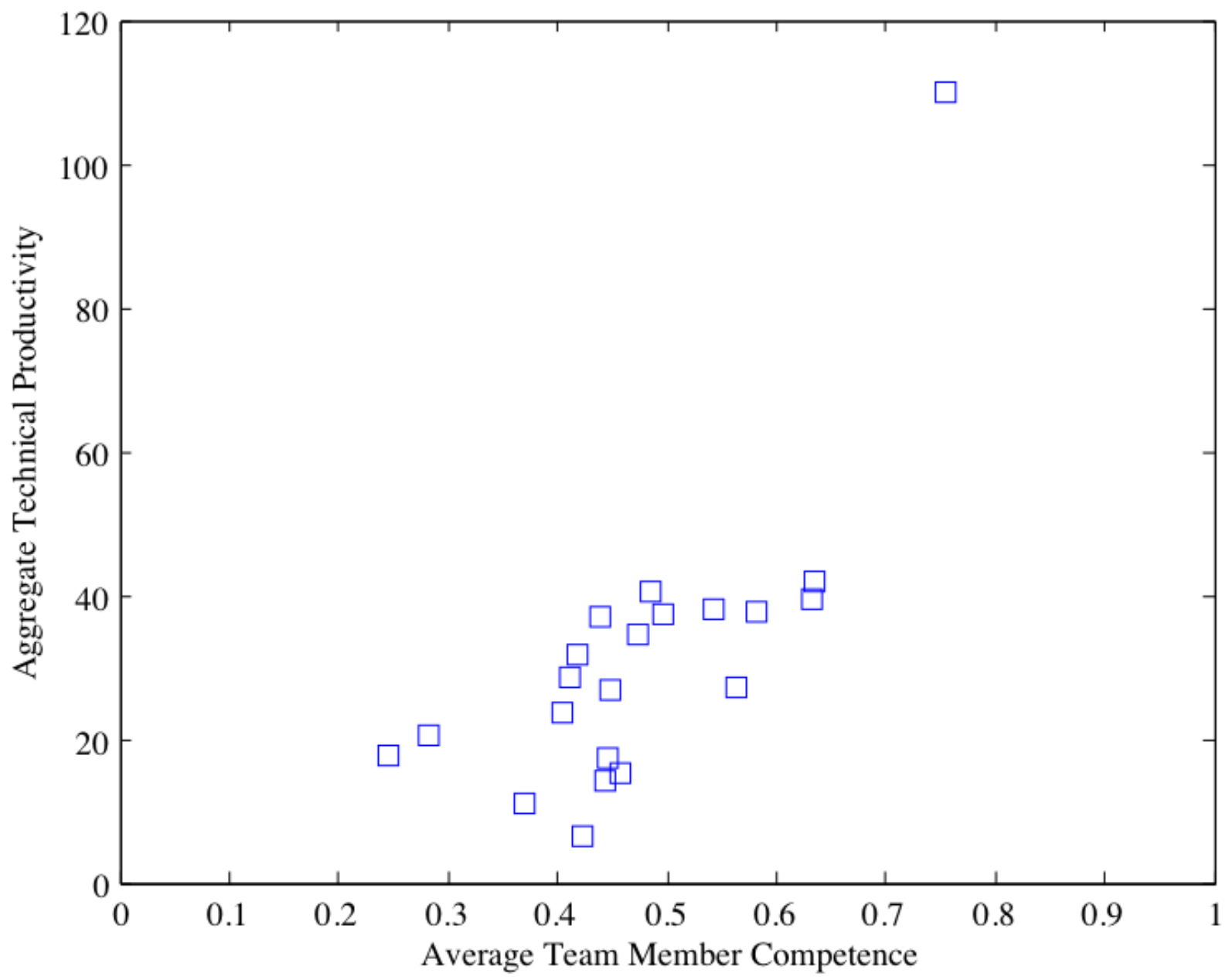

Figure 42. The distribution of teams' capabilities in the study

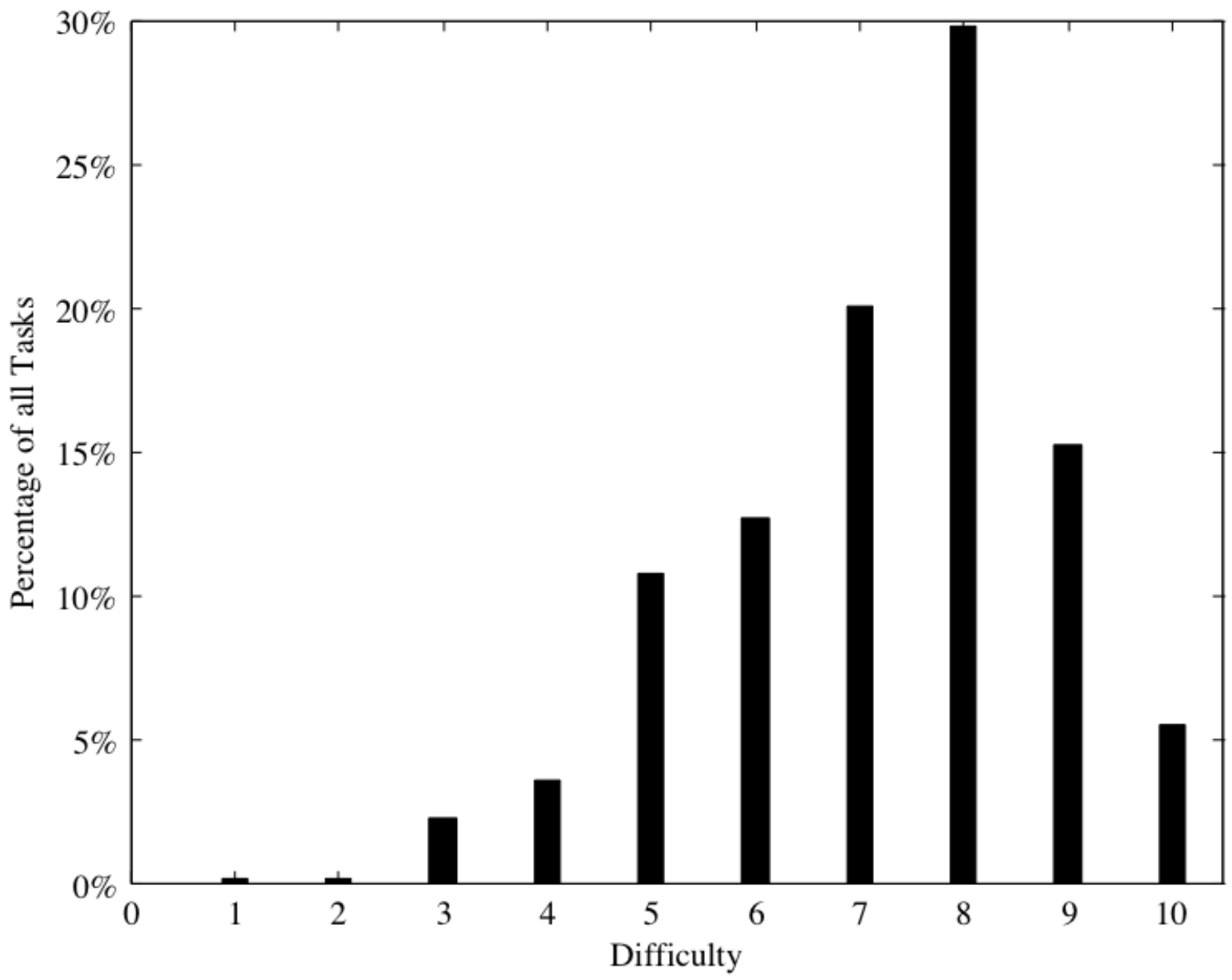

Figure 43. The distribution of task difficulty values

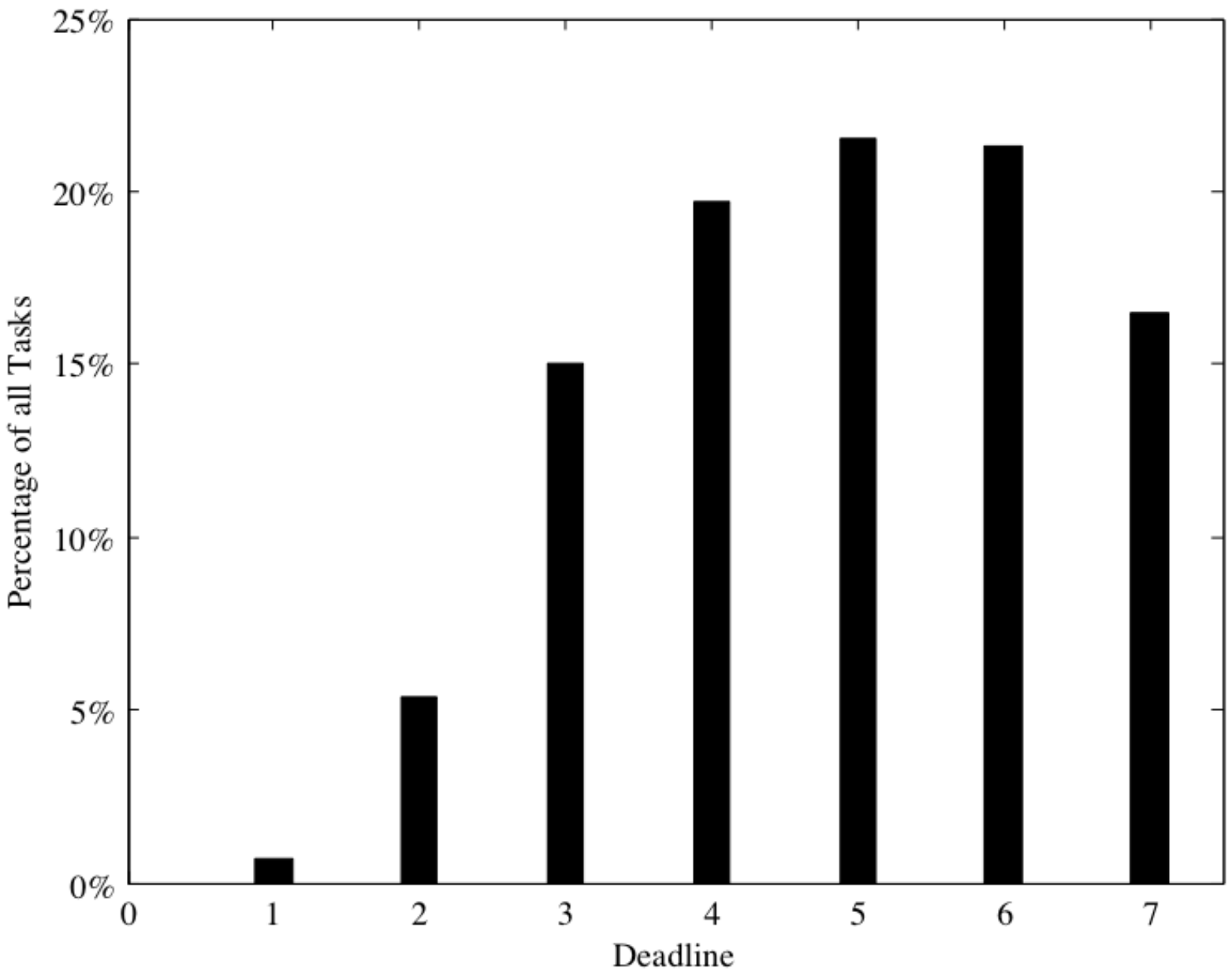

Figure 44. The distribution of task deadlines

Figure 42 and Figure 43 show the distributions of the difficulty values and the deadlines of the 893 tasks proposed by students during this study. The task deadline values range from 1 to 7 days. The task difficulty and deadline roughly follow bell curves centered around 8 and 5, respectively. The PCC between task difficulty and deadline is ($r = 0.4086$, $p < 0.01$), indicating a statistically significant positive correlation.

## 6.4 Results and Analysis

In this section, we present the results from preliminary analysis of the data collected from this study. We focus on three human factor related aspects of the *Scrum* methodology which are important to understanding the agile process and have not been well studied by previous research. They are: 1) task allocation decision-making, 2) collaboration, and 3) team morale.

### *6.4.1 Task Allocation*

Decision-making in [213], the ratio between participants' competence and the normalized difficulty values of the tasks assigned to them has been shown to positively correlate to the timeliness of task completion. In the data collected, no task was rated by participants as having a difficulty value of 0. Thus, the ratio $\frac{\text{Comp}_i}{\text{D}_\tau} \in (0,10)$. If $\frac{\text{Comp}_i}{\text{D}_\tau} > 1$, it means that a student $i$ is assigned a task $\tau$ with a normalized difficulty value lower than $i$'s competence value. If $\frac{\text{Comp}_i}{\text{D}_\tau} \leq 1$, $i$ is assigned a task with a normalized difficulty value higher than or equal to $i$'s competence value. In this study, we investigate whether students take the $\frac{\text{Comp}_i}{\text{D}_\tau}$ ratio into account when allocating tasks among themselves.

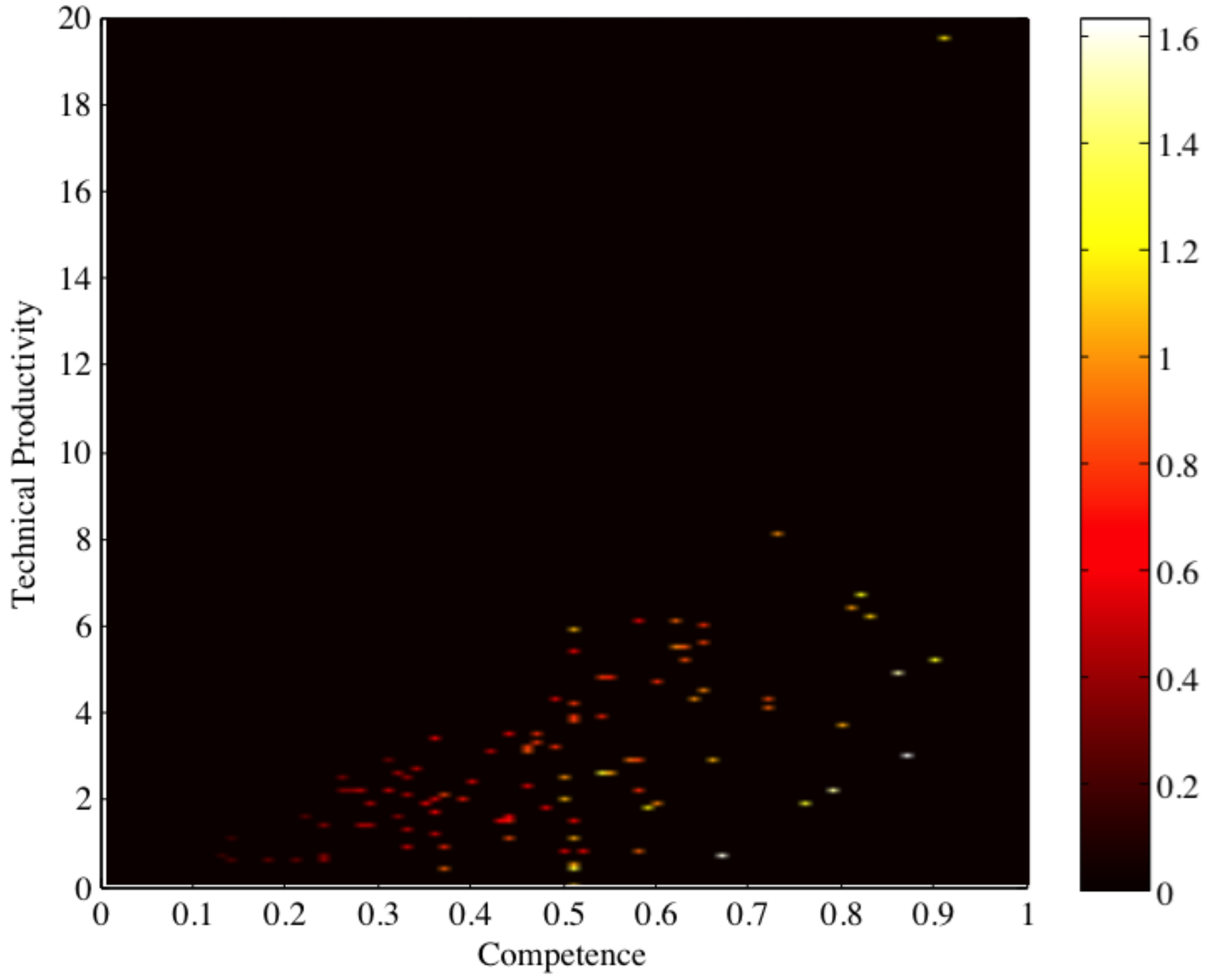

Figure 45. The $\text{Comp}_i/\text{D}_t$ distribution among students in the study

Figure 44 illustrates the average $\frac{\text{Comp}_i}{\text{D}_\tau}$ ratios for all students against their competence and technical productivity. The colour scale corresponds to different $\frac{\text{Comp}_i}{\text{D}_\tau}$ ratios. It can be observed that students showing higher competence values tend to be assigned tasks with normalized difficulty values lower than their competence values. The PCC between students' competence values and their $\frac{\text{Comp}_i}{\text{D}_\tau}$ ratios in this study is ($r = 0.6567$, $p < 0.01$), indicating a statistically significant positive correlation. The PCC between students' technical productivity values and their $\frac{\text{Comp}_i}{\text{D}_\tau}$ ratios in this study is ($r = 0.2992$, $p < 0.01$), indicating a statistically significant, albeit weak, positive correlation. Thus, from this study, it appears students indeed attempt to allocate tasks within the assignees' competence when following the Scrum methodology.

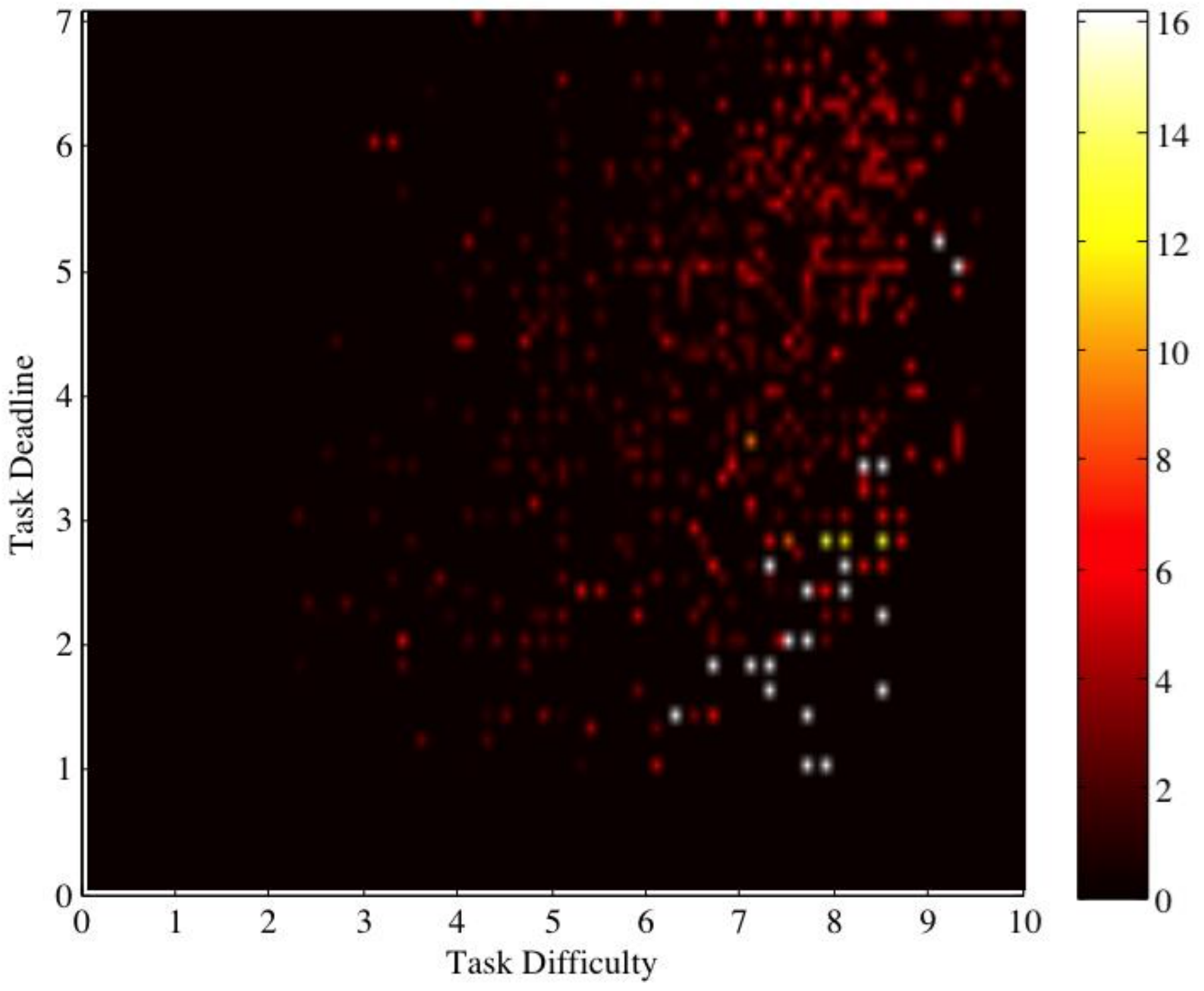

Figure 46. Task allocation vs. students' technical productivity

Apart from the $\frac{\text{Comp}_i}{\text{D}_\tau}$ ratio, students' technical productivity also plays an important part in task allocation decision-making. Figure 45 shows the assignees' technical productivity each task is allocated to. The colour scale corresponds to different technical productivity values. It can be observed that students with high technical productivity are generally allocated more difficult tasks. Tasks with high difficulty values and short deadlines tend to be allocated to students with high technical productivity. The PCC between the $\frac{\text{D}_\tau}{\text{T}_\tau^{est}}$ ratio of the tasks and the technical productivity values of the students assigned the tasks is ($r = 0.4397$, $p < 0.01$), indicating a statistically significant positive correlation.

### *6.4.2 Collaboration*

In this part of the study, we investigate two research questions: 1) what is the relationship between collaboration and team characteristics? and 2) what is the relationship between collaboration and team performance?

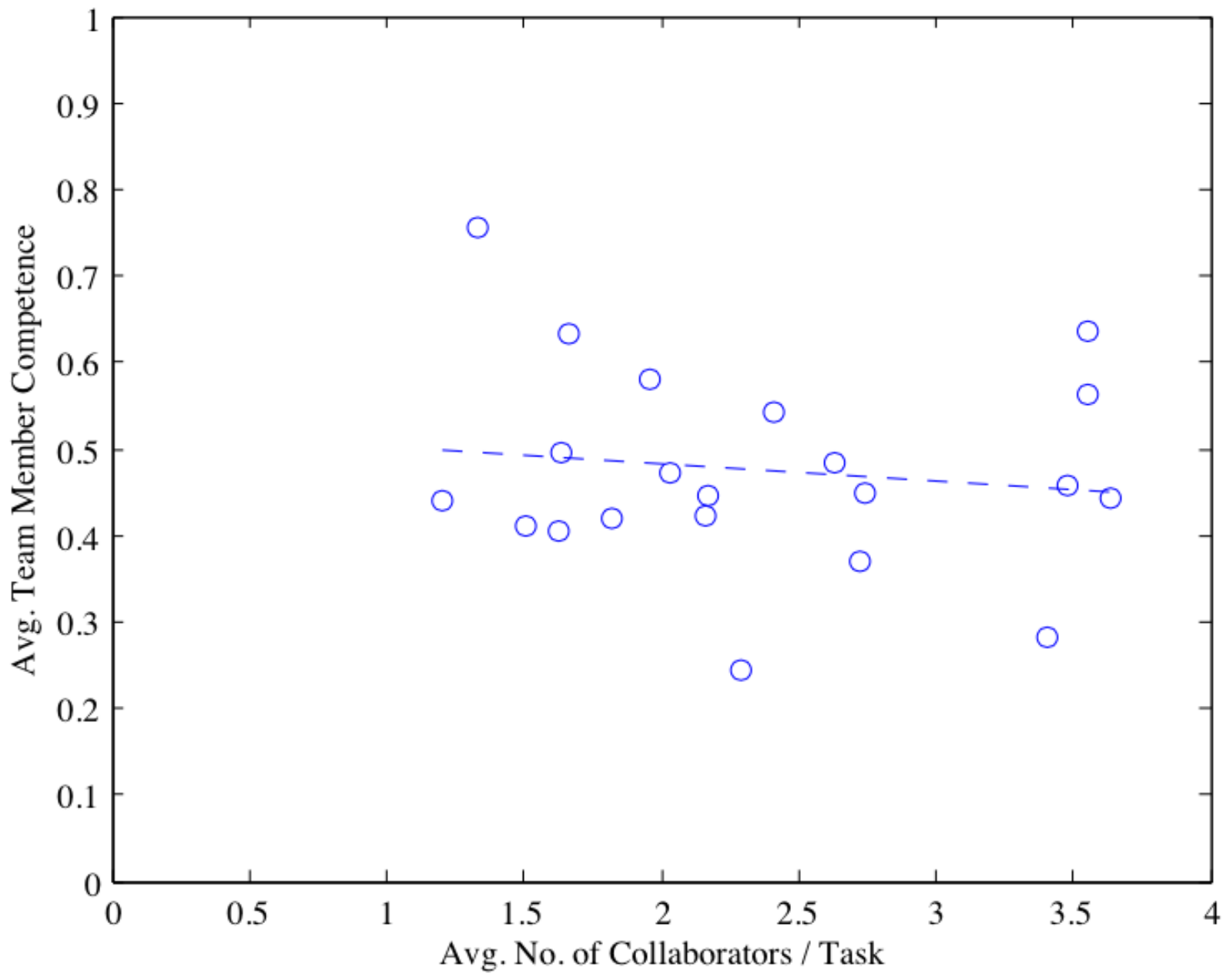

Figure 47. Collaboration vs. team members' competence

With regard to the first research question, we look into the relationship between each team's capabilities and their collaboration behaviors. Figure 46 shows the relationship between the average team competence (which is calculated by averaging team members' competence values) and the average number of collaborators per task in each team. The PCC between the average team competence and the average number of collaborators per task is (r = -0.1376, p = 0.5537), which is not statistically significant. This result favours the null hypothesis that there is no correlation between these two factors.

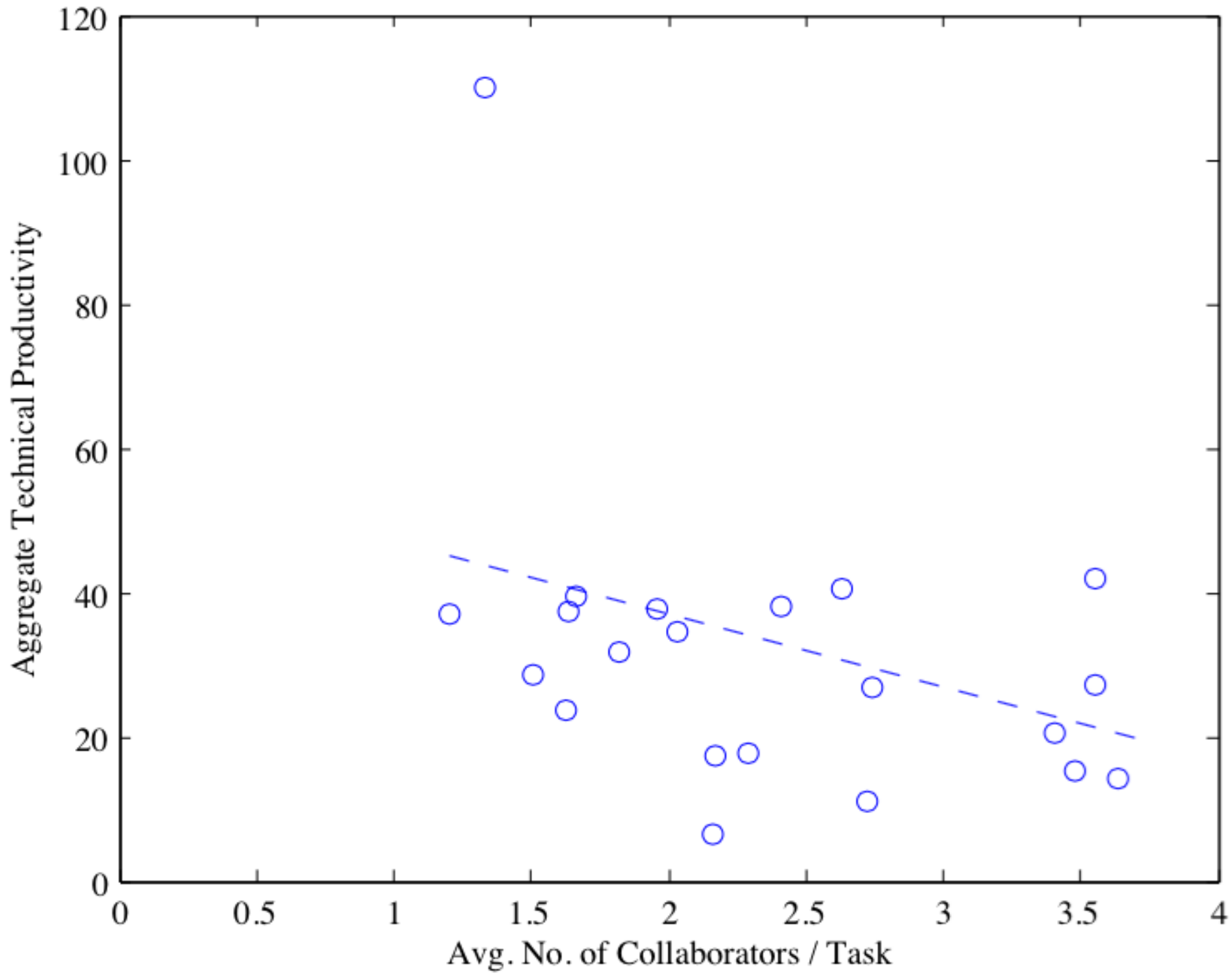

Figure 48. Collaboration vs. team members' task processing capacity

Figure 47 shows the relationship between the aggregate team technical productivity (which is calculated by summing team members' technical productivity values) and the average number of collaborators per task in each team. The PCC between the aggregate team technical productivity and the average number of collaborators per task is (r = -0.4064, p < 0.1), indicating a statistically significant negative correlation. Thus, it appears students in teams with low aggregate team technical productivity values tend to engage in collaborations more often, which is a rational strategy from the students' perspective.

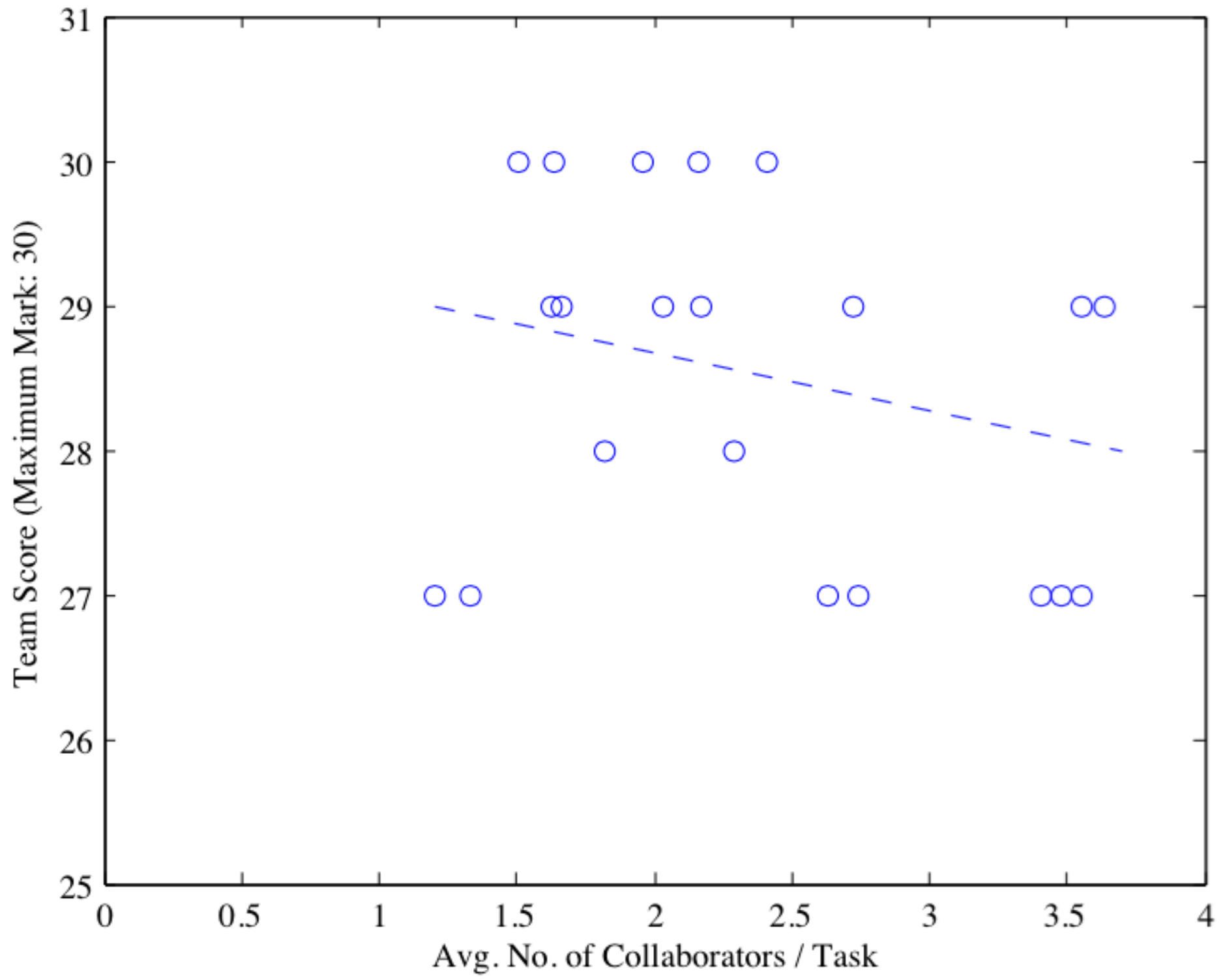

Figure 49. Collaboration vs. team score

In this study, we use the team score to measure the performance of a team. Figure 48 illustrates the relationship between collaboration and team score. The PCC between the team scores and the average number of collaborators per task is ($r = -0.3721$, $p < 0.1$), indicating a statistically significant negative correlation. This result contradicts the popular preconception that collaboration improves team performance. Therefore, we conducted an interview with the course instructor to obtain his opinions on the possible reasons for such a negative correlation. The course instructor suggested that due to the lack of team-based project development experience among the students as well as possible differences in terms of competence and technical productivity among team members, collaboration might not have occurred in an effective manner. Another reason for this result can be the incompatibility between collaboration and modularity in software development. In another study from [222], it has been shown that groups with small file-wise collaborative editing ratio tend to score higher grades for the software developed.

### *6.4.3 Team Morale*

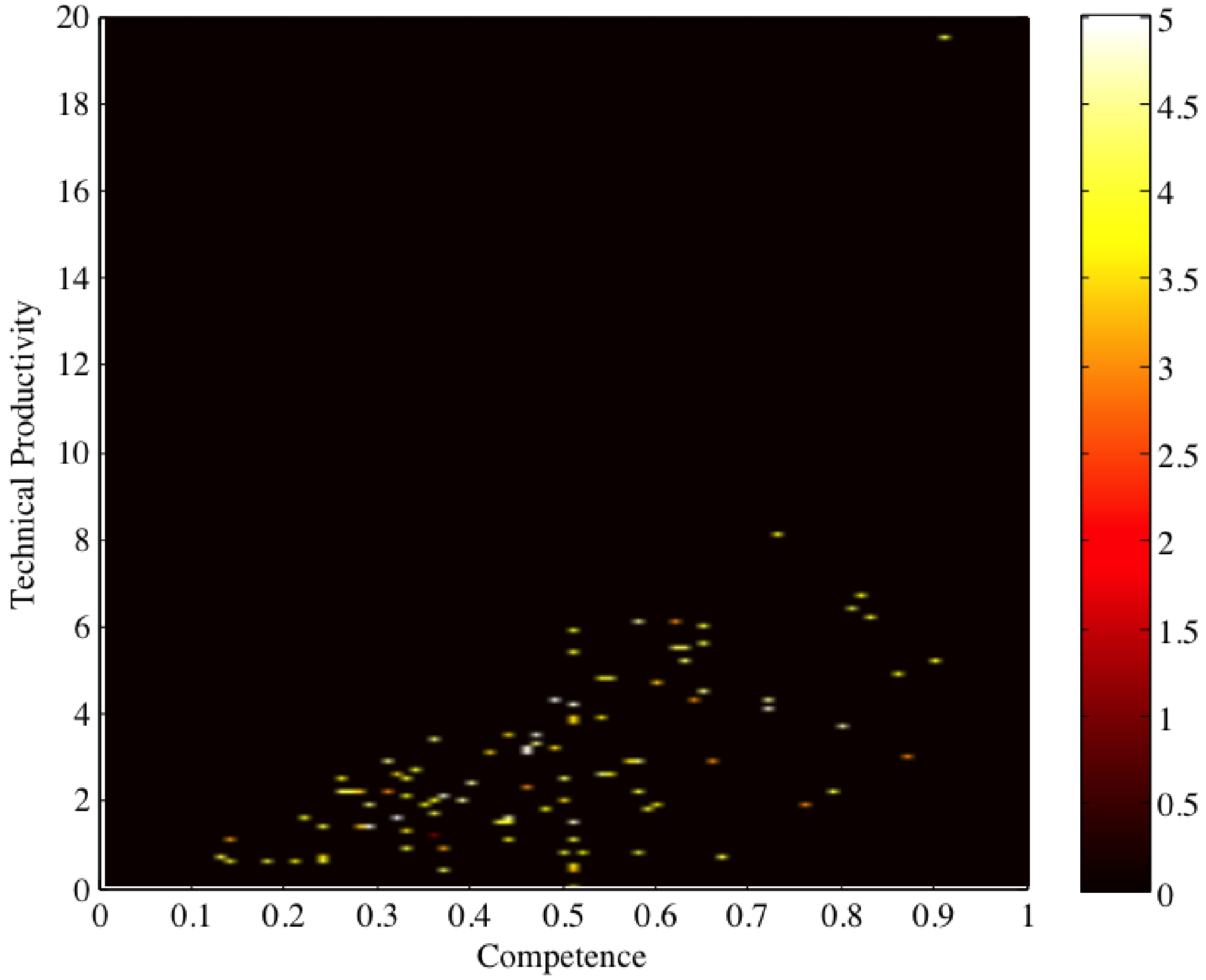

Figure 50. Students' average morale before a Sprint

Figure 49 shows the distribution of students' average self-reported mood values during the sprint planning meeting at the start of each sprint. The colour scale represents the average self-reported mood values. The average mood value is 3.86 out of 5. The PCC between students' mood during the sprint planning meetings and their competence values is (r = - 0.0025, p = 0.9394), indicating no statistically significant correlation. The PCC between students' mood during the sprint planning meetings and their technical productivity values is (r = 0.1505, p < 0.01), indicating a statistically significant albeit weak positive correlation.

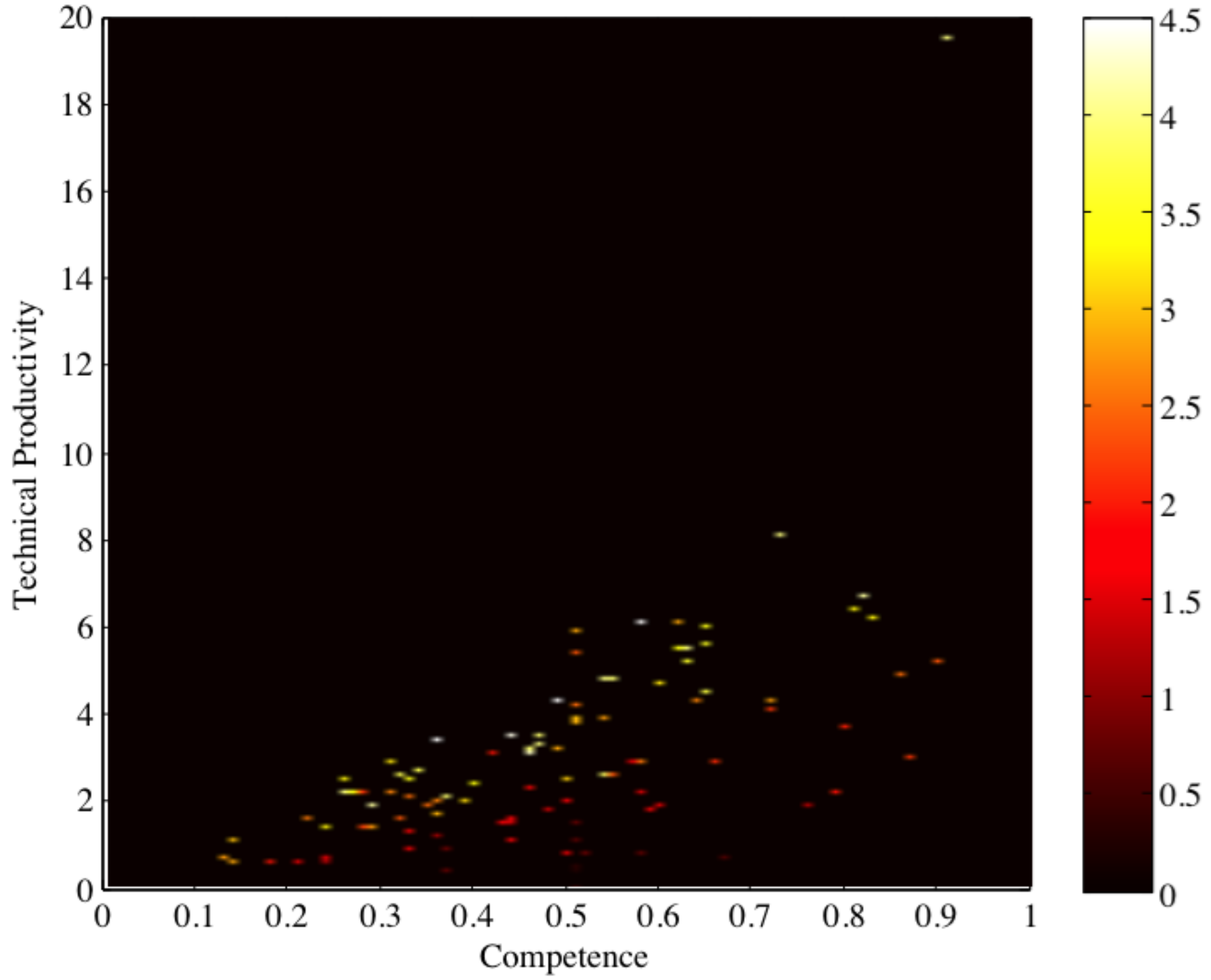

Figure 51. Students' average morale after a Sprint

Figure 50 shows the distribution of students' average self-reported mood values during the sprint review/retrospective meeting at the end of each sprint. The colour scale represents the average self-reported mood values. The average mood value is 3.80 out of 5 which is slightly lower than at the beginning of the sprint. The PCC between students' mood during the sprint review/retrospective meetings and their competence values is (r = -0.0148, p = 0.5946), indicating no statistically significant correlation. The PCC between students' mood during the sprint review/retrospective meetings and their technical productivity values is (r = 0.4207, p <

0.01), indicating a statistically significant positive correlation. Therefore, based on these analyses, team members with high technical productivity tend to have high morale, especially at the end of a sprint after completing the tasks allocated to them. In addition to investigating the relationship between students' capabilities and their morale, we also investigate the relationship between collaboration and team morale. The morale value of a team is the average of the mood values reported by its members over the 12 week period.

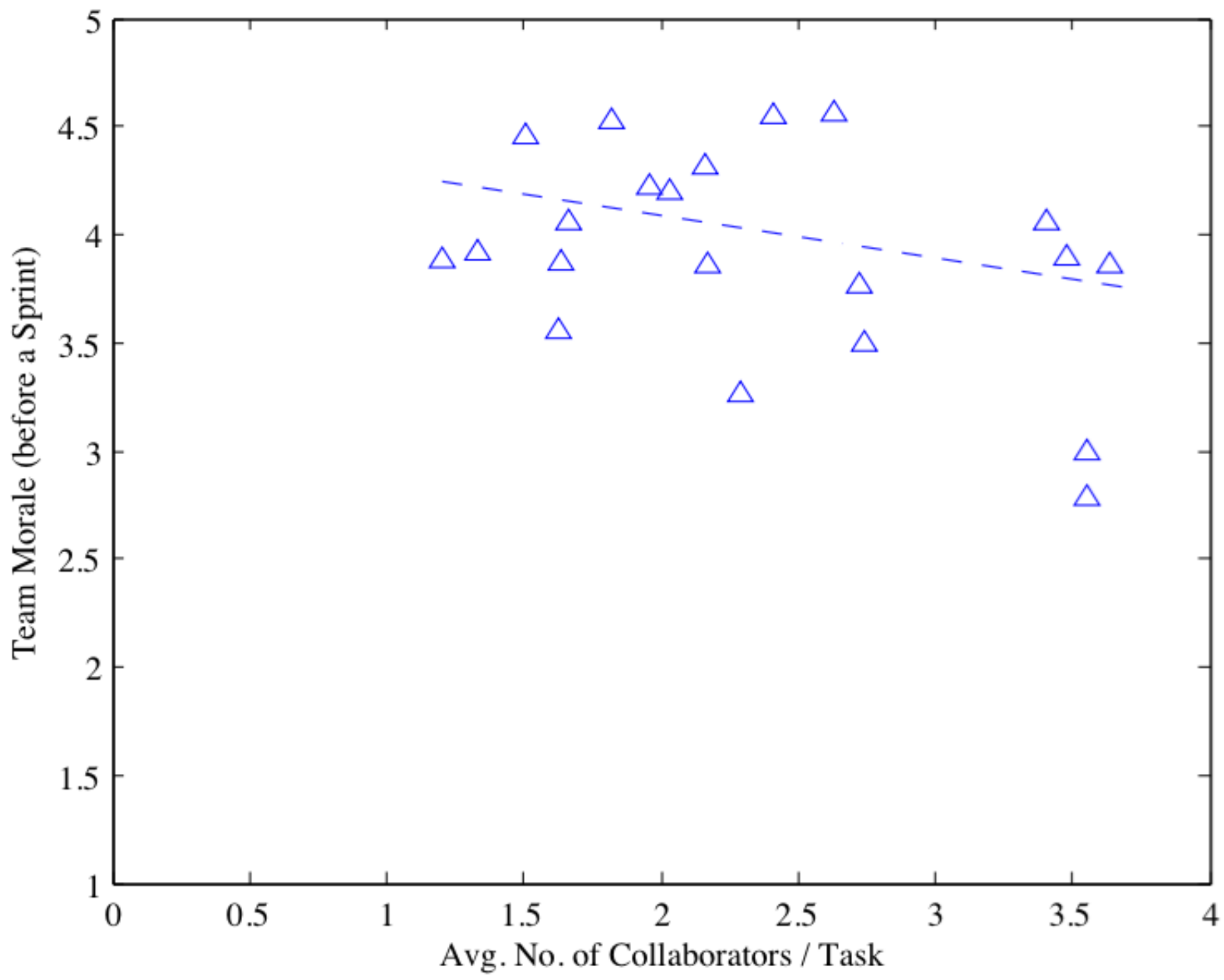

Figure 52. Collaboration vs. average team morale before a Sprint

Figure 51 shows the relationship between the team morale values during the sprint planning meetings and the average number of collaborators per task in each team. The PCC between team morale during the sprint planning meetings and the average number of collaborators per task is ($r = -0.4135$, $p < 0.1$), indicating a statistically significant negative correlation.

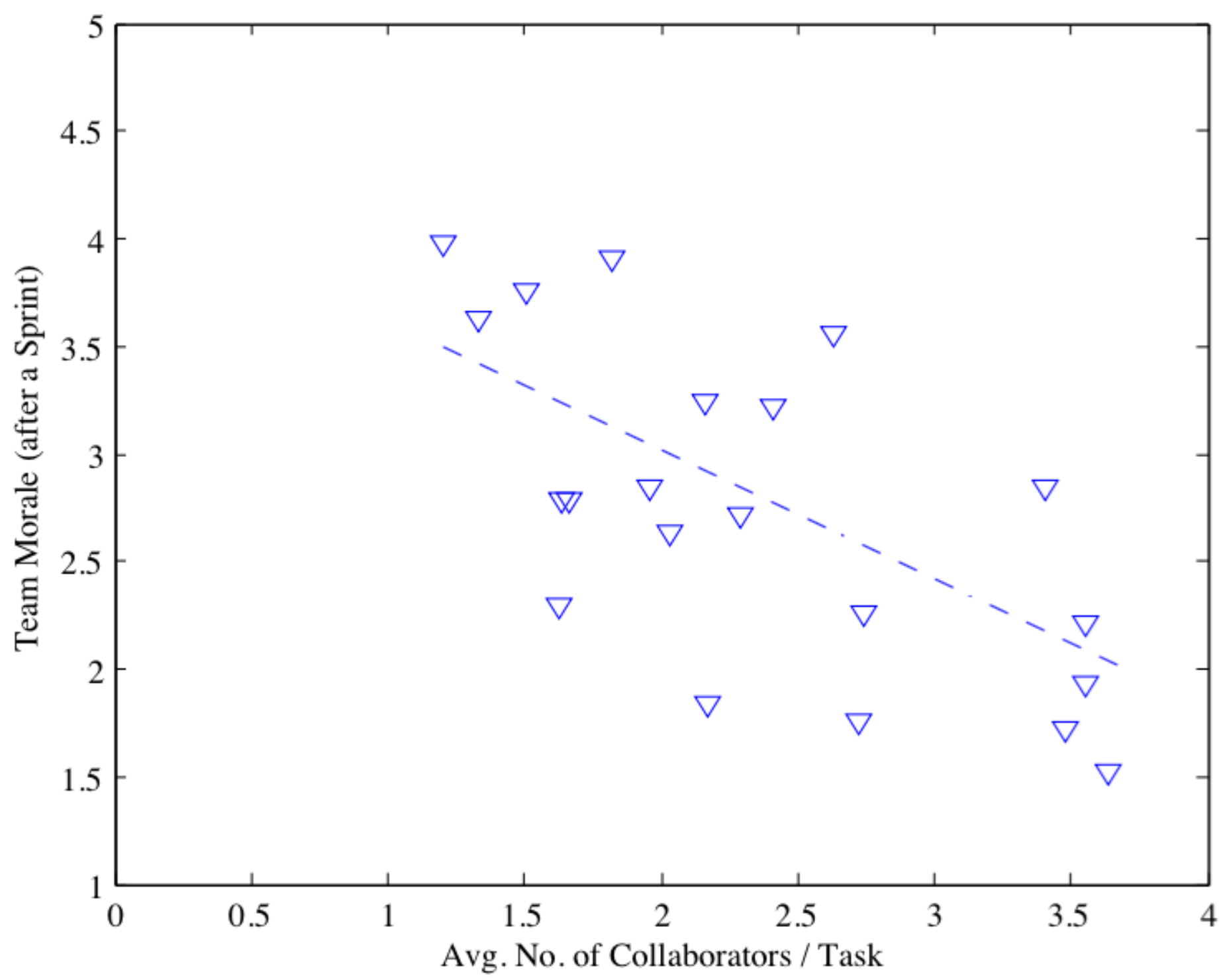

Figure 53. Collaboration vs. average team morale after a Sprint

Figure 52 shows the relationship between the team morale values during the sprint review/retrospective meetings and the average number of collaborators per task in each team. The PCC between team morale during the sprint review/retrospective meetings and the average number of collaborators per task is ($r = -0.6632$, $p < 0.01$), indicating a statistically significant negative correlation. Therefore, the results suggest that collaboration by team members who are inexperienced in software development in Scrum teams negatively affects team morale.

In summary, the key findings about the Scrum-based ASD process practiced by novice teams from this study are: 1) task allocation in agile teams positively correlate to students technical productivity; 2) collaboration is negatively correlated with team technical productivity, team morale, and team score; 3) team morale is positively correlated to their technical productivity.

## 6.5 Study Limitations

This section discusses the study limitations based on three categories of threats to validity described in [223]. For each category, we list all possible threats, measures taken to reduce the risks, and suggestions for improvements in future studies.

### *6.5.1 Internal Validity*

Internal validity threats that may have affected our study are the lack of control of the following variables: 1) the students competence (other than all being in the same semester of the course); and 2) how the teams are formed (students decide on their own which teams they wish to join, the instructor only controls the sizes of the teams). With respect to Threat 1, the repeated *Scrum* activity data logging over a 12 week period of time decreases the probability of this threat affecting our outcomes to some extent as students are provided with many opportunities to demonstrate their competence on different tasks. Threat 2 has affected the study. In regard to it, we believe the sample size, which was not small (125 students performing close to 900 tasks over 12 weeks), reduces the extent of the effect of this threat (i.e., different types of pairings with respect to student competence and task difficulty have occurred). In future studies, we suggest that pre-testing should be organized to assess the competence of the students. The results can be used to guide instructors in organizing the students into teams.

### *6.5.2 External Validity*

A factor that might reduce the external validity of our study is the use of students as subjects. Nevertheless, according to [224, 225], students can play an important role in experimentation in the field of software engineering. Some attempts to replicate the same studies with both student and professional subjects even produced similar results [226]. However, to be conservative, we refrain from generalizing our results to professionals practicing ASD process. In future studies, we plan to replicate the experiment with professionals. Another threat to the external validity of our study is the representativeness of the tasks proposed for the students to work on. As tasks are proposed by each team during their Sprint planning meetings based on the objectives of their respective projects, we do not have control over the types, priorities, difficulties, and expected deadlines of the tasks involved in the study. However, such a situation is similar to what happens in real world Agile programming, and, thus, its impact on the validity of the study should not be overemphasized. In any case, the large scale and long period of time of this study is one way to reduce the effect of this threat. In the future, we plan to replicate our experiments with more well defined tasks of various complexities (possibly from open source software projects).

### *6.5.3 Construct Validity*

A characteristic of our study that might affect its construct validity is that students had limited previous experience with the Scrum ASD approach during the course work. In addition, in most cases, students in the same team had not worked with their team mates before. Therefore, similar to [49], our results might be conservative with respect to the effects of collaboration. In subsequent studies, we will consider involving programmers who have more experience with this development approach and who have worked together before.

## 6.6 Summary

In this chapter, we firstly outline related work in studying the ASD process through both survey-based approaches and activity data-based approaches, and then we discuss the design of our experimental design, the metrics we measure, and the characteristics of the student participants. Through the data we collected from the experiment, we present key empirical findings on the human factors in Scrum ASD process, including decision-making, collaboration, and team morale etc. Finally, the limitations of the study in terms of internal, external, and construct validity are discussed.

# CHAPTER 7

# DISCUSSION OF FUTURE DIRECTIONS

This book sets out to investigate how human factors influence software development process in agile teams based on the analysis of empirical data collected from student's ASD process in four years' time. In this final chapter, we will summarize the research contributions of this work and discuss potential directions for future research.

## 7.1 Discussion

In the traditional plan-based methodologies, people usually follow the plans to participate in the development process, where human factors are usually not a key consideration during the development process. The Agile methodologies firstly introduce human factors into the software development process, which highlight the importance of collaboration, communication and team morale for such processes. However, existing agile methodologies lack good quantitative methods to analyze the impacts of human factors on the process, which rely too much on project managers and team members' intuition and feeling.

This book makes a number of important contributions to human factor research in agile software development. We proposed a set of new models and tools for ASD teams, and conducted a series of empirical data analysis that brings new insights into this field of research. The contributions of this work can be summarized as follows:

1) **In Chapter 3, we integrated Goal Net model into current ASD process.** The proposed light weight Goal Net-based method can be used to model structured stakeholders' requirements based on current user stories collection in ASD process. By three simple steps: 1) defining high level goals by top-down approach, 2) identifying middle level hidden goals by bottom-up approach, and 3) Modeling goal structure by Goal Net approach. *Product Owner* can model all user stories into a Goal Net diagram that provide the whole team a clear picture and a new perspective to see the project requirements. We have conducted some experiments in SE education environment to evaluate the proposed approach against past approach. The results showed that the number and quality of the user stories produced by student teams are significantly improved with the usage of the Goal Net diagrams.

2) **In Chapter 4, we proposed the *SMART* Approach for task allocation in ASD process.** We firstly formulate the ASD process as a distributed constraint optimization problem, and then propose a SMART algorithm that assesses individual developers' situations based on data collected from the development process, and helps them to make situation-aware decisions on which tasks from the backlog to select in real-time. The SMART approach can help ASD automatically streamline the tasks distribution and management process to more efficiently utilize the collective capacity of an agile team, which balances the considerations for quality and timeliness to improve the overall utility derived from an ASD project. Our theoretical analysis and experimental simulation show that the personal task allocation agent equipped with SMART can achieve close to optimally efficient utilization of the developers' collective capacity.

3) **In Chapter 5, we proposed a *FCM*-based affective model for predicting developer's mood swing in ASD process.** As we know, developers' emotion and the mood swing will affect the software development process and software production.

Traditional *Happiness Chart* method depends on people's intuitions but not process data. Our proposed method is based on OCC emotional theory and *Fuzzy Cognitive Maps* (FCMs) model, which predict developer's mood swing curve according to the causal relationship between mood status and consequences of task execution. According to our two simulated case studies, the approach may be used to help team leader to find right developer to implement tasks when some event occurred or some condition is triggered during ASD process.

4) **In Chapter 6, we discussed more empirical insights into the Scrum process for novice Agile teams based on our HASE collaboration development platform and its new data collection method.** By analyzing and visualizing student ASD team members' activity trajectory data collected unobtrusively during normal ASD processes via our HASE APM platform, we discussed the key empirical findings on the human factors in Scrum ASD process, including task allocation decision-making, team member collaboration, personal mood and team morale etc. These results offer new insights into the aspects of agile team collaboration and team morale which have not yet been well studied by existing research. Furthermore, we identified ASD process as a human computation system, which uses human effort to perform tasks that computers are not good at solving. On the other hand, the computers can assist human to make decision more efficiently.

## 7.2 Future Research Directions

In the future, we will continue to build a more powerful APM platform based on current HASE and conduct large-scale studies in university and industry. With APM tools increasingly adopted by ASD teams, they offer researchers a new source of unobtrusively collected behavior data that provide insight into the characteristics and effects of decision-making which cannot be extracted from traditional survey and interview based study methods. In this research, we have demonstrated the scientific research potential of APM tools. This manner of data collection occurs during normal usage of the APM tool and does not introduce additional overhead to the ASD teams. In addition, as team members are not put into a formal situation in which they know they are being studied, the behavior patterns captured by such data is more likely to reflect their real behavior. Through this study, we discovered effects of novice ASD team members' competence and the difficulty of the tasks assigned to them on their workload variation, their confidence, their morale, and the timeliness of completion of these tasks. The findings based on unobtrusively collected behavior data from APM tools are quantitative in nature and can help future research construct computational task allocation decision models useful for building automated ASD task allocation decision support systems.

To our best knowledge, we were the first to study these problems from the perspective of human aspects for agile methodologies. Nevertheless, this is only a start for the development of a full-fledged methodology and many issues remain to be addressed in future work. The following several areas can be extended from our current research:

**1) Develop the goal net chart for requirement management in ASD process**

Combining *User Story* and *Goal Net*, product owner can both efficiently manage specific requirements and accurately grasp high-level real goals behind tasks. GET card method also give team leader the ability to make decisions according to environment variables and goals. We have applied the *Goal Net* based goals/requirements management method to educational practice for students who are in the program of master of software engineering at *College of Software, Beihang University, China.* The result shows that the work productivity and artifact quality can be improved by infusing our method into their course and software development process. Liking very useful and popular burn down chart as a graphical representation of work left to do versus time in ASD, we can develop a new type of goal net chart to represent the stakeholders' hierarchical goals aiming at user stories, which should be automatically generated by APM system with slight configurations by product owner.

**2) Infuse the SMART approach for automatically task allocation in ASD process**

Now the SMART approach we proposed in chapter 4 can be used to recommend tasks to each developer in APM. In the future, we will further develop an evaluation framework or matrix to assess and verify the proposed approach, and some large-scale studies might also be conducted on agile teams from university students and professional developers, the final target is to infuse the approach into HASE to automatically assign task to developer, especially for whom in a distributed agile team.

**3) Develop the mood swing chart for team morale management in ASD process**

Based on OCC theory and FCM model, we have proposed the mood-events affective model for simulating team members' personal EQ status and its trends. The model might be refined from two aspects in the future: 1) adding more emotional concept nodes; and 2) adding more events and activities concept nodes. With more nodes added into the FCM model, it can accurately reflect the real development situation. However, a complex model might also bring more errors and too much effort to understand and use it. So our future work on this direction is to verify the effectiveness of this model or find out some easier way to help the team to manage human emotions and team morale according to development process situations and contexts. In the future, a new type of mood swing chart is also considered to be developed for project manager monitoring each member's mood swing curve, as well as the team morale to make decision when special event or consequence occurs in the process.

On the other hand, currently the weights of FCM models are determined by team members' self-assessment, in which developers are asked to describe relationships among nodes and use simple *IF-THEN* rules [227] to calculated the value. How to learn the weights by system automatically according developers' history behaves and process data can be a new research direction for future work.

**4) Further understanding the impact of more human factors to ASD process**

With this study, we see the start of a series of research on agile software development with ASD activity trajectory data. In future research, we plan to conduct surveys/interviews to understand more in-depth how students in each Scrum team collaborate. We will continue using the HASE platform to collect agile programming activity data over subsequent semesters and expand our data collection effort to include more universities so as to investigate the possible effects of sociocultural factors. And we plan to enhance the HASE Platform to enable more detailed and larger scale studies about other important human factors of the ASD process to be effectively carried out. With more in-depth understanding of the dynamic factors and characteristics in ASD process, we can design more useful and powerful situation-aware decision support system for Agile.